\documentclass[twocolumn,aps,superscriptaddress,prl]{revtex4-1}

\usepackage[caption=false]{subfig}
\usepackage{dcolumn}
\usepackage{amsmath}
\usepackage{bm}
\usepackage{color}
\usepackage[english]{babel}
\usepackage{calrsfs}
\DeclareMathAlphabet{\pazocal}{OMS}{zplm}{m}{n}

\usepackage{graphicx,psfrag} 
\usepackage{natbib}
\usepackage{epstopdf}
\usepackage{appendix}
\usepackage{changes}

\definecolor{mygrey}{gray}{0.35}
\definecolor{myblue}{rgb}{0.2,0.2,0.8}
\definecolor{myzard}{cmyk}{0,0,0.05,0}
\definecolor{mywhite}{rgb}{1,1,1}
\definecolor{myred}{rgb}{1,0.,0.3}

\usepackage[colorlinks=true,citecolor=myblue,linkcolor=myred]{hyperref}

\def\be{\begin{equation}}
\def\ee{\end{equation}}
\def\ba{\begin{align}}
\def\enda{\end{align}}
\def\bi{\begin{itemize}}
\def\ei{\end{itemize}}

\def\beq{\begin{equation}}
\def\beq{\begin{equation}}
\def\eeq{\end{equation}}

 \newcommand{\ket}[1]{|#1\rangle}

\graphicspath{{./Figures/}}

\begin{document}
\title[Short Title]{Confined nano-NMR spectroscopy using NV centers}
\author{D. Cohen}
\email{email: daniel.cohen7@mail.huji.ac.il}
\affiliation{Racah Institute of Physics, The Hebrew University of Jerusalem, Jerusalem 
	91904, Givat Ram, Israel}
\author{R. Nigmatullin}
\affiliation{Center for Engineered Quantum Systems, Dept. of Physics \& Astronomy, Macquarie University, 2109 NSW, Australia}
\author{M. Eldar}
\author{A. Retzker}
\email{email: retzker@phys.huji.ac.il}
\affiliation{Racah Institute of Physics, The Hebrew University of Jerusalem, Jerusalem 
	91904, Givat Ram, Israel}

	\begin{abstract}
	Nano-NMR spectroscopy with nitrogen-vacancy centers holds the potential to provide high resolution spectra  of minute samples. This is likely to have important implications for chemistry, medicine and pharmaceutical engineering.  One of the main hurdles facing the technology is that diffusion of unpolarized liquid samples broadens the spectral lines thus limiting resolution. Experiments in the field are therefore impeded by the efforts involved in achieving high polarization of the sample which is a challenging endeavor. Here we examine a scenario where the liquid is confined to a small volume. We show that the confinement 'counteracts' the effect of diffusion, thus overcoming a major obstacle to the resolving abilities of the NV-NMR spectrometer.
	
	\end{abstract}
	\maketitle
	\section{INTRODUCTION}
	Nano Nuclear Magnetic Resonance (nano-NMR) spectroscopy is a new field that aimes to implement methods similar to classic NMR techniques on minute samples.
	The new technology may outperform microscopy methods and enable high resolution imaging at the single molecule level. Therefore, advances in the field will have major implications for fundamental science and pave the way for drug and medicine development. 
	
\begin{figure*}
\subfloat[]{\includegraphics[width=0.48\textwidth]{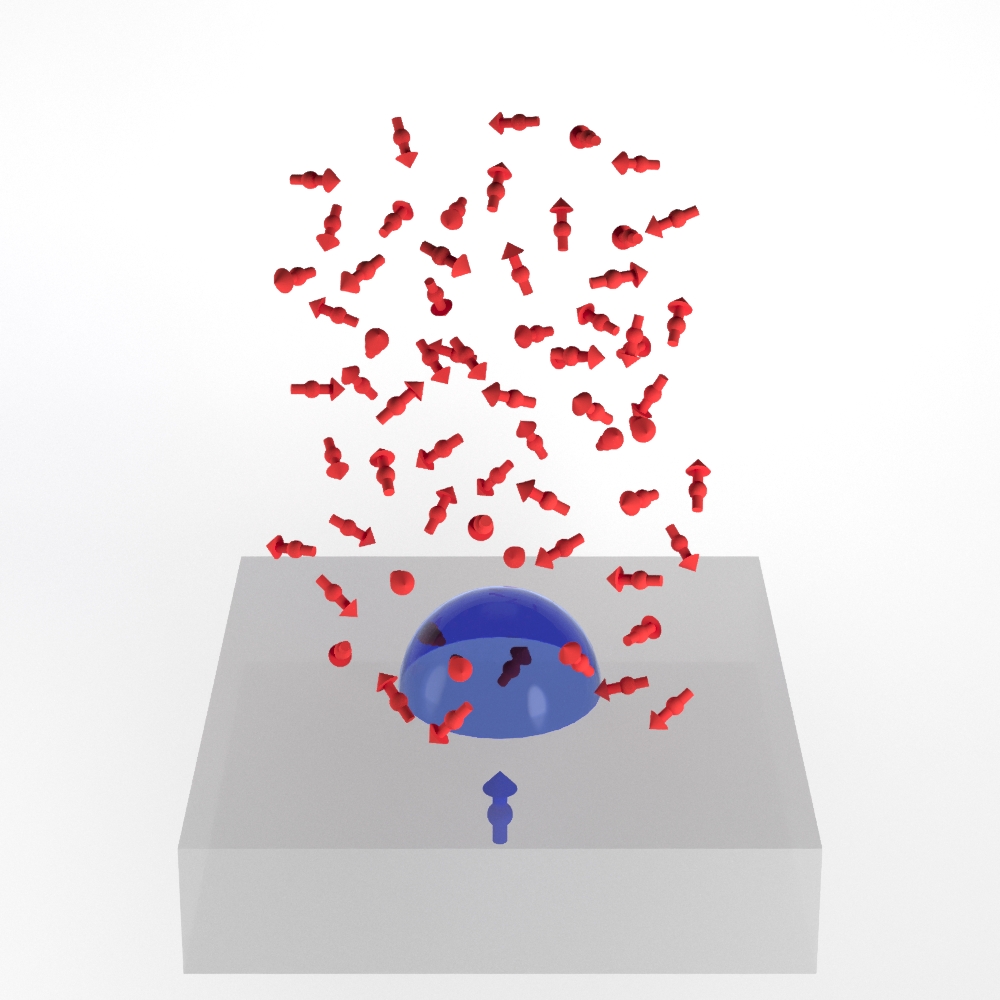} \label{System_classic}}
\hfill
\subfloat[]{\includegraphics[width=0.48\textwidth]{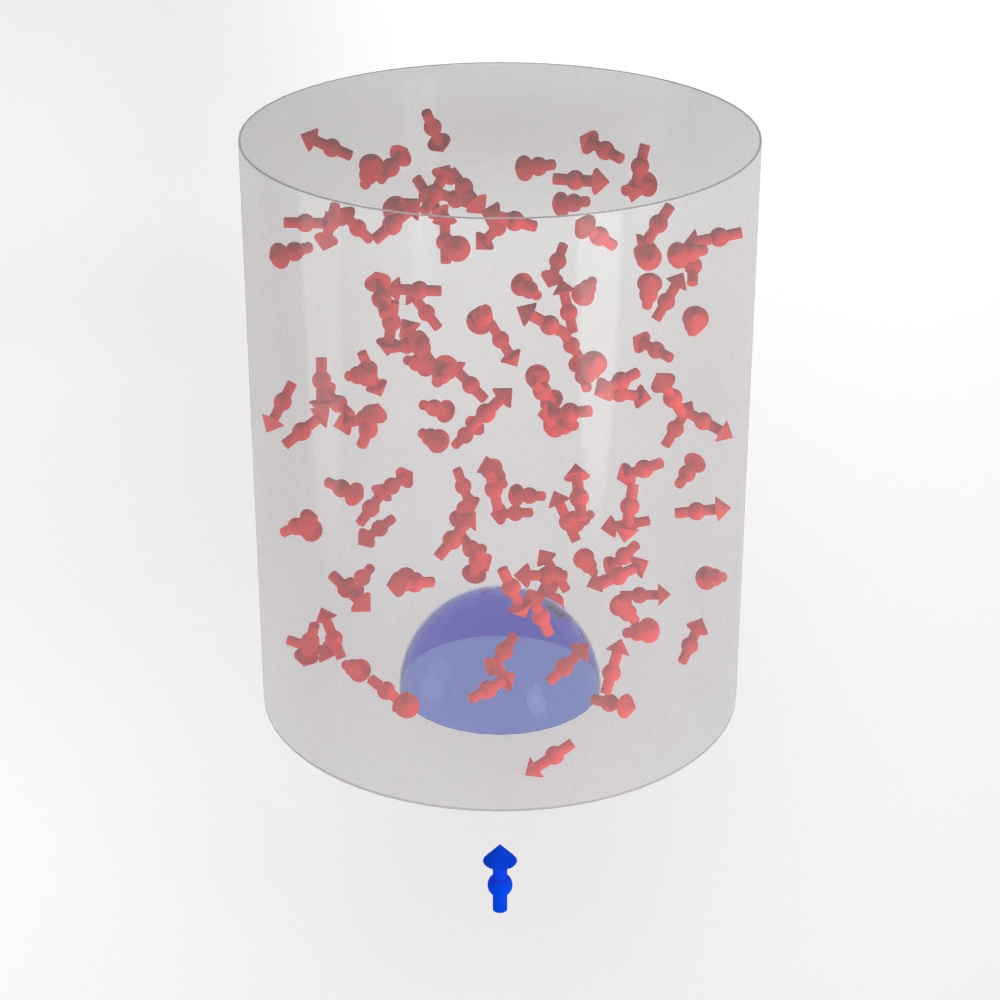} \label{Confined_system}}
\caption{(a) Nano-NMR using NV centers - the nuclear spin sample (red arrows) is placed on top of a diamond surface. The ensemble interacts with the NV center (blue arrow), which is situated at a distance $d$ below the surface. The dioplar interaction creates an effective cutoff, such that only spins that are found in a hemisphere of radius $\sim d$ (blue dome) contribute to the magnetic field at the NV's position. While the generated field of the statistically polarized spins provides information about their Larmor frequency, after interrogation times $t>\tau_D\equiv\frac{d^2}{D}$, the diffusion of spins outside the effective interaction region causes the field to fluctuate, which eventually leads to the decohernce of the sensor. This results in a resolution problem in nano-NMR using shallow sensors.
(b) Confined nano-NMR - the nuclear spin ensemble is confined to a small space (gray cylinder). For interrogation times $t<\tau_V\equiv \frac{V^{2/3}}{D}$ the dynamics is similar to the non-confined scenario, since the interaction with the walls in negligible. For times $t>\tau_V$, however, the nuclei have a finite probability of returning to the interaction volume, which leads to non-vanishing correlations of the magnetic field. These figures were created by Bar Soffer, Mirage studio.}
\end{figure*}

	Scaling down the sample size requires a different measurement device, because small samples generate magnetic fields that are undetectable by classical sensors.
	The Nitrogen-Vacancy (NV) center based spectrometer is a promising platform for nano-NMR, since the NV is a superb magnetometer of small fields \cite{NV_Mag1,Review_NV}.
	Although many successful experiments using the NV spectrometer have been carried out in recent years \cite{NanoNMRHarward,NanoNMRIBM,NanoNMRMeriles,NanoNMRQdyne,NanoNMRstutgart}, nearly all of them used a large sample size to generate a reasonable signal. 
	The gain from using large volumes has to do with the dipolar interaction between the sensor and the nuclear spin ensemble. This creates an effective cut-off distance which is proportional to the NV's depth $d$, such that only nuclei within a radius of $\sim d$ from the surface contribute to the magnetic field. 
	This limitation suggests that a resolution problem occurs when using shallow NVs since the diffusion time across the detection volume which sets the resolution capability is shorter (see Fig. \ref{System_classic}).

	One way of avoiding the resolution problem is by using an ultra-polarized sample. While some experiments rely solely on statistical polarization, other types of polarizations; e.g. thermal polarization or hyperpolarized non-equilibrium states can be used to amplify the signal and circumvent the effects of diffusion \cite{glenn2018high,NV_ensemble}. It was also shown that given high polarization, the entanglement formed between the shallow sensors and the nuclear spin ensemble can be utilized to further increase precision \cite{Laura_arxiv}.  However, despite considerable efforts to enhance the sample's polarization, current state-of-the art achieves about $1\%$. Going beyond this requires efficient hyperpolarization schemes which are technically hard to implement.
	Recent works suggest practical alternatives by showing that contrary to common assumptions, diffusion noise can be managed by implementing an appropriate measurement protocol. These methods rely on the long lasting correlation of magnetic field noise in liquid state nano-NMR with NV centers \cite{microfludics,Amit-Santi}.
	
	Another elegant and practical approach that is currently being explored by the community is to confine the nuclei in a small volume, thus limiting the effects of diffusion (see Fig. \ref{Confined_system}). 
	Here we provide a detailed theoretical analysis of confined nano-NMR spectroscopy. We calculate the expected average magnetic field and correlations in three different geometries: a cylinder, a hemisphere and a full sphere. We show that for any geometry of volume $V$ the correlations decay to a constant value, resulting in unlimited resolution and an SNR which is a fraction of  $\sim\frac{d^3}{V}$ from the SNR of the power spectrum at low frequency.
		
\begin{figure}
	    \centering
	    \includegraphics[width=0.48\textwidth]{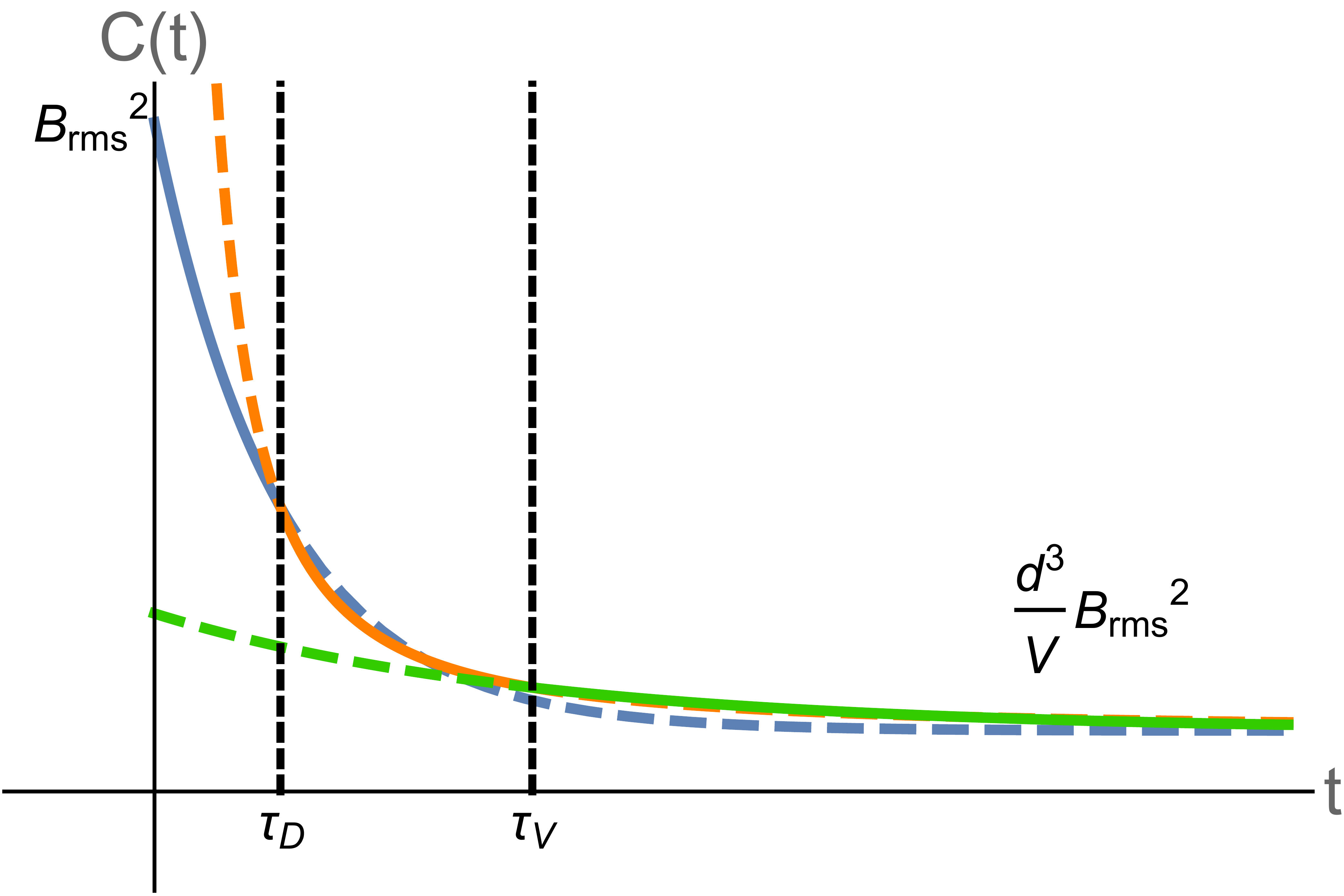}
	    \caption{A sketch of the expected behavior of the correlation function. Solid lines represent the approximate behavior in each regime, and dashed lines are extensions of the approximations to regimes where they are invalid.
	    The value of the correlation at $t=0$ is defined as $B_{rms}^2$. For times $t\ll\tau_D$ the correlation decays exponentially, which is approximately linear for small arguments (blue line). As diffusion becomes dominant at times $t>\tau_D$ the exponential scaling is no longer valid (dashed blue line) and it changes to a power law $\propto t^{-3/2}$ (orange line). When the nuclei interact with the walls at times $t>\tau_V$ the correlation decays exponentially to a constant $\sim\frac{d^3}{V}B_{rms}^2$. }
	    \label{corr_sketch}
	\end{figure}

	
	The advantage of confinement can easily be understood by the following argument.
	Denoting the liquid's diffusion coefficient by $D$, keep in mind that the system has two characteristic time scales: the time it takes a nucleus to diffuse out of the effective interaction region, $\tau_D=\frac{d^2}{D}$, and the time it takes a nucleus to reach the volume's boundary, $\tau_V=\frac{V^{2/3}}{D}$.  
	Assuming that $d<V^{1/3}$, we expect that for short times, $t<\tau_D=\frac{d^2}{D}$, when nuclei can rarely leave the interaction region, the magnetic field at the NV's position will be approximately constant. Therefore, the correlation is also constant. For intermidiate times $\tau_D<t  $, as diffusion becomes dominant, the fluctuations in the magnetic field will cause the correlations to decay, while for $\tau_D \ll t \ll \tau_V$ the correlation will decay as a power law \cite{microfludics}.  However, for longer times $t>\tau_V$, nuclei are reflected from the boundary and therefore have a finite probability to return to the interaction region, which causes the correlation to decay to a constant value. The probability can be estimated by the ratio of the effective interaction region and the total volume $\sim \frac{d^3}{V}$. The expected form of the correlation function is sketched in Fig. \ref{corr_sketch}.
	In the following, we show in detail that these simple arguments indeed describe the correlation of confined nano-NMR.           
	
	\section{RESULTS}
	\subsection{Measurement protocol}
	We explore two measurement schemes: correlation spectroscopy \cite{Corr_spect,kong2015towards,laraoui2013high,Corr_spect_memory1,Corr_spect_memory2} and a phase sensitive measurement (Qdyne) \cite{qdyne,boss2017quantum,glenn2018high}. In the following we provide an outline of the two.  
	We consider the NV center as a two level system (spin $\frac{1}{2}$) with an energy gap of $\gamma_e B_{ext}$, where $B_{ext}$ is the externally applied magnetic field and $\gamma_{e/n}$ is the electron/nuclei gyromagntic ratio \cite{Review_NV}.
    We denote $S_i\ (I_i)$ as the spin operator of the NV (nuclear spin) at the $i$ direction, $S_{\pm}\ (I_\pm)$ as the NV's (nuclear spin's) raising/lowering operators and $\theta_i$ as the angle between the NV's magnetization axis and the vector connecting the NV to the $i$-th nucleus.
    
   In correlation spectroscopy, the NV interacts with nuclear spins for a period of time $\tau$ via the dipolar interaction with the ensemble at the end of which the NV state is written onto a memory qubit or alternatively the relevant information found in the relative phase is mapped to the population difference of the NV. After a waiting time $t$ the experiment is carried out again for the same period of time $\tau$. At the interrogation times the NV is driven with an external pulse sequence at frequency $\omega_p$, which is close to the Larmor frequency of the nuclei $\omega_N$, such that $\left(\omega_N-\omega_p\right) t \equiv \delta t \ll 1$. The Hamiltonian is effectively 
    \beq\label{corr1}
    H_{eff}=\gamma_e B(t)S_z\cos\left(\delta t+\phi\right),
  \eeq
  where $\phi$ is a random phase deriving from the unpolarized dynamics.
  Initializing the NV in the state $\ket{\psi_0}=\frac{1}{\sqrt{2}}\left(\ket{\uparrow}+\ket{\downarrow}\right)$, after both interrogation periods, the probability of measuring the NV at the $\ket{\psi_0}$ state is
 	\begin{align}\label{corr2}
	&P_{\ket{\psi_0}}\approx 1-\left(2\gamma_e^2\tau\right)^2\left[B^2\left(0\right)\cos^2\left(\phi\right)\right.\\\nonumber
	&\left.+B^2\left(t\right)\cos^2\left(\delta t+\phi\right)+2B\left(0\right)B\left(t\right)\cos\left(\phi\right)\cos\left(\delta t+\phi\right)\right],
	\end{align}
	where we took the weak back-action limit, namely $\gamma_e B(t) \tau \ll 1$.
	Averaging \eqref{corr2} over realizations yields
	\beq\label{corr_spec_prob}
		P_{\ket{\psi_0}}\approx 1-\left(2\gamma_e B_{rms}\tau\right)^2\left[1+\cos\left(\delta t\right)\frac{C^{(1)}\left(t\right)}{B_{rms}^2}\right],
	\eeq 	
	where $C^{(1)}\left(t\right)=\left<B\right(t\left)B\right(0\left)\right>$ is the correlation function of the magnetic field at time $t$
	thus $C\left(0\right)\equiv B_{rms}^2$. The correlation is given by \cite{microfludics,Supp}
	\beq\label{corr_general}
	C^{(m)}\left(t\right)=\tilde{\zeta}_m^2 \int\frac{d^3r}{r^3}\int\frac{d^3r_0}{r_0^3}Y_2^{(m)}\left(\Omega\right)Y_2^{(m)*}\left(\Omega_0\right)P\left(\bar{r},\bar{r}_0,t\right),
	\eeq
	where $Y_l^{(m)}$ are the spherical harmonics,  $P\left(\bar{r},\bar{r}_0,t\right)$ is the diffusion propagator from $\bar{r}$ to $\bar{r}_0$ with time difference $t$ and $\tilde{\zeta}_m$ are defined in \cite{Supp}. For $m=1$, for example, $\tilde{\zeta}_1=\frac{3}{2}\left(2\sqrt{\frac{2\pi}{15}}\right)\left(\frac{\hbar\mu_0\gamma_N}{4\pi}\right)$. For a detailed derivation of Eq. \eqref{corr_spec_prob} see \cite{Supp}.  
	We note that different values of $m$ can be sampled by changing the dynamical decoupling pulse frequency. For example, $m=0$ can be sensed by not applying any pulses ($\omega_p=0$). 
	
	In a phase sensitive measurement we employ a similar protocol, where the main difference stems from the fact that we measure multiple times in one realization; every period $\tau$ the NV is measured and then initialized at the state $\ket{\psi_0}$.
	The evolution of the system between times $t_n\equiv n\tau$ and $t_{n+1}=\left(n+1\right)\tau$ for some integer $n$ is
	\beq
	\ket{\psi_{t_{n+1}}}=\frac{1}{\sqrt{2}}\left[\exp\left(-2i\gamma_e\tau B\left(t_n \right)\cos\left(\delta t_n +\phi\right)\right)\ket{\uparrow}+\ket{\downarrow}\right],
	\eeq
	where we assumed $\delta\tau\ll1$. The probability of measuring $\ket{\uparrow_y}=\frac{1}{\sqrt{2}}\left(\ket{\uparrow}+i\ket{\downarrow}\right)$ is therefore
	\beq\label{Qdyne_prob1}
	P_{\ket{\uparrow_y}}\left(t_n\right)=\frac{1}{2}\left(1+\sin\left[\Phi\left(t_n\right)\right]\right)
	\eeq
	with $\Phi\left(t_n\right)=2\gamma_e \tau  B\left(t_n \right)\cos\left(\delta t_n+\phi\right)$. Assuming $\Phi\left(t_n\right)\ll1$ we can approximate \eqref{Qdyne_prob1} by
	\beq\label{Qdyne_prob2}
	P_{\ket{\uparrow_y}}\left(t_n\right)\approx\frac{1}{2}\left[1+\Phi\left(t_n\right)\right].
	\eeq
	The joint probability is then 
	\begin{align}\label{Qdyne_prob3}
	P_{\ket{\uparrow_y}}\left(t_n\right)P_{\ket{\uparrow_y}}\left(t_k\right)\approx\frac{1}{4}\left[1+\Phi\left(t_n\right)+\Phi\left(t_k\right)+\Phi\left(t_n\right)\Phi\left(t_k\right)\right].
	\end{align}
	When combined with the probability of both results being $\vert\downarrow_y\rangle$, an extra factor of 2 is gained.
	Averaging \eqref{Qdyne_prob3} over the different measurements gives
	\begin{align}\label{Qdyne_prob4}
	&\left<P_{\ket{\uparrow_y}}\left(t_n\right)P_{\ket{\uparrow_y}}\left(t_k\right)\right>\approx\\\nonumber
	&\frac{1}{4}\left[1+\left(2\gamma_e\tau\right)^2 C^{(m)}\left(t_{n-k}\right)\cos\left(\delta t_{n-k}\right)\right].
	\end{align}
	This can be used to define a new Bernoulli variable by 
	\begin{align}\label{qudyne_prob_final}
	&P=\left<P_{\ket{\uparrow_y}}\left(t_n\right)P_{\ket{\uparrow_y}}\left(t_k\right)\right>+\left<P_{\ket{\downarrow_y}}\left(t_n\right)P_{\ket{\downarrow_y}}\left(t_k\right)\right>\\\nonumber
	&=\frac{1}{2}\left[1+\left(2\gamma_e\tau\right)^2 C^{(m)}\left(t_{n-k}\right)\cos\left(\delta t_{n-k}\right)\right],
	\end{align}
	where the value of $m$ in Eqs. \eqref{Qdyne_prob4} and \eqref{qudyne_prob_final} is determined by the external pulse frequency. 
	In the following we present the calculation magnetic field correlations in three different geometries. We provide analytic calculations for the asymptotic behavior and numerical assessments for arbitrary times. Since the asymptotic behavior of the correlation functions is universal, we focus on the case $m=0$. The calculation of the other correlation functions can be found in \cite{Supp}.
	
	\subsubsection{Cylinder}
	Given a statistically polarized ensemble which is confined to a cylinder with height $L$ and radius $R$ and an NV that is located at a depth $d$ below the cylinder's base. The NV's position coincides with the cylinder's symmetry axis denoted as $\hat{z}$.  Assuming the NV's magnetization axis is along the $\hat{z}$ axis, the magnetic field correlation is given by \eqref{corr_general},
	where the different $m$ values can be probed by changing the pulse frequency mentioned in the previous section, see details in \cite{microfludics}. The instantaneous correlation, $B_{rms}^2=C^{(0)}\left(t=0\right)$ can be calculated analytically as the propagator approaches a delta function, yielding:
	\begin{align}\label{Brms_cylinder0}
    &\frac{B_{rms}^2}{J^2}=\frac{\pi}{64}\left\{\frac{1}{R^{2}}\left[\frac{23d}{d^{2}+R^{2}}+\frac{24d^{5}}{\left(d^{2}+R^{2}\right)^{3}}\right.\right.\\\nonumber
    &+\frac{16LR^{2}\left(3d^{2}+3dL+L^{2}\right)}{d^{3}(d+L)^{3}}-\frac{38d^{3}}{\left(d^{2}+R^{2}\right)^{2}}\\\nonumber
    &\left.-\frac{23(d+L)}{(d+L)^{2}+R^{2}}+\frac{38(d+L)^{3}}{\left((d+L)^{2}+R^{2}\right)^{2}}-\frac{24(d+L)^{5}}{\left((d+L)^{2}+R^{2}\right)^{3}}\right]\\\nonumber
    &+\frac{1}{R^{3}}\left[105\tan^{-1}\left(\frac{R}{d+L}\right)+96\tan^{-1}\left(\frac{d+L}{R}\right)\right.\\\nonumber
    &\left.\left.-105\tan^{-1}\left(\frac{R}{d}\right)-96\tan^{-1}\left(\frac{d}{R}\right)\right]\right\},
\end{align}
where we denote the physical coupling constant by $J=\left(\frac{\hbar\mu_0\gamma_N}{4\pi}\right)$. 
Eq. \eqref{Brms_cylinder0} reproduces the well-known scaling, $B^2_{rms}\propto J^2d^{-3}$, of the semi-infinite volume in the limit $d\ll R,L$, as expected. In the other limit, $d\gg R,L$, Eq. \eqref{Brms_cylinder0} can be approximated by $B^2_{rms}\propto J^2\frac{V}{d^6}$. This is because as the statistical polarization scales as $\sqrt{V}$, the dipole-dipole interaction decays as $\frac{1}{d^3}$  and ${B_{RMS}}$ scales with both.   

The correlation \eqref{corr_general} decays asymptotically to a constant as the propagator approaches the uniform distribution for times $t\gg\tau_V$,
\begin{align}\label{longtimes_cylinder0}
 \frac{C^{(0)}(t\gg \tau_V)}{J^2}\approx\frac{4\pi^2}{V} \left(\frac{L+d}{\sqrt{\left(L+d\right)^2+R^2}}-\frac{d}{\sqrt{d^2+R^2}}\right)^2.
 \end{align}
 For a sufficiently shallow sensor, $d\ll R,L$ Eq. \eqref{longtimes_cylinder0} approaches $C^{(0)}\propto B^2_{rms}\frac{d^3}{V} = \frac{J^2}{V}$. The factor of $\frac{d^3}{V}$ in this limit can be interpreted as the probability that a nucleus will be found in the effective interaction region, because at long times, diffusion spreads out the nuclei uniformly. Since the signal $B_{rms}^2\propto d^{-3}$,  
 remarkably, the correlation does not depend on the NV's depth. This regime ($t\gg\tau_V$), where the signal remains constant, is reminiscent of the situation of the unbounded region with a polarized sample, wherein the constant correlation is replaced by the average magnetic field.
 
At arbitrary times, \eqref{corr_general} is given by
\begin{align}\label{correlation_cylinder_temporal_decay}
 &\tilde{C}^{(m)}=\frac{d^3}{V}\sum_{s=1}^{\infty}\sum_{k=0}^{\infty}\frac{2}{J_{m}^{2}\left(\nu_{s,m}\right)\left(1-\left(\frac{m}{\nu_{s,m}}\right)^{2}\right)\left(1+\delta_{k,0}\right)}\times\\\nonumber
 &e^{-t/\tau_{s,k,m}}\left[\int\frac{d^3r}{r^3}\cos\left[\pi k\left(\frac{z-d}{L}\right)\right]J_m\left(\nu_{s,m}\frac{\rho}{R}\right)\left|Y_2^{(m)}\left(\Omega\right)\right|\right]^2,
 \end{align}
where $\tilde{C}^{(m)}=\frac{\left[C^{(m)}(t)-C^{(m)} (t\gg\tau_V)\right]}{C^{(m)(0)}}$, $J_m$ is the $m$th Bessel function of the first kind, $\nu_{k,m}$ is the $k$th zero of $J_m'$, $\delta_{i,j}$ is Kronecker's delta and the characteristic decay time $\tau_{s,k,m}=\left[D\left(\frac{\nu_{s,m}}{R}\right)^{2}+D\left(\frac{\pi k}{L}\right)^{2}\right]^{-1}$.
Note that while Eq. \eqref{corr_general} might be very hard to estimate even numerically, Eq. \eqref{correlation_cylinder_temporal_decay} only contains a sum over two dimensional integrals (since the azimuthal integration is trivial) which are easy to calculate.

\begin{figure*}
	\subfloat[]{\includegraphics[width=0.32\textwidth]{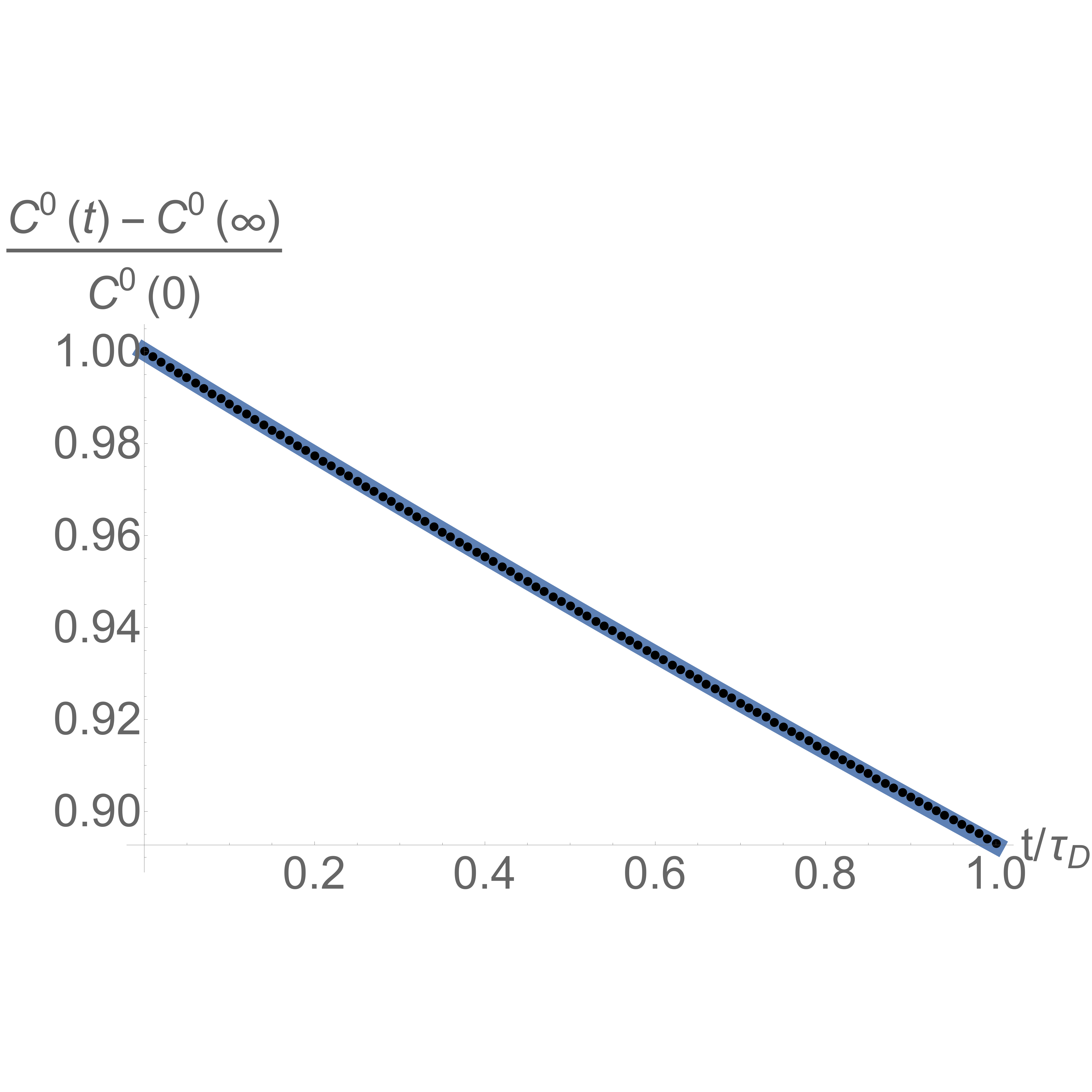} \label{cylinder200short}}
	\subfloat[]{\includegraphics[width=0.32\textwidth]{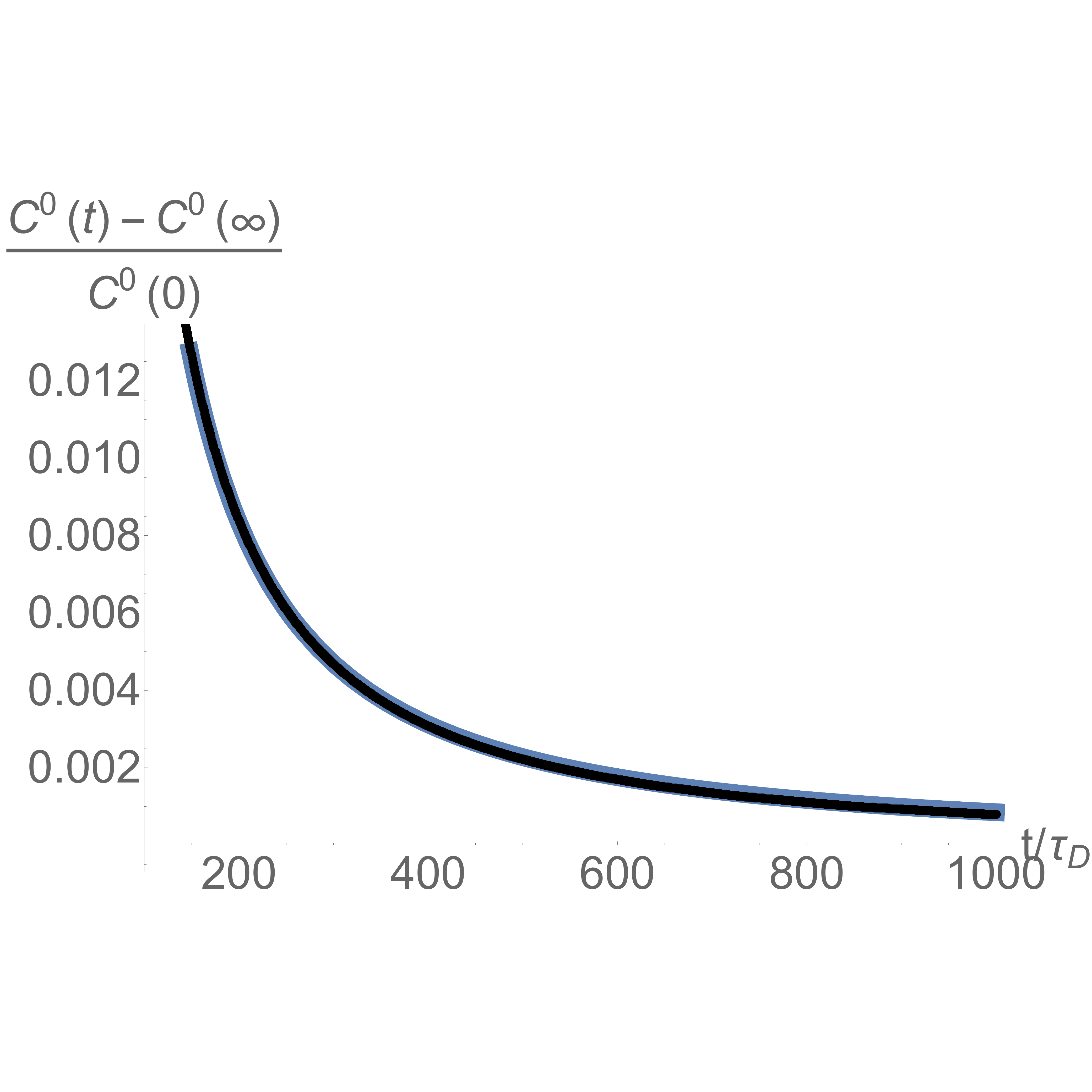} \label{cylinder200inter}}
	\subfloat[]{\includegraphics[width=0.32\textwidth]{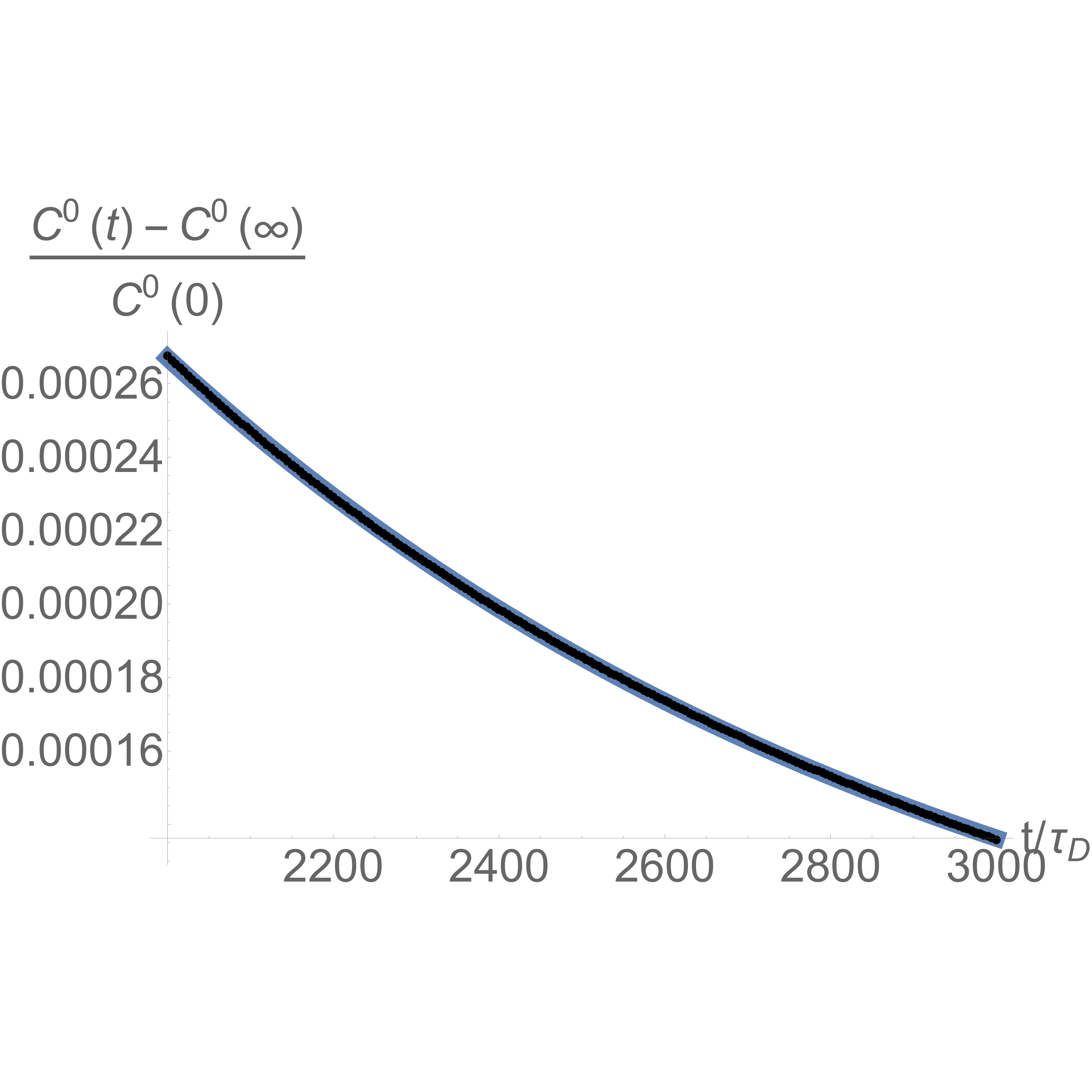} \label{cylinder200long}}
	\caption{The correlation function in \textit{cylindrical} geometry \eqref{correlation_cylinder_temporal_decay}  (black points) for $m=0$ in the three time regimes - (a) $t\ll\tau_D$, (b) $\tau_D\ll t\ll \tau_V$ and (c) $\tau\gg\tau_V$ (right) . The correlation was calculated for $d=1\ \textrm{nm}, \ D=0.5\ \frac{\textrm{nm}^2}{\mu\textrm{s}}$ and $R=L=200 \ \textrm{nm}$. The short times limit was fitted to a linear function, the intermediate times to a power law  and long times were fitted to an exponential function (blue lines). 
	The power in the intermediate regime fit the expected $-1.5$ scaling predicted in previous works, 
	and the decay rate at long times fit the slowest decaying mode of \eqref{correlation_cylinder_temporal_decay}. \label{corr_cylinder_m0}}
\end{figure*}

We evaluated \eqref{correlation_cylinder_temporal_decay} for $k,s\in[1,25]$, the diffusion coefficient of immersion oil $D=0.5\ \frac{\textrm{nm}^2}{\mu \textrm{s}}$, $d=1\ \textrm{nm}$ and different volumes \cite{Supp}. In Fig \ref{corr_cylinder_m0} we present a representative case where $R=L=200\ \textrm{nm}$. The correlation at short times decays linearly, which agrees with an exponential decay for small arguments. In the intermediate regime the decays changes to a power law $\sim t^{-1.5}$ as predicted in previous works \cite{microfludics}, while  
at long times the decay rate fits the slowest decaying mode of \eqref{correlation_cylinder_temporal_decay}.

The numerical integration was done with a diffusion coefficient fitting immersion oil, since state-of-the-art experiments are conducted with high viscous fluids primarily for their long diffusion time, which results in a long correlation time and thus a high resolution capability.  For water, for example, resolution is extremely limited, since the correlation vanishes rapidly. In a confined setting, however, the correlation decays rapidly to a constant value; therefore, diffusion poses no real limitation on the interrogation time thus eliminating the need to use high viscous fluids.

	\subsubsection{Hemisphere}
The $B_{rms}$ of a hemisphere with radius $R$, given by the limit $t\rightarrow0$ of Eq. \eqref{corr_general} is
\begin{align}\label{Brms_hemisphere0}
 &\frac{B_{rms}^2}{J^2}=  \frac{\pi}{32d^{5}}\left[\left(d^{2}-9R^{2}\right)\ln\left(\frac{d^{2}+R^{2}}{\left(d+R\right)^{2}}\right)\right.\\\nonumber
 &+2d^{3}\left(-\frac{5d}{d^{2}+R^{2}}-\frac{4d^{2}}{(d+R)^{3}}-\frac{6d^{5}}{\left(d^{2}+R^{2}\right)^{3}}\right.\\\nonumber
 &\left.\left.+\frac{2d^{3}}{\left(d^{2}+R^{2}\right)^{2}}+\frac{2d}{(d+R)^{2}}-\frac{2}{d+R}\right)+26d^{2}-18dR\right].
 \end{align}
 The limits of \eqref{Brms_hemisphere0} are similar to those of the cylindrical case.
 The asymptotic behavior at long times is
 \begin{align}\label{longtimes_hemisphere0}
 &\frac{C^{(0)}(t\gg \tau_V)}{J^2}\approx\frac{4\pi^2}{9V}\left[2-\frac{2d}{\sqrt{d^{2}+R^{2}}}\right.\\\nonumber
 &\left.-\frac{R^{2}}{d\sqrt{d^{2}+R^{2}}}-\frac{2R^{3}}{d^{3}}\left(\frac{R}{\sqrt{d^{2}+R^{2}}}-1\right)\right]^2.
 \end{align}
 Taking $d\ll R$, Eq. \eqref{longtimes_hemisphere0} reduces to $\frac{16 \pi^2}{9V}\propto B_{rms}^2\frac{d^3}{V}$.
 In general the correlation can be expressed by the sum
 \begin{align}\label{corr_hemisphere_temporal_decay}
&\tilde{C}^{(m)}=\frac{d^3}{V}\sum_{l=0}^{\infty}\sum_{k=1}^{\infty}\frac{4\pi\left[1+\left(-1\right)^{l+m}\right]}{3j_{l}^{2}\left(\tilde{\nu}_{k,l}\right)\left(1-\left(\frac{l}{\tilde{\nu}_{k,l}}\right)^{2}\right)}\times\\\nonumber
&\left[\int \frac{d^3r}{\tilde{r}^3}j_{l}\left(\frac{\tilde{\nu}_{k,l}}{R}r\right)Y_2^{(m)*}\left(\tilde{\theta},\tilde{\phi}\right)Y_{l}^{(m)}\left(\theta,\phi\right)\right]^2\exp\left(-\frac{\tilde{\nu}_{k,l}^{2}}{\tau_{V}}t\right),
 \end{align}
 where $j_m$ is the $m$th spherical Bessel function of the first kind and $\tilde{\nu}_{k,m}$ is the $k$th zero that does not equal zero of $j_m'$ (namely, $\tilde{\nu}_{1,m}\neq0$). The coordinates $\bar{r} / \bar{\tilde{r}}$ are the spherical coordinates where the origin is found at the center of the hemisphere's base \ at the NV's position.
 The results are similar to those of the cylindrical geometry and are found in \cite{Supp}.
\subsection{Full sphere}
\begin{figure*}
	\subfloat[]{\includegraphics[width=0.32\textwidth]{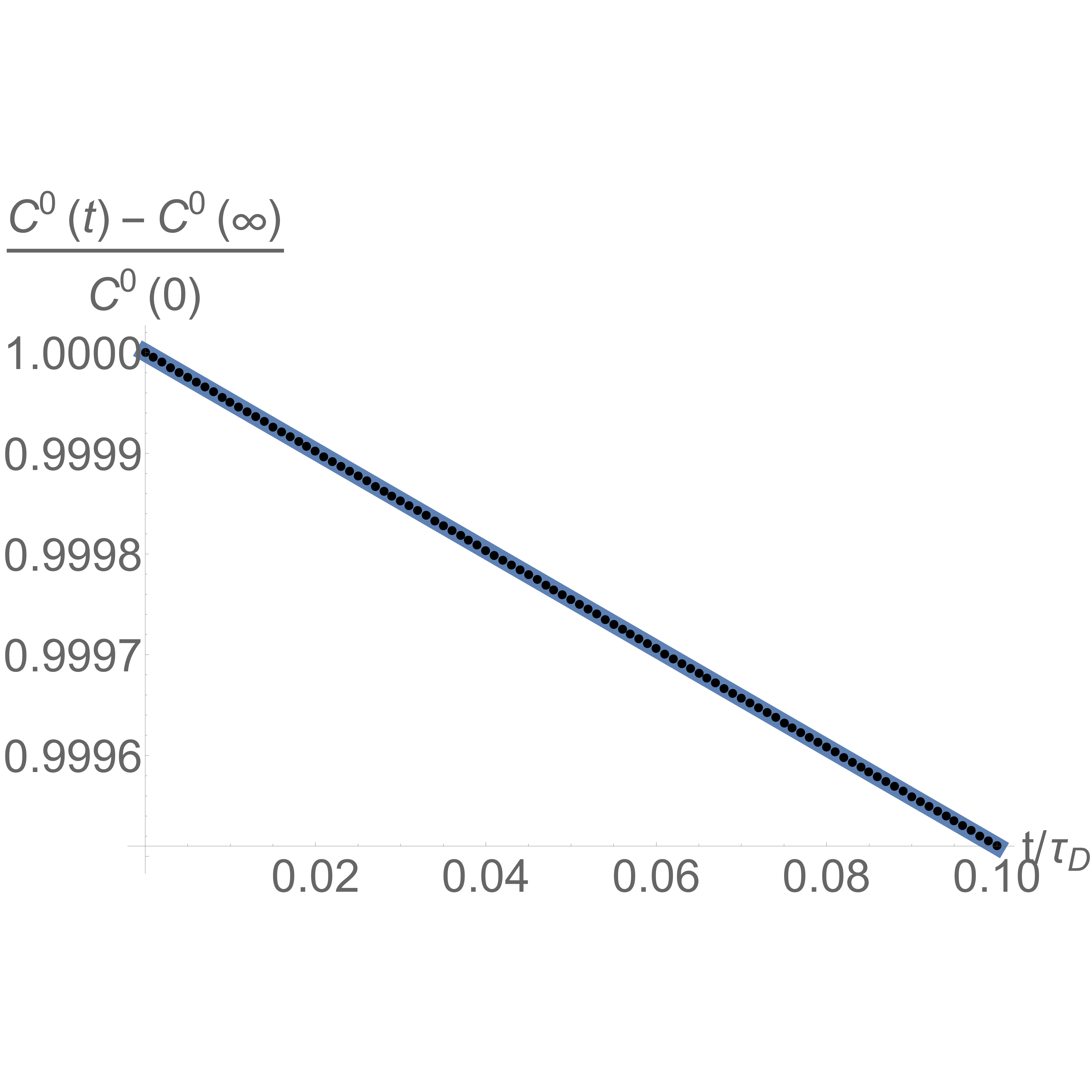} \label{sphere200short}}
	\subfloat[]{\includegraphics[width=0.32\textwidth]{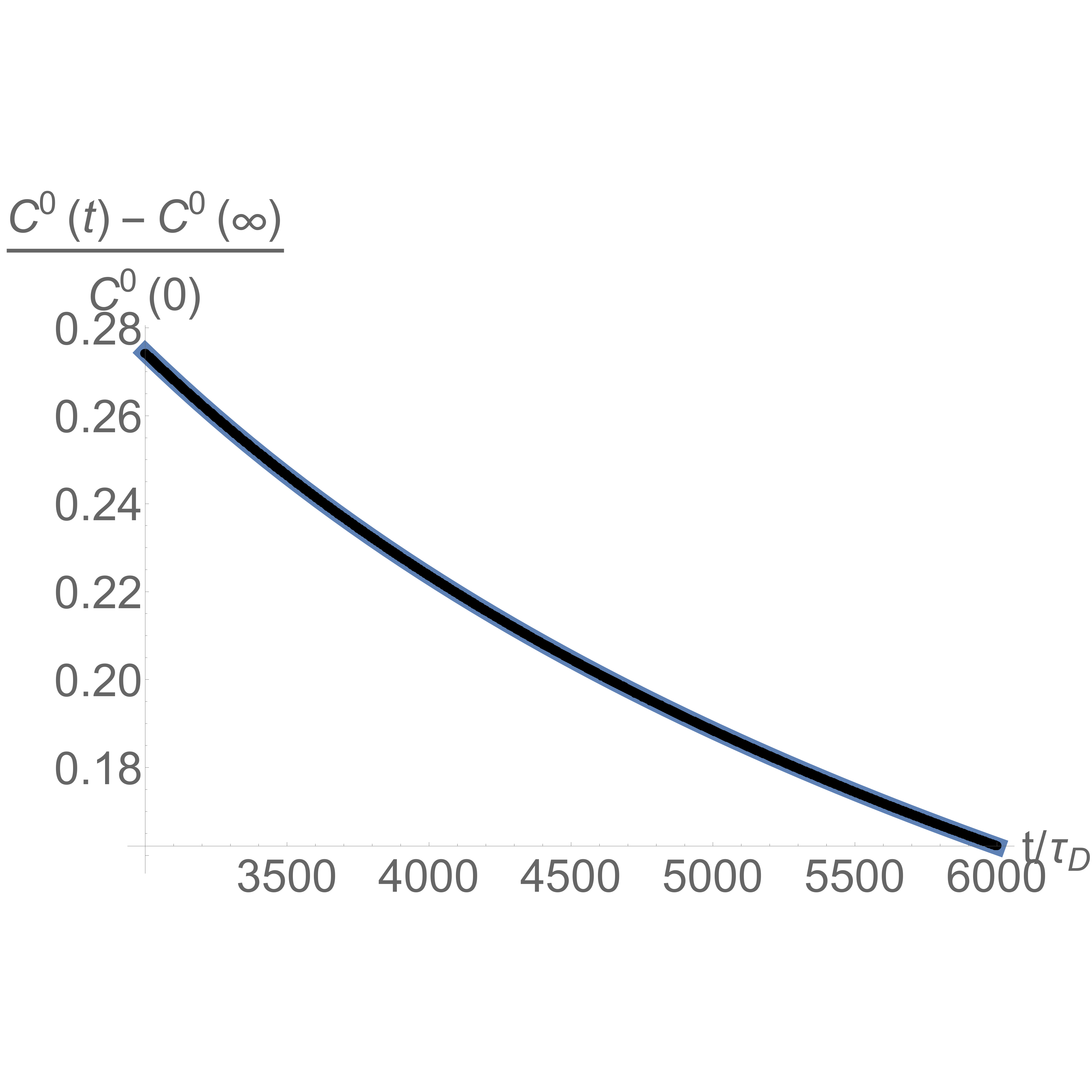} \label{sphere200inter}}
	\subfloat[]{\includegraphics[width=0.32\textwidth]{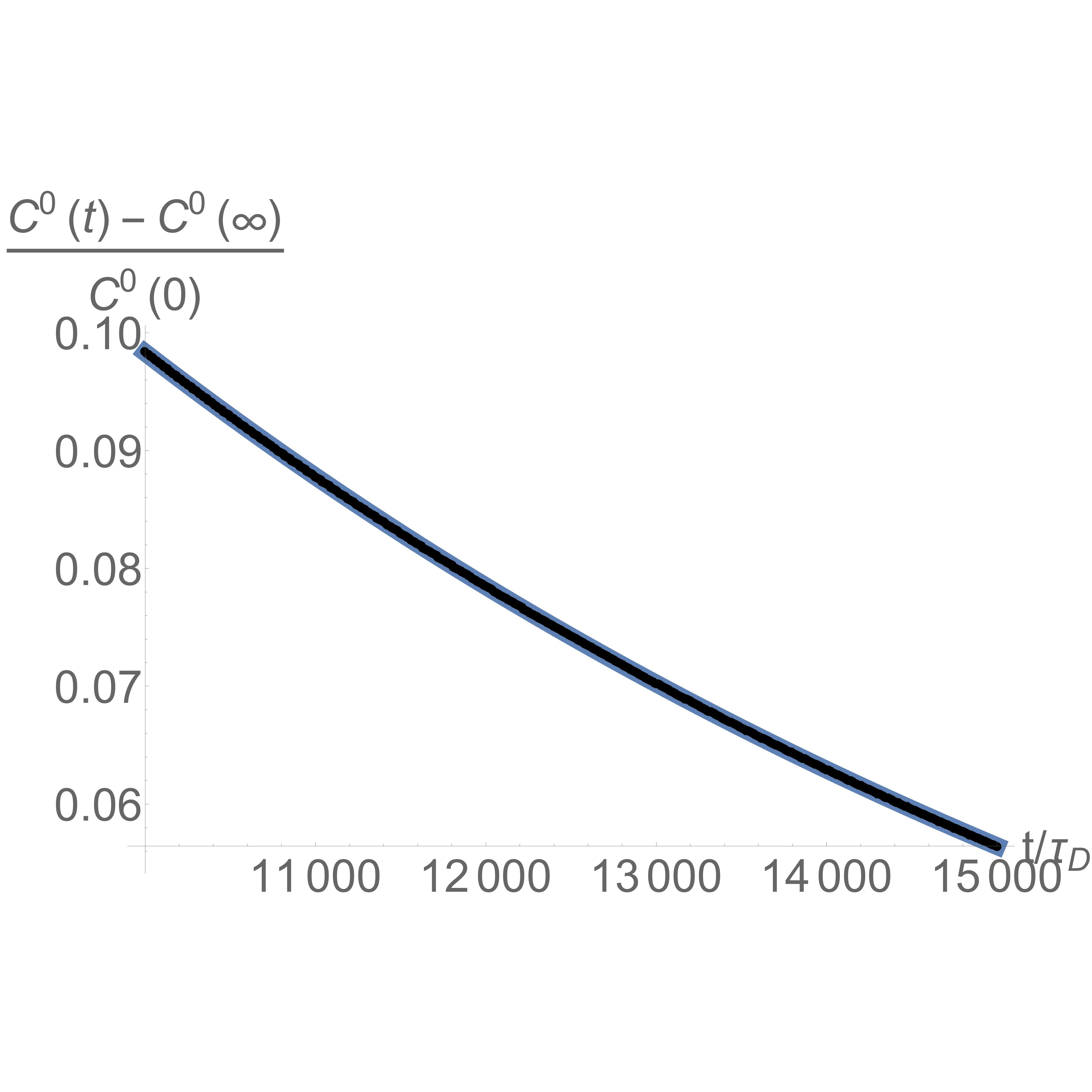} \label{sphere200long}}
	\caption{The correlation function in \textit{spherical} geometry \eqref{corr_sphere_temporal_decay}  (black points) for $m=0$ in the three time regimes - (a) $t\ll\tau_D$, (b) $\tau_D\ll t\ll \tau_V$ and (c) $\tau\gg\tau_V$ (right) . The correlation was calculated for $d=1\ \textrm{nm}, \ D=0.5\ \frac{\textrm{nm}^2}{\mu\textrm{s}}$ and $R=L=200 \ \textrm{nm}$. The short times limit was fitted to a linear function, the intermediate times to a power law  and long times were fitted to an exponential function (blue lines). 
	The results for short times and long times are similar to those of the previous geometries, while the power in the intermediate regime fits a $-0.5$ scaling. \eqref{corr_sphere_temporal_decay}. \label{corr_sphere_m0}}
\end{figure*}

Given a spherical geometry of radius $R$, the instantaneous correlation is
\begin{align}
&\frac{B_{rms}^2}{J^2}=\frac{\pi}{8d^{3}(d+R)^{5}(d+2R)^{3}}\left[d^{3}\left(-d^{5}-8d^{4}R-16d^{3}R^{2}\right.\right.\\\nonumber
&\left.+16d^{2}R^{3}+80dR^{4}+64R^{5}\right)\tanh^{-1}\left(\frac{R}{d+R}\right)+d^{7}R\\\nonumber
&+7d^{6}R^{2}+52d^{5}R^{3}+190d^{4}R^{4}+320d^{3}R^{5}+256d^{2}R^{6}\\\nonumber
&\left.+96dR^{7}+16R^{8}\right].
 \end{align}
The asymptotic behavior of \eqref{corr_general} is then given by
\begin{align}\label{longtimes_sphere0}
  &\frac{C^{(0)}(t\gg \tau_V)}{J^2}\approx\frac{64 \pi ^2}{9V}\frac{R^6}{(d+R)^6}.
  \end{align}
  For arbitrary times, the correlation is 
 \begin{align}\label{corr_sphere_temporal_decay}
&\tilde{C}^{(m)}=\frac{1}{V}\sum_{l=0}^{\infty}\sum_{k=1}^{\infty}\frac{8\pi}{3j_{l}^{2}\left(\tilde{\nu}_{k,l}\right)\left(1-\left(\frac{l}{\tilde{\nu}_{k,l}}\right)^{2}\right)}\times\\\nonumber
&\left[\int \frac{d^3r}{\tilde{r}^3}j_{l}\left(\frac{\tilde{\nu}_{k,l}}{R}r\right)Y_2^{(m)*}\left(\tilde{\theta},\tilde{\phi}\right)Y_{l}^{(m)}\left(\theta,\phi\right)\right]^2\exp\left(-\frac{\tilde{\nu}_{k,l}^{2}}{\tau_{V}}t\right),
 \end{align}
We evaluated \eqref{corr_sphere_temporal_decay} for $l\in[0,30],\ k\in[1,30]$, the diffusion coefficient of immersion oil $D=0.5\ \frac{\textrm{nm}^2}{\mu \textrm{s}}$, $d=1\ \textrm{nm}$ and different volumes \cite{Supp}. In Fig \ref{corr_sphere_m0} we present a representative case where $R=200\ \textrm{nm}$. The correlation at short times and long times agrees with the results of the cylindrical geometry. In the intermediate regime the decay changes to a power law $\sim t^{-0.5}$. 
	
	\subsection{Sensitivity analysis}
	In our sensitivity analysis we used the tools of classic parameter estimation. We calculated the Fisher Information (FI), denoted by $\pazocal{I}$; hence, the sensitivity can then be estimated with the Cramer-Rao bound \cite{Information_theory}
	\beq\label{CR_bound}
	\Delta\delta\geq\frac{1}{\sqrt{\pazocal{I}}}.
	\eeq
	
	We start by calculating the asymptotic behavior of the FI in either correlation spectroscopy or a phase sensitive measurement. In the limits $t\le \tau_D$ and $t\gg\tau_V$ the correlation is approximately a constant. Note that the regime $t\le \tau_D$ is similar  to the unconfined case. The FI of correlation spectroscopy is given by
	\begin{eqnarray}\label{corr_spec_fisher}
	\pazocal{I}&=&\frac{\left(\frac{d P_{\ket{\Psi_0}}}{d\delta}\right)^2}{P_{\ket{\Psi_0}}\left(1-P_{\ket{\Psi_0}}\right)}\frac{T}{t}\nonumber \\&=&\left(2\gamma_e t \tau\right)^2\left(\frac{C}{B_{rms}}\right)^2\frac{\sin^2\left(\delta t\right)}{1+\frac{C}{B_{rms}^2}\cos\left(\delta t\right)}\frac{T}{t},
	\end{eqnarray}
	where $T$ is the total time, and we used Eq. \eqref{corr_spec_prob} for this derivation. Since $C\left(t\ll\tau_D\right)\approx B_{rms}^2$ we can deduce that for $t\approx \tau_D$,
	\beq\label{corr_spec_fisher_short}
	\pazocal{I}_{UCL}\approx2\left(\gamma_e B_{rms} \delta \tau \tau_D^2\right)^2 \frac{T}{\tau_D}.
	\eeq
	Eq. \eqref{corr_spec_fisher_short} is the unconfined limit (UCL), since without confinement the measurement time is bounded by $\tau_D$.
	For times $t\gg\tau_V$, on the other hand, $C\left(t\gg\tau_V\right)\approx \frac{d^3}{V} B_{rms}^2$ and therefore
	\beq\label{corr_spec_fisher_long}
	\pazocal{I}_{C}\approx\left(2\gamma_e B_{rms} T_m \tau\right)^2\left(\frac{d^3}{V}\right)^2\sin^2\left(\delta T_m\right) \frac{T}{T_m},
	\eeq
    where $T_m$ is the memory time.
	In the case where $\delta T_m > \frac{\pi}{2}$
    Eq. \eqref{corr_spec_fisher_long} can be simplified to
	\beq\label{corr_spec_fisher_long_max}
	\pazocal{I}_{C}\approx\left(2\gamma_e B_{rms}  \tau\right)^2\left(\frac{d^3}{V}\right)^2 T_m T.
	\eeq
	Comparing Eq. \eqref{corr_spec_fisher_long_max} to Eq. \eqref{corr_spec_fisher_short} shows that we gained a factor of 
	\beq\label{enhancment}
	\frac{\pazocal{I}_{C}}{\pazocal{I}_{UCL}}=\frac{T_m}{\tau_D}\frac{1}{\left(\delta\tau_D\right)^2} \left(\frac{d^3}{V}\right)^2.
	\eeq
	Since $T_m$ can be orders of magnitude larger than $\tau_D$ and $\delta \tau_D$ might be very small, this can lead to considerable enhancement in sensitivity. Numerical evaluations of Eq. \eqref{enhancment} are found at the last section.
	
	In a Qdyne type measurement the probability \eqref{qudyne_prob_final} leads to
	\beq\label{single_time_qdyne}
	\pazocal{I}_s=16\left(\gamma_e\tau\right)^4 C^2\left(t_s\right)t_s^2\sin^2\left(\delta t_s\right).
	\eeq
	For times $t_s\gg\tau_V$, the Fisher information \eqref{single_time_qdyne} is 
	\beq\label{single_time_qdyne_long}
	\pazocal{I}_s=\frac{16\left(\gamma_e J\tau\right)^4 }{V^2} t_s^2 \sin^2\left(\delta t_s\right).
	\eeq
	Since the Fisher information is additive and the correlation is always larger than the asymptotic value we can bound the total Fisher information by
	\beq\label{Qdyne_Fisher}
	\pazocal{I}\geq \frac{16\left(\gamma_e J\tau\right)^4 }{V^2}\sum_{s=0}^{T/\tau} t_s^2 \sin^2\left(\delta t_s\right)\approx\frac{8\left(\gamma_e J\right)^4 }{3V^2} \tau^3 T^3,
	\eeq
	where $T$ is the total measurement time. The result \eqref{Qdyne_Fisher} is a major improvement over \eqref{corr_spec_fisher_short}, since it scales with $T^3$ which is not limited by the diffusion time. 
    The bound \eqref{Qdyne_Fisher} is quite crude, since for viscous fluids a great deal of information can be gained from the slow temporal decay. A comprehensive analysis can be found in \cite{Amit-Santi}. 

\subsection{Molecular dynamics}
\begin{figure}
\centering

\includegraphics[scale=0.6]{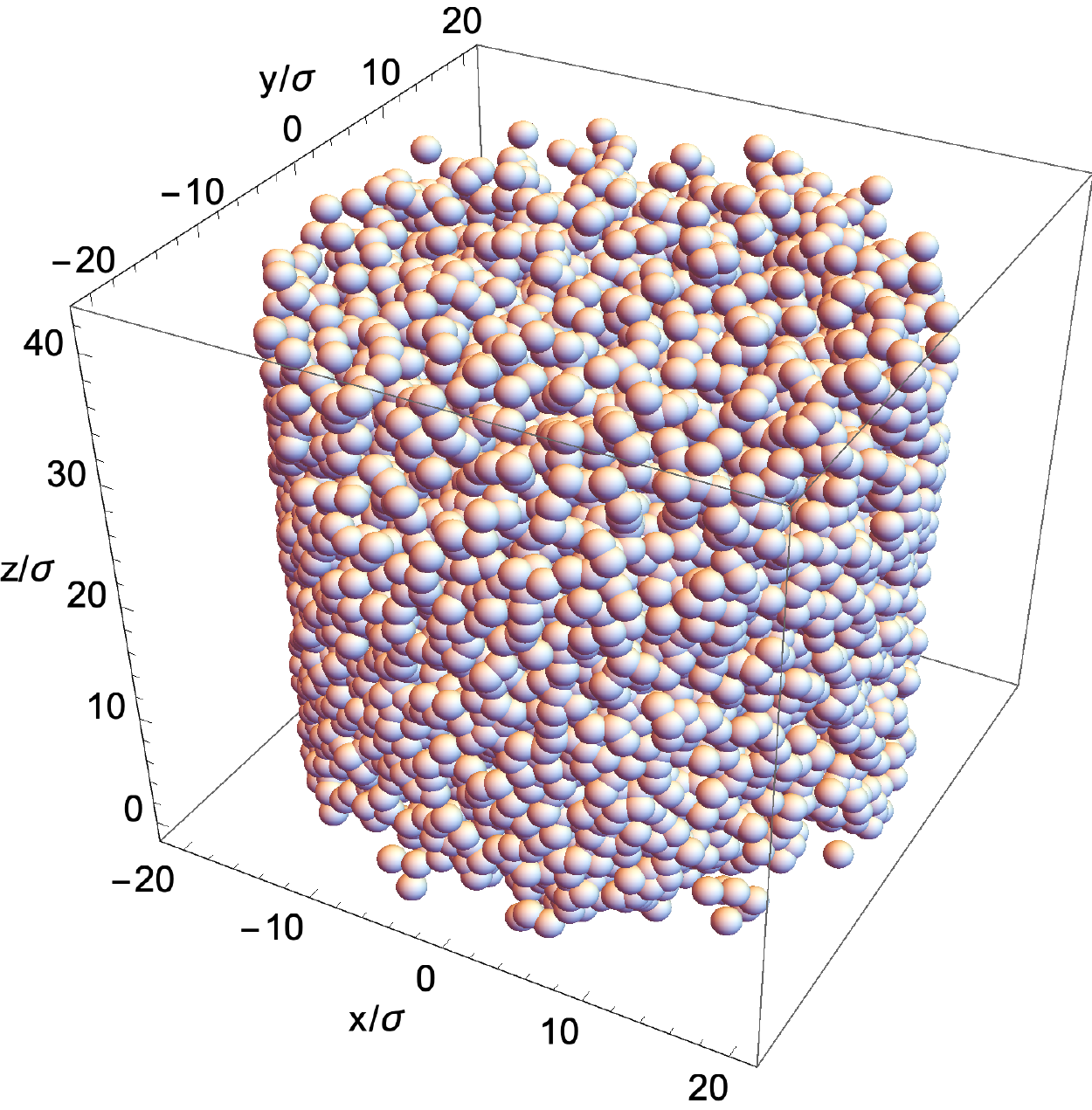}

\caption{The simulated Lennard-Jones fluid consisted of 22091 particles moving
in a closed cylindrical pillbox of radius $R=16.45\sigma$ and height
$L=16.45\sigma$.  \label{fig:simulation_box}}
\end{figure}

We corroborated the analytic prediction of the correlation function
of the magnetic field as a result of the fluid motion in the closed geometry
using molecular dynamics simulations, focusing on the case of the
cylindrical geometry. The set-up is illustrated in figure \ref{fig:simulation_box}.
The system contains $N$ particles, which are confined to a cylinder
of radius $R$ and height $L$. The particles interact via
the Lennard-Jones (LJ) potential $4\epsilon[(\sigma/r)^{12}-(\sigma/r)^{6}]$
with an interaction cutoff distance of $r_{c}=2.5\sigma$. 
Specular reflections are applied on the top and bottom walls of the cylinder and the Lennard-Jones 9/3 potential is applied at the curved walls to confine the particles to the interior of the cylinder. The system
is initialized into a thermal state at temperature $T$ by running
a Langevin dynamics simulation until the system reaches thermal equilibrium.
Each particle is assigned a random value of spin $I_{z}\in\{-1,1\}$.
After initializing the system, we ran a deterministic molecular dynamics
simulation, integrating Newton's laws using the Velocity-Verlet
method with step size $\Delta t=0.005\sqrt{\epsilon/(m\sigma^{2})}$.
During the simulation we computed and stored the $z$ component of the
magnetic field induced by the particles at the position of the NV

\beq
B(t)=\sum_{j=1}^{N}\frac{1}{r_{i}^{3}(t)}\left[3\cos^{2}(\theta_{i}(t))-1\right]I_{z}^{i}.
\eeq
where $r_{i}$ is the distance between particle $i$ and the NV center
and $\theta_{i}$ is the angle between $r_{i}$ and the NV quantization
axis. The NV is placed along the axis of rotational symmetry of the
cylinder, at a distance $d$ below the bottom of the cylinder. 

\begin{figure}
\centering

\includegraphics[scale=0.55]{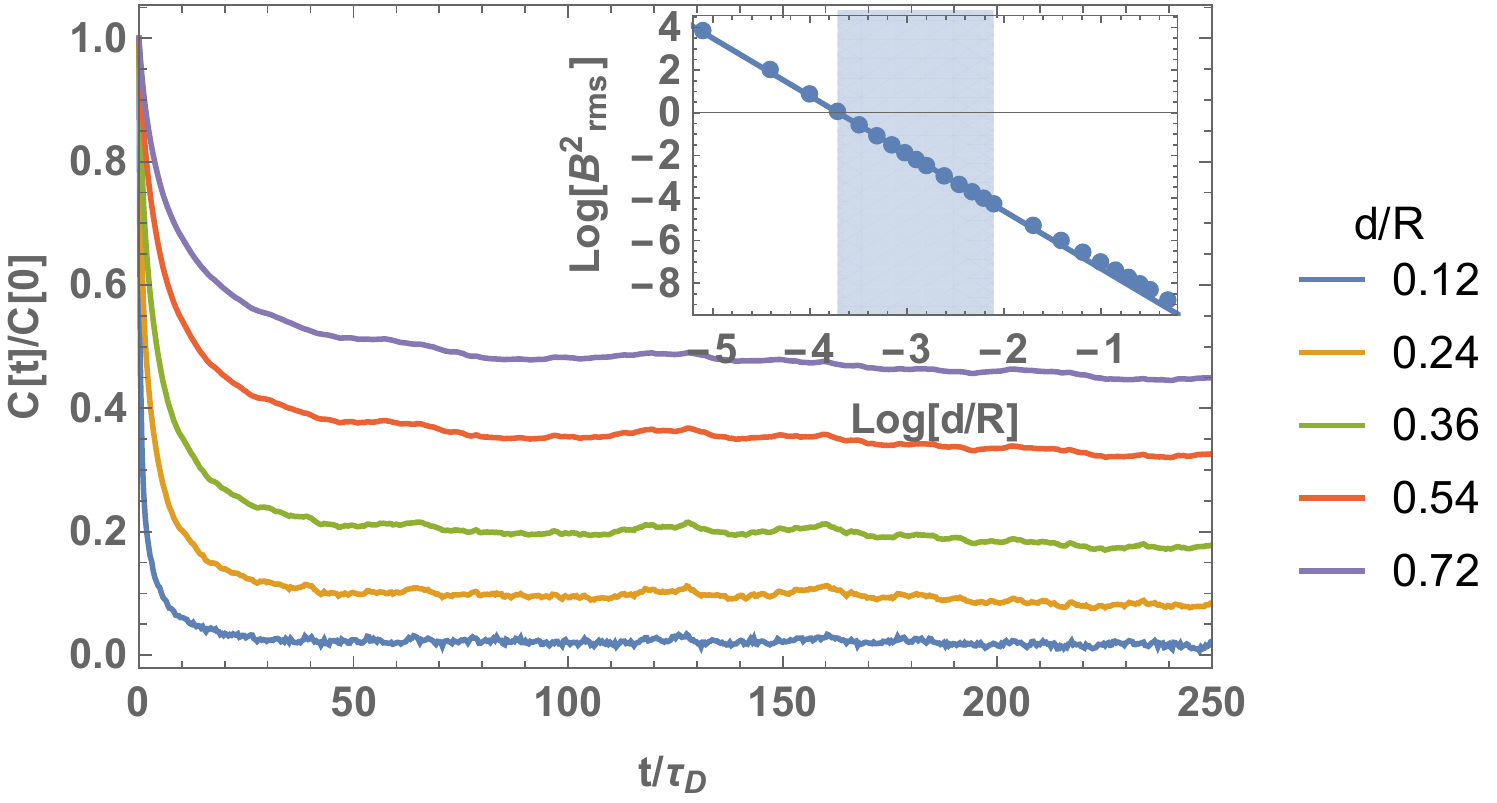}

\caption{Magnetic field correlation function $C(t)$ at the NV's position resulting from the dynamics
of LJ particles confined to a cylinder. The field is evaluated
at several values of NV implantation depth $d/R$. The dimensions
of the cylinder were $R=L=16.45$. The particle density
was $\rho=0.79\sigma^{-3}$. LJ fluid parameters in the reduced units
were $\epsilon=1$, $\sigma=1$ and temperature $T=1$. The dynamics
were integrated for $4\times10^{6}$ timesteps of size
$0.005\sqrt{\epsilon/(m\sigma^{2})}$. The correlation length was
obtained by computing the power spectrum $S(\omega)=\left\langle |\mathcal{F}[B(t)]|^{2}\right\rangle $ and
then performing the inverse Fourier transformation. The averaging
was over 16 simulation runs. The correlation function at $t=0$ is presented at the inset. The colored region fits a linear line with a slope of $-2.7$. This scaling is similar to the $d^{-3}$ scaling predicted by Eq. \eqref{Brms_cylinder0} at small values of $d/R$.   \label{fig:corr_scaled}}
\end{figure}

\begin{figure}
\centering

\includegraphics[scale=0.48]{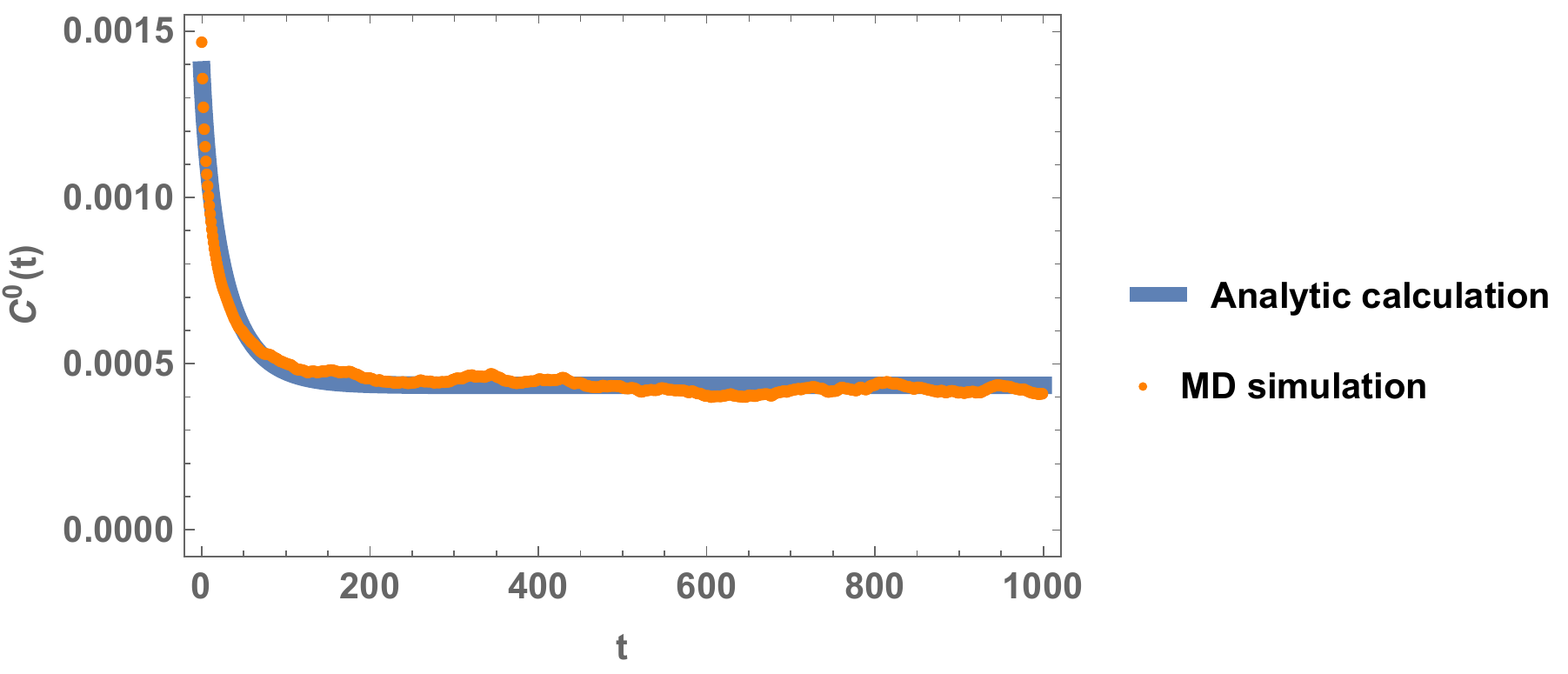}

\caption{The magnetic field correlation function in a \textit{cylindrical} geometry with $R=L=16.44\sigma$ and $d=8\sigma$. The correlation is given in arbitrary units, whereas the time is given in LJ units. The analytic curve was estimated by numerically calculating the first terms of Eq. \eqref{correlation_cylinder_temporal_decay} ($k,s\in[1,25]$) and adding the asymptotic constant given by Eq. \eqref{longtimes_cylinder0}. The simulation agrees with the analytic results apart from the value at $t=0$. This is because the limited integration time of the simulation is insufficient for estimating the power spectrum at $\omega=0$, which translates to an error in the estimation of $B_{rms}$.   \label{fig:Cylinder_overlay}}
\end{figure}

\begin{figure}
\centering

\includegraphics[scale=0.48]{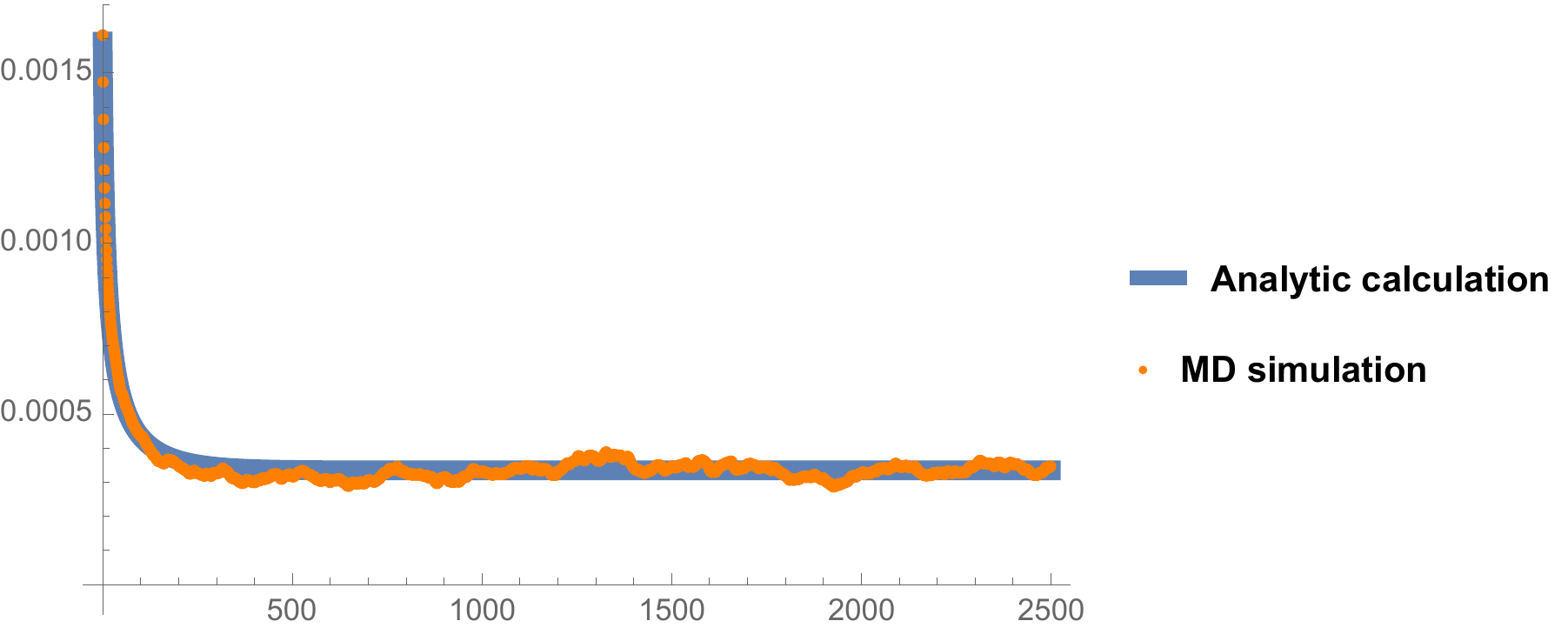}

\caption{The magnetic field correlation function in a \textit{spherical} geometry with $R=L=20.47\sigma$ and $d=8\sigma$. 
The correlation is given in arbitrary units, whereas the time is given in LJ units. The analytic curve was estimated by numerically calculating the first terms of Eq. \eqref{corr_sphere_temporal_decay} ($l\in[0,30],k\in[1,30]$) and adding the asymptotic constant given by Eq. \eqref{longtimes_sphere0}. The simulation agrees well with the analytic results. \label{fig:Sphere_overlay}}
\end{figure}

Figure \ref{fig:corr_scaled} shows the correlation function for the
chosen cylindrical geometry at several values of $d$.
At $t=0$ the correlation decays as $d^{-2.7}$ (shown at the inset), which is close to the $d^{-3}$ scaling predicted by \eqref{Brms_cylinder0} at small values of $d/R$. The change in the decay is possibly due to the interaction between the particles, which is not taken into account in our model. At long times the the correlation decays to a constant value that increases with $d/R$ as expected. A comparison between the asymptotic values of the MD simulation and the analytic prediction \eqref{longtimes_cylinder0} is found at \cite{Supp}.
We compared the simulation results to the analytic correlation function (Eqs. \eqref{longtimes_cylinder0} and \eqref{correlation_cylinder_temporal_decay}) by rescaling the asymptotic value of the former to match the one in Eq. \eqref{longtimes_cylinder0} and choosing an appropriate diffusion coefficient. A representative case is shown in Fig. \ref{fig:Cylinder_overlay}. 
We verified the results of the spherical geometry with a second set of simulations. The spherical geometry simulations follow a very similar protocol to the cylindrical geometry simulations. 
The number of particles was set to $N=28371$ and the radius of the sphere was set to $R=20.47\sigma$  resulting in particle density of $\rho=0.79\sigma^{-3}$. The temperature, integration timestep and the total integration time was identical to the cylindrical geometry simulation. The LJ 9/3 potential is applied across the whole surface of the sphere to confine the 
A comparison between the simulation results and the analytic prediction is presented at Fig. \ref{fig:Sphere_overlay}.


%
%


\subsection{Limits of the model}
	In the calculation of the correlation function we performed an average over realizations of the magnetic field noise. While this averaging holds for correlation spectroscopy, in a phase sensitive measurement which is based on a single realization it does not. 
	The main difference is that in a single realization, the statistical polarization will cause the amplitude of the signal to fluctuate as different nuclei diffuse in and out of the effective interaction region. 
    Assuming the phase of the signal remains constant, the FI is bounded by that of the average amplitude, which is given by the correlation function.
	As long as the spins retain their initial polarization, the phase will remain stable.
	The spin orientation will eventually fluctuate, mostly due to the interaction between the nuclei and walls.
	The total measurement time $T$ in \eqref{Qdyne_Fisher} is therefore limited by the characteristic time of the interaction, $T_{walls}$. 
    
	\section{CONCLUSION}
	We provided a theoretical basis for confined nano-NMR spectroscopy based on NV centers. We showed that the confinement indeed mitigates the deleterious effects of diffusion and therefore can be used to enhance the resolution abilities of both correlation spectroscopy and phase sensitive measurements. 
	As advances are made with hyperpolarization\cite{NV_ensemble,shagieva2018microwave,Pelayo2018toward,broadway2018quantum,abrams2014dynamic,scheuer2016optically,london2013detecting}, it might prove more beneficial to combine polarization and confinement. The relevant calculations can be found in \cite{Supp}. 
	
	\section{Acknowledgments}
	We thank Carlos Meriles for pointing out the possibility of confinement to us. A. R. acknowledges the support of ERC grant QRES, project No. 770929, grant agreement No 667192(Hyperdiamond) and the ASTERIQS collaborative project.

\bibliography{Confined_nano_NMR_spectroscopy_refs}

\begin{thebibliography}{27}%
\makeatletter
\providecommand \@ifxundefined [1]{%
 \@ifx{#1\undefined}
}%
\providecommand \@ifnum [1]{%
 \ifnum #1\expandafter \@firstoftwo
 \else \expandafter \@secondoftwo
 \fi
}%
\providecommand \@ifx [1]{%
 \ifx #1\expandafter \@firstoftwo
 \else \expandafter \@secondoftwo
 \fi
}%
\providecommand \natexlab [1]{#1}%
\providecommand \enquote  [1]{``#1''}%
\providecommand \bibnamefont  [1]{#1}%
\providecommand \bibfnamefont [1]{#1}%
\providecommand \citenamefont [1]{#1}%
\providecommand \href@noop [0]{\@secondoftwo}%
\providecommand \href [0]{\begingroup \@sanitize@url \@href}%
\providecommand \@href[1]{\@@startlink{#1}\@@href}%
\providecommand \@@href[1]{\endgroup#1\@@endlink}%
\providecommand \@sanitize@url [0]{\catcode `\\12\catcode `\$12\catcode
  `\&12\catcode `\#12\catcode `\^12\catcode `\_12\catcode `\%12\relax}%
\providecommand \@@startlink[1]{}%
\providecommand \@@endlink[0]{}%
\providecommand \url  [0]{\begingroup\@sanitize@url \@url }%
\providecommand \@url [1]{\endgroup\@href {#1}{\urlprefix }}%
\providecommand \urlprefix  [0]{URL }%
\providecommand \Eprint [0]{\href }%
\providecommand \doibase [0]{http://dx.doi.org/}%
\providecommand \selectlanguage [0]{\@gobble}%
\providecommand \bibinfo  [0]{\@secondoftwo}%
\providecommand \bibfield  [0]{\@secondoftwo}%
\providecommand \translation [1]{[#1]}%
\providecommand \BibitemOpen [0]{}%
\providecommand \bibitemStop [0]{}%
\providecommand \bibitemNoStop [0]{.\EOS\space}%
\providecommand \EOS [0]{\spacefactor3000\relax}%
\providecommand \BibitemShut  [1]{\csname bibitem#1\endcsname}%
\let\auto@bib@innerbib\@empty
\bibitem [{\citenamefont {Jensen}\ \emph {et~al.}(2017)\citenamefont {Jensen},
  \citenamefont {Kehayias},\ and\ \citenamefont {Budker}}]{NV_Mag1}%
  \BibitemOpen
  \bibfield  {author} {\bibinfo {author} {\bibfnamefont {K.}~\bibnamefont
  {Jensen}}, \bibinfo {author} {\bibfnamefont {P.}~\bibnamefont {Kehayias}}, \
  and\ \bibinfo {author} {\bibfnamefont {D.}~\bibnamefont {Budker}},\ }\enquote
  {\bibinfo {title} {Magnetometry with nitrogen-vacancy centers in diamond},}\
  in\ \href@noop {} {\emph {\bibinfo {booktitle} {High Sensitivity
  Magnetometers}}},\ \bibinfo {editor} {edited by\ \bibinfo {editor}
  {\bibfnamefont {A.}~\bibnamefont {Grosz}}, \bibinfo {editor} {\bibfnamefont
  {M.}~\bibnamefont {Haji-Sheikh}}, \ and\ \bibinfo {editor} {\bibfnamefont
  {S.}~\bibnamefont {Mukhopadhyay}}}\ (\bibinfo  {publisher} {Springer, Cham},\
  \bibinfo {address} {Berlin, Heidelberg},\ \bibinfo {year} {2017})\BibitemShut
  {NoStop}%
\bibitem [{\citenamefont {Doherty}\ \emph {et~al.}(2013)\citenamefont
  {Doherty}, \citenamefont {B.Manson}, \citenamefont {Delaney}, \citenamefont
  {Jelezko}, \citenamefont {Wrachtrup},\ and\ \citenamefont
  {Hollenberg}}]{Review_NV}%
  \BibitemOpen
  \bibfield  {author} {\bibinfo {author} {\bibfnamefont {M.~W.}\ \bibnamefont
  {Doherty}}, \bibinfo {author} {\bibfnamefont {N.}~\bibnamefont {B.Manson}},
  \bibinfo {author} {\bibfnamefont {P.}~\bibnamefont {Delaney}}, \bibinfo
  {author} {\bibfnamefont {F.}~\bibnamefont {Jelezko}}, \bibinfo {author}
  {\bibfnamefont {J.}~\bibnamefont {Wrachtrup}}, \ and\ \bibinfo {author}
  {\bibfnamefont {L.~C.~L.}\ \bibnamefont {Hollenberg}},\ }\href {\doibase
  https://doi.org/10.1016/j.physrep.2013.02.001} {\bibfield  {journal}
  {\bibinfo  {journal} {Phys. Rep.}\ }\textbf {\bibinfo {volume} {528}},\
  \bibinfo {pages} {1} (\bibinfo {year} {2013})}\BibitemShut {NoStop}%
\bibitem [{\citenamefont {DeVience}\ \emph {et~al.}(2015)\citenamefont
  {DeVience}, \citenamefont {Pham}, \citenamefont {Lovchinsky}, \citenamefont
  {Sushkov}, \citenamefont {Bar-Gill}, \citenamefont {Belthangady},
  \citenamefont {Casola}, \citenamefont {Corbett}, \citenamefont {Zhang},
  \citenamefont {Lukin}, \citenamefont {Park}, \citenamefont {Yacoby},\ and\
  \citenamefont {Walsworth}}]{NanoNMRHarward}%
  \BibitemOpen
  \bibfield  {author} {\bibinfo {author} {\bibfnamefont {S.~J.}\ \bibnamefont
  {DeVience}}, \bibinfo {author} {\bibfnamefont {L.~M.}\ \bibnamefont {Pham}},
  \bibinfo {author} {\bibfnamefont {I.}~\bibnamefont {Lovchinsky}}, \bibinfo
  {author} {\bibfnamefont {A.~O.}\ \bibnamefont {Sushkov}}, \bibinfo {author}
  {\bibfnamefont {N.}~\bibnamefont {Bar-Gill}}, \bibinfo {author}
  {\bibfnamefont {C.}~\bibnamefont {Belthangady}}, \bibinfo {author}
  {\bibfnamefont {F.}~\bibnamefont {Casola}}, \bibinfo {author} {\bibfnamefont
  {M.}~\bibnamefont {Corbett}}, \bibinfo {author} {\bibfnamefont
  {H.}~\bibnamefont {Zhang}}, \bibinfo {author} {\bibfnamefont
  {M.}~\bibnamefont {Lukin}}, \bibinfo {author} {\bibfnamefont
  {H.}~\bibnamefont {Park}}, \bibinfo {author} {\bibfnamefont {A.}~\bibnamefont
  {Yacoby}}, \ and\ \bibinfo {author} {\bibfnamefont {R.~L.}\ \bibnamefont
  {Walsworth}},\ }\href {\doibase https://doi.org/10.1038/nnano.2014.313}
  {\bibfield  {journal} {\bibinfo  {journal} {Nat. Nanotech}\ }\textbf
  {\bibinfo {volume} {10}},\ \bibinfo {pages} {129} (\bibinfo {year}
  {2015})}\BibitemShut {NoStop}%
\bibitem [{\citenamefont {Mamin}\ \emph {et~al.}(2013)\citenamefont {Mamin},
  \citenamefont {Kim}, \citenamefont {Sherwood}, \citenamefont {Rettner},
  \citenamefont {Ohno}, \citenamefont {Awschalom},\ and\ \citenamefont
  {Rugar}}]{NanoNMRIBM}%
  \BibitemOpen
  \bibfield  {author} {\bibinfo {author} {\bibfnamefont {H.~J.}\ \bibnamefont
  {Mamin}}, \bibinfo {author} {\bibfnamefont {M.}~\bibnamefont {Kim}}, \bibinfo
  {author} {\bibfnamefont {M.~H.}\ \bibnamefont {Sherwood}}, \bibinfo {author}
  {\bibfnamefont {C.~T.}\ \bibnamefont {Rettner}}, \bibinfo {author}
  {\bibfnamefont {K.}~\bibnamefont {Ohno}}, \bibinfo {author} {\bibfnamefont
  {D.~D.}\ \bibnamefont {Awschalom}}, \ and\ \bibinfo {author} {\bibfnamefont
  {D.}~\bibnamefont {Rugar}},\ }\href {\doibase 10.1126/science.1231540}
  {\bibfield  {journal} {\bibinfo  {journal} {SCIENCE}\ }\textbf {\bibinfo
  {volume} {339}} (\bibinfo {year} {2013}),\
  10.1126/science.1231540}\BibitemShut {NoStop}%
\bibitem [{\citenamefont {Shagieva}\ \emph
  {et~al.}(2018{\natexlab{a}})\citenamefont {Shagieva}, \citenamefont {Zaiser},
  \citenamefont {Neumann}, \citenamefont {Dasari}, \citenamefont {St{\"o}hr},
  \citenamefont {Denisenko}, \citenamefont {Reuter}, \citenamefont {Meriles},\
  and\ \citenamefont {Wrachtrup}}]{NanoNMRMeriles}%
  \BibitemOpen
  \bibfield  {author} {\bibinfo {author} {\bibfnamefont {F.}~\bibnamefont
  {Shagieva}}, \bibinfo {author} {\bibfnamefont {S.}~\bibnamefont {Zaiser}},
  \bibinfo {author} {\bibfnamefont {P.}~\bibnamefont {Neumann}}, \bibinfo
  {author} {\bibfnamefont {D.~B.~R.}\ \bibnamefont {Dasari}}, \bibinfo {author}
  {\bibfnamefont {R.}~\bibnamefont {St{\"o}hr}}, \bibinfo {author}
  {\bibfnamefont {A.}~\bibnamefont {Denisenko}}, \bibinfo {author}
  {\bibfnamefont {R.}~\bibnamefont {Reuter}}, \bibinfo {author} {\bibfnamefont
  {C.~A.}\ \bibnamefont {Meriles}}, \ and\ \bibinfo {author} {\bibfnamefont
  {J.}~\bibnamefont {Wrachtrup}},\ }\href {\doibase
  10.1021/acs.nanolett.8b00925} {\bibfield  {journal} {\bibinfo  {journal}
  {Nano Lett.}\ }\textbf {\bibinfo {volume} {18}},\ \bibinfo {pages} {3731}
  (\bibinfo {year} {2018}{\natexlab{a}})}\BibitemShut {NoStop}%
\bibitem [{\citenamefont {Pfender}\ \emph {et~al.}(2019)\citenamefont
  {Pfender}, \citenamefont {Wang}, \citenamefont {Sumiya}, \citenamefont
  {Onoda}, \citenamefont {Yang}, \citenamefont {Dasari}, \citenamefont
  {Neumann}, \citenamefont {Pan}, \citenamefont {Isoya}, \citenamefont {Liu},\
  and\ \citenamefont {Wrachtrup}}]{NanoNMRQdyne}%
  \BibitemOpen
  \bibfield  {author} {\bibinfo {author} {\bibfnamefont {M.}~\bibnamefont
  {Pfender}}, \bibinfo {author} {\bibfnamefont {P.}~\bibnamefont {Wang}},
  \bibinfo {author} {\bibfnamefont {H.}~\bibnamefont {Sumiya}}, \bibinfo
  {author} {\bibfnamefont {S.}~\bibnamefont {Onoda}}, \bibinfo {author}
  {\bibfnamefont {W.}~\bibnamefont {Yang}}, \bibinfo {author} {\bibfnamefont
  {D.~B.~R.}\ \bibnamefont {Dasari}}, \bibinfo {author} {\bibfnamefont
  {P.}~\bibnamefont {Neumann}}, \bibinfo {author} {\bibfnamefont {X.-Y.}\
  \bibnamefont {Pan}}, \bibinfo {author} {\bibfnamefont {J.}~\bibnamefont
  {Isoya}}, \bibinfo {author} {\bibfnamefont {R.-B.}\ \bibnamefont {Liu}}, \
  and\ \bibinfo {author} {\bibfnamefont {J.}~\bibnamefont {Wrachtrup}},\ }\href
  {\doibase https://doi.org/10.1038/s41467-019-08544-z} {\bibfield  {journal}
  {\bibinfo  {journal} {Nat. Commun}\ }\textbf {\bibinfo {volume} {10}},\
  \bibinfo {pages} {594} (\bibinfo {year} {2019})}\BibitemShut {NoStop}%
\bibitem [{\citenamefont {Aslam}\ \emph {et~al.}(2017)\citenamefont {Aslam},
  \citenamefont {Pfender}, \citenamefont {Neumann}, \citenamefont {Reuter},
  \citenamefont {Zappe}, \citenamefont {de~Oliveira}, \citenamefont
  {Denisenko}, \citenamefont {Sumiya}, \citenamefont {Onoda}, \citenamefont
  {Isoya},\ and\ \citenamefont {Wrachtrup}}]{NanoNMRstutgart}%
  \BibitemOpen
  \bibfield  {author} {\bibinfo {author} {\bibfnamefont {N.}~\bibnamefont
  {Aslam}}, \bibinfo {author} {\bibfnamefont {M.}~\bibnamefont {Pfender}},
  \bibinfo {author} {\bibfnamefont {P.}~\bibnamefont {Neumann}}, \bibinfo
  {author} {\bibfnamefont {R.}~\bibnamefont {Reuter}}, \bibinfo {author}
  {\bibfnamefont {A.}~\bibnamefont {Zappe}}, \bibinfo {author} {\bibfnamefont
  {F.~F.}\ \bibnamefont {de~Oliveira}}, \bibinfo {author} {\bibfnamefont
  {A.}~\bibnamefont {Denisenko}}, \bibinfo {author} {\bibfnamefont
  {H.}~\bibnamefont {Sumiya}}, \bibinfo {author} {\bibfnamefont
  {S.}~\bibnamefont {Onoda}}, \bibinfo {author} {\bibfnamefont
  {J.}~\bibnamefont {Isoya}}, \ and\ \bibinfo {author} {\bibfnamefont
  {J.}~\bibnamefont {Wrachtrup}},\ }\href {\doibase 10.1126/science.aam8697}
  {\bibfield  {journal} {\bibinfo  {journal} {SCIENCE}\ }\textbf {\bibinfo
  {volume} {357}} (\bibinfo {year} {2017}),\
  10.1126/science.aam8697}\BibitemShut {NoStop}%
\bibitem [{\citenamefont {Glenn}\ \emph {et~al.}(2018)\citenamefont {Glenn},
  \citenamefont {Bucher}, \citenamefont {Lee}, \citenamefont {Lukin},
  \citenamefont {Park},\ and\ \citenamefont {Walsworth}}]{glenn2018high}%
  \BibitemOpen
  \bibfield  {author} {\bibinfo {author} {\bibfnamefont {D.~R.}\ \bibnamefont
  {Glenn}}, \bibinfo {author} {\bibfnamefont {D.~B.}\ \bibnamefont {Bucher}},
  \bibinfo {author} {\bibfnamefont {J.}~\bibnamefont {Lee}}, \bibinfo {author}
  {\bibfnamefont {M.~D.}\ \bibnamefont {Lukin}}, \bibinfo {author}
  {\bibfnamefont {H.}~\bibnamefont {Park}}, \ and\ \bibinfo {author}
  {\bibfnamefont {R.~L.}\ \bibnamefont {Walsworth}},\ }\href@noop {} {\bibfield
   {journal} {\bibinfo  {journal} {Nature}\ }\textbf {\bibinfo {volume}
  {555}},\ \bibinfo {pages} {351} (\bibinfo {year} {2018})}\BibitemShut
  {NoStop}%
\bibitem [{\citenamefont {Bucher}\ \emph {et~al.}(2018)\citenamefont {Bucher},
  \citenamefont {Glenn}, \citenamefont {Park}, \citenamefont {Lukin},\ and\
  \citenamefont {Walsworth}}]{NV_ensemble}%
  \BibitemOpen
  \bibfield  {author} {\bibinfo {author} {\bibfnamefont {D.~B.}\ \bibnamefont
  {Bucher}}, \bibinfo {author} {\bibfnamefont {D.~R.}\ \bibnamefont {Glenn}},
  \bibinfo {author} {\bibfnamefont {H.}~\bibnamefont {Park}}, \bibinfo {author}
  {\bibfnamefont {M.~D.}\ \bibnamefont {Lukin}}, \ and\ \bibinfo {author}
  {\bibfnamefont {R.~L.}\ \bibnamefont {Walsworth}},\ }\href
  {https://arxiv.org/abs/1810.02408} {\enquote {\bibinfo {title}
  {Hyperpolarization-enhanced nmr spectroscopy with femtomole sensitivity using
  quantum defects in diamond},}\ } (\bibinfo {year} {2018})\BibitemShut
  {NoStop}%
\bibitem [{\citenamefont {Cohen}\ \emph {et~al.}(2019)\citenamefont {Cohen},
  \citenamefont {Gefen}, \citenamefont {Ortiz},\ and\ \citenamefont
  {Retzker}}]{Laura_arxiv}%
  \BibitemOpen
  \bibfield  {author} {\bibinfo {author} {\bibfnamefont {D.}~\bibnamefont
  {Cohen}}, \bibinfo {author} {\bibfnamefont {T.}~\bibnamefont {Gefen}},
  \bibinfo {author} {\bibfnamefont {L.}~\bibnamefont {Ortiz}}, \ and\ \bibinfo
  {author} {\bibfnamefont {A.}~\bibnamefont {Retzker}},\ }\href
  {http://arxiv.org/abs/1912.09062} {\enquote {\bibinfo {title} {Achieving the
  ultimate precision limit in quantum nmr spectroscopy},}\ } (\bibinfo {year}
  {2019})\BibitemShut {NoStop}%
\bibitem [{\citenamefont {Cohen}\ \emph {et~al.}(2020)\citenamefont {Cohen},
  \citenamefont {Nigmatullin}, \citenamefont {Kenneth}, \citenamefont
  {Jelezko}, \citenamefont {Khodas},\ and\ \citenamefont
  {Retzker}}]{microfludics}%
  \BibitemOpen
  \bibfield  {author} {\bibinfo {author} {\bibfnamefont {D.}~\bibnamefont
  {Cohen}}, \bibinfo {author} {\bibfnamefont {R.}~\bibnamefont {Nigmatullin}},
  \bibinfo {author} {\bibfnamefont {O.}~\bibnamefont {Kenneth}}, \bibinfo
  {author} {\bibfnamefont {F.}~\bibnamefont {Jelezko}}, \bibinfo {author}
  {\bibfnamefont {M.}~\bibnamefont {Khodas}}, \ and\ \bibinfo {author}
  {\bibfnamefont {A.}~\bibnamefont {Retzker}},\ }\href {\doibase
  10.1038/s41598-020-61095-y} {\bibfield  {journal} {\bibinfo  {journal} {Sci.
  Rep.}\ }\textbf {\bibinfo {volume} {10}},\ \bibinfo {pages} {5298} (\bibinfo
  {year} {2020})}\BibitemShut {NoStop}%
\bibitem [{\citenamefont {Rotem}\ \emph {et~al.}()\citenamefont {Rotem},
  \citenamefont {Oviedo-Casado},\ and\ \citenamefont {Retzker}}]{Amit-Santi}%
  \BibitemOpen
  \bibfield  {author} {\bibinfo {author} {\bibfnamefont {A.}~\bibnamefont
  {Rotem}}, \bibinfo {author} {\bibfnamefont {S.}~\bibnamefont
  {Oviedo-Casado}}, \ and\ \bibinfo {author} {\bibfnamefont {A.}~\bibnamefont
  {Retzker}},\ }\href@noop {} {\bibinfo  {journal} {In preperation}\
  }\BibitemShut {NoStop}%
\bibitem [{\citenamefont {Staudacher}\ \emph {et~al.}(2015)\citenamefont
  {Staudacher}, \citenamefont {Raatz}, \citenamefont {Pezzagna}, \citenamefont
  {Meijer}, \citenamefont {Reinhard}, \citenamefont {Meriles},\ and\
  \citenamefont {Wrachtrup}}]{Corr_spect}%
  \BibitemOpen
\bibfield  {journal} {  }\bibfield  {author} {\bibinfo {author} {\bibfnamefont
  {T.}~\bibnamefont {Staudacher}}, \bibinfo {author} {\bibfnamefont
  {N.}~\bibnamefont {Raatz}}, \bibinfo {author} {\bibfnamefont
  {S.}~\bibnamefont {Pezzagna}}, \bibinfo {author} {\bibfnamefont
  {J.}~\bibnamefont {Meijer}}, \bibinfo {author} {\bibfnamefont
  {F.}~\bibnamefont {Reinhard}}, \bibinfo {author} {\bibfnamefont {C.~A.}\
  \bibnamefont {Meriles}}, \ and\ \bibinfo {author} {\bibfnamefont
  {J.}~\bibnamefont {Wrachtrup}},\ }\href {\doibase 10.1038/ncomms9527}
  {\bibfield  {journal} {\bibinfo  {journal} {Nat. Commun.}\ }\textbf {\bibinfo
  {volume} {6}},\ \bibinfo {pages} {8527} (\bibinfo {year} {2015})}\BibitemShut
  {NoStop}%
\bibitem [{\citenamefont {Kong}\ \emph {et~al.}(2015)\citenamefont {Kong},
  \citenamefont {Stark}, \citenamefont {Du}, \citenamefont {McGuinness},\ and\
  \citenamefont {Jelezko}}]{kong2015towards}%
  \BibitemOpen
  \bibfield  {author} {\bibinfo {author} {\bibfnamefont {X.}~\bibnamefont
  {Kong}}, \bibinfo {author} {\bibfnamefont {A.}~\bibnamefont {Stark}},
  \bibinfo {author} {\bibfnamefont {J.}~\bibnamefont {Du}}, \bibinfo {author}
  {\bibfnamefont {L.~P.}\ \bibnamefont {McGuinness}}, \ and\ \bibinfo {author}
  {\bibfnamefont {F.}~\bibnamefont {Jelezko}},\ }\href@noop {} {\bibfield
  {journal} {\bibinfo  {journal} {Physical Review Applied}\ }\textbf {\bibinfo
  {volume} {4}},\ \bibinfo {pages} {024004} (\bibinfo {year}
  {2015})}\BibitemShut {NoStop}%
\bibitem [{\citenamefont {Laraoui}\ \emph {et~al.}(2013)\citenamefont
  {Laraoui}, \citenamefont {Dolde}, \citenamefont {Burk}, \citenamefont
  {Reinhard}, \citenamefont {Wrachtrup},\ and\ \citenamefont
  {Meriles}}]{laraoui2013high}%
  \BibitemOpen
  \bibfield  {author} {\bibinfo {author} {\bibfnamefont {A.}~\bibnamefont
  {Laraoui}}, \bibinfo {author} {\bibfnamefont {F.}~\bibnamefont {Dolde}},
  \bibinfo {author} {\bibfnamefont {C.}~\bibnamefont {Burk}}, \bibinfo {author}
  {\bibfnamefont {F.}~\bibnamefont {Reinhard}}, \bibinfo {author}
  {\bibfnamefont {J.}~\bibnamefont {Wrachtrup}}, \ and\ \bibinfo {author}
  {\bibfnamefont {C.~A.}\ \bibnamefont {Meriles}},\ }\href@noop {} {\bibfield
  {journal} {\bibinfo  {journal} {Nature communications}\ }\textbf {\bibinfo
  {volume} {4}},\ \bibinfo {pages} {1} (\bibinfo {year} {2013})}\BibitemShut
  {NoStop}%
\bibitem [{\citenamefont {Rosskopf}\ \emph {et~al.}(2017)\citenamefont
  {Rosskopf}, \citenamefont {Zopes}, \citenamefont {Boss},\ and\ \citenamefont
  {Degen}}]{Corr_spect_memory1}%
  \BibitemOpen
  \bibfield  {author} {\bibinfo {author} {\bibfnamefont {T.}~\bibnamefont
  {Rosskopf}}, \bibinfo {author} {\bibfnamefont {J.}~\bibnamefont {Zopes}},
  \bibinfo {author} {\bibfnamefont {J.~M.}\ \bibnamefont {Boss}}, \ and\
  \bibinfo {author} {\bibfnamefont {C.~L.}\ \bibnamefont {Degen}},\ }\href
  {\doibase 10.1038/s41534-017-0030-6} {\bibfield  {journal} {\bibinfo
  {journal} {npj Quantum Inf.}\ }\textbf {\bibinfo {volume} {3}} (\bibinfo
  {year} {2017}),\ 10.1038/s41534-017-0030-6}\BibitemShut {NoStop}%
\bibitem [{\citenamefont {Zaiser}\ \emph {et~al.}(2016)\citenamefont {Zaiser},
  \citenamefont {Rendler}, \citenamefont {Jakobi}, \citenamefont {Wolf},
  \citenamefont {Lee}, \citenamefont {Wagner}, \citenamefont {Bergholm},
  \citenamefont {Schulte-Herbr{\"u}ggen}, \citenamefont {Neumann},\ and\
  \citenamefont {Wrachtrup}}]{Corr_spect_memory2}%
  \BibitemOpen
  \bibfield  {author} {\bibinfo {author} {\bibfnamefont {S.}~\bibnamefont
  {Zaiser}}, \bibinfo {author} {\bibfnamefont {T.}~\bibnamefont {Rendler}},
  \bibinfo {author} {\bibfnamefont {I.}~\bibnamefont {Jakobi}}, \bibinfo
  {author} {\bibfnamefont {T.}~\bibnamefont {Wolf}}, \bibinfo {author}
  {\bibfnamefont {S.-Y.}\ \bibnamefont {Lee}}, \bibinfo {author} {\bibfnamefont
  {S.}~\bibnamefont {Wagner}}, \bibinfo {author} {\bibfnamefont
  {V.}~\bibnamefont {Bergholm}}, \bibinfo {author} {\bibfnamefont
  {T.}~\bibnamefont {Schulte-Herbr{\"u}ggen}}, \bibinfo {author} {\bibfnamefont
  {P.}~\bibnamefont {Neumann}}, \ and\ \bibinfo {author} {\bibfnamefont
  {J.}~\bibnamefont {Wrachtrup}},\ }\href {\doibase 10.1038/ncomms12279}
  {\bibfield  {journal} {\bibinfo  {journal} {Nat. Commun.}\ }\textbf {\bibinfo
  {volume} {7}} (\bibinfo {year} {2016}),\ 10.1038/ncomms12279}\BibitemShut
  {NoStop}%
\bibitem [{\citenamefont {Schmitt}\ \emph {et~al.}(2017)\citenamefont
  {Schmitt}, \citenamefont {Gefen}, \citenamefont {St{\"u}rner}, \citenamefont
  {Unden}, \citenamefont {Wolff}, \citenamefont {M{\"u}ller}, \citenamefont
  {Scheuer}, \citenamefont {Naydenov}, \citenamefont {Markham}, \citenamefont
  {Pezzagna}, \citenamefont {Meijer}, \citenamefont {Schwarz}, \citenamefont
  {Plenio}, \citenamefont {Retzker}, \citenamefont {McGuinness},\ and\
  \citenamefont {Jelezko}}]{qdyne}%
  \BibitemOpen
  \bibfield  {author} {\bibinfo {author} {\bibfnamefont {S.}~\bibnamefont
  {Schmitt}}, \bibinfo {author} {\bibfnamefont {T.}~\bibnamefont {Gefen}},
  \bibinfo {author} {\bibfnamefont {F.~M.}\ \bibnamefont {St{\"u}rner}},
  \bibinfo {author} {\bibfnamefont {T.}~\bibnamefont {Unden}}, \bibinfo
  {author} {\bibfnamefont {G.}~\bibnamefont {Wolff}}, \bibinfo {author}
  {\bibfnamefont {C.}~\bibnamefont {M{\"u}ller}}, \bibinfo {author}
  {\bibfnamefont {J.}~\bibnamefont {Scheuer}}, \bibinfo {author} {\bibfnamefont
  {B.}~\bibnamefont {Naydenov}}, \bibinfo {author} {\bibfnamefont
  {M.}~\bibnamefont {Markham}}, \bibinfo {author} {\bibfnamefont
  {S.}~\bibnamefont {Pezzagna}}, \bibinfo {author} {\bibfnamefont
  {J.}~\bibnamefont {Meijer}}, \bibinfo {author} {\bibfnamefont
  {I.}~\bibnamefont {Schwarz}}, \bibinfo {author} {\bibfnamefont
  {M.}~\bibnamefont {Plenio}}, \bibinfo {author} {\bibfnamefont
  {A.}~\bibnamefont {Retzker}}, \bibinfo {author} {\bibfnamefont {L.~P.}\
  \bibnamefont {McGuinness}}, \ and\ \bibinfo {author} {\bibfnamefont
  {F.}~\bibnamefont {Jelezko}},\ }\href {\doibase 10.1126/science.aam5532}
  {\bibfield  {journal} {\bibinfo  {journal} {SCIENCE}\ }\textbf {\bibinfo
  {volume} {356}},\ \bibinfo {pages} {6340} (\bibinfo {year}
  {2017})}\BibitemShut {NoStop}%
\bibitem [{\citenamefont {Boss}\ \emph {et~al.}(2017)\citenamefont {Boss},
  \citenamefont {Cujia}, \citenamefont {Zopes},\ and\ \citenamefont
  {Degen}}]{boss2017quantum}%
  \BibitemOpen
  \bibfield  {author} {\bibinfo {author} {\bibfnamefont {J.}~\bibnamefont
  {Boss}}, \bibinfo {author} {\bibfnamefont {K.}~\bibnamefont {Cujia}},
  \bibinfo {author} {\bibfnamefont {J.}~\bibnamefont {Zopes}}, \ and\ \bibinfo
  {author} {\bibfnamefont {C.}~\bibnamefont {Degen}},\ }\href@noop {}
  {\bibfield  {journal} {\bibinfo  {journal} {Science}\ }\textbf {\bibinfo
  {volume} {356}},\ \bibinfo {pages} {837} (\bibinfo {year}
  {2017})}\BibitemShut {NoStop}%
\bibitem [{Sup()}]{Supp}%
  \BibitemOpen
  \href@noop {} {\enquote {\bibinfo {title} {See supporting information},}\
  }\BibitemShut {NoStop}%
\bibitem [{\citenamefont {DeGroot}\ and\ \citenamefont
  {Schervish}(2012)}]{Information_theory}%
  \BibitemOpen
  \bibfield  {author} {\bibinfo {author} {\bibfnamefont {M.~H.}\ \bibnamefont
  {DeGroot}}\ and\ \bibinfo {author} {\bibfnamefont {M.~J.}\ \bibnamefont
  {Schervish}},\ }\href@noop {} {\emph {\bibinfo {title} {Probability and
  statistics}}}\ (\bibinfo  {publisher} {Pearson Education},\ \bibinfo {year}
  {2012})\BibitemShut {NoStop}%
\bibitem [{\citenamefont {Shagieva}\ \emph
  {et~al.}(2018{\natexlab{b}})\citenamefont {Shagieva}, \citenamefont {Zaiser},
  \citenamefont {Neumann}, \citenamefont {Dasari}, \citenamefont {St~{\"o}hr},
  \citenamefont {Denisenko}, \citenamefont {Reuter}, \citenamefont {Meriles},\
  and\ \citenamefont {Wrachtrup}}]{shagieva2018microwave}%
  \BibitemOpen
  \bibfield  {author} {\bibinfo {author} {\bibfnamefont {F.}~\bibnamefont
  {Shagieva}}, \bibinfo {author} {\bibfnamefont {S.}~\bibnamefont {Zaiser}},
  \bibinfo {author} {\bibfnamefont {P.}~\bibnamefont {Neumann}}, \bibinfo
  {author} {\bibfnamefont {D.}~\bibnamefont {Dasari}}, \bibinfo {author}
  {\bibfnamefont {R.}~\bibnamefont {St~{\"o}hr}}, \bibinfo {author}
  {\bibfnamefont {A.}~\bibnamefont {Denisenko}}, \bibinfo {author}
  {\bibfnamefont {R.}~\bibnamefont {Reuter}}, \bibinfo {author} {\bibfnamefont
  {C.}~\bibnamefont {Meriles}}, \ and\ \bibinfo {author} {\bibfnamefont
  {J.}~\bibnamefont {Wrachtrup}},\ }\href@noop {} {\bibfield  {journal}
  {\bibinfo  {journal} {Nano letters}\ }\textbf {\bibinfo {volume} {18}},\
  \bibinfo {pages} {3731} (\bibinfo {year} {2018}{\natexlab{b}})}\BibitemShut
  {NoStop}%
\bibitem [{\citenamefont {Fern{\'a}ndez-Acebal}\ \emph
  {et~al.}(2018)\citenamefont {Fern{\'a}ndez-Acebal}, \citenamefont {Rosolio},
  \citenamefont {Scheuer}, \citenamefont {M{\"u}ller}, \citenamefont
  {M{\"u}ller}, \citenamefont {Schmitt}, \citenamefont {McGuinness},
  \citenamefont {Schwarz}, \citenamefont {Chen}, \citenamefont {Retzker} \emph
  {et~al.}}]{Pelayo2018toward}%
  \BibitemOpen
  \bibfield  {author} {\bibinfo {author} {\bibfnamefont {P.}~\bibnamefont
  {Fern{\'a}ndez-Acebal}}, \bibinfo {author} {\bibfnamefont {O.}~\bibnamefont
  {Rosolio}}, \bibinfo {author} {\bibfnamefont {J.}~\bibnamefont {Scheuer}},
  \bibinfo {author} {\bibfnamefont {C.}~\bibnamefont {M{\"u}ller}}, \bibinfo
  {author} {\bibfnamefont {S.}~\bibnamefont {M{\"u}ller}}, \bibinfo {author}
  {\bibfnamefont {S.}~\bibnamefont {Schmitt}}, \bibinfo {author} {\bibfnamefont
  {L.~P.}\ \bibnamefont {McGuinness}}, \bibinfo {author} {\bibfnamefont
  {I.}~\bibnamefont {Schwarz}}, \bibinfo {author} {\bibfnamefont
  {Q.}~\bibnamefont {Chen}}, \bibinfo {author} {\bibfnamefont {A.}~\bibnamefont
  {Retzker}},  \emph {et~al.},\ }\href@noop {} {\bibfield  {journal} {\bibinfo
  {journal} {Nano letters}\ }\textbf {\bibinfo {volume} {18}},\ \bibinfo
  {pages} {1882} (\bibinfo {year} {2018})}\BibitemShut {NoStop}%
\bibitem [{\citenamefont {Broadway}\ \emph {et~al.}(2018)\citenamefont
  {Broadway}, \citenamefont {Tetienne}, \citenamefont {Stacey}, \citenamefont
  {Wood}, \citenamefont {Simpson}, \citenamefont {Hall},\ and\ \citenamefont
  {Hollenberg}}]{broadway2018quantum}%
  \BibitemOpen
  \bibfield  {author} {\bibinfo {author} {\bibfnamefont {D.~A.}\ \bibnamefont
  {Broadway}}, \bibinfo {author} {\bibfnamefont {J.-P.}\ \bibnamefont
  {Tetienne}}, \bibinfo {author} {\bibfnamefont {A.}~\bibnamefont {Stacey}},
  \bibinfo {author} {\bibfnamefont {J.~D.}\ \bibnamefont {Wood}}, \bibinfo
  {author} {\bibfnamefont {D.~A.}\ \bibnamefont {Simpson}}, \bibinfo {author}
  {\bibfnamefont {L.~T.}\ \bibnamefont {Hall}}, \ and\ \bibinfo {author}
  {\bibfnamefont {L.~C.}\ \bibnamefont {Hollenberg}},\ }\href@noop {}
  {\bibfield  {journal} {\bibinfo  {journal} {Nature communications}\ }\textbf
  {\bibinfo {volume} {9}},\ \bibinfo {pages} {1246} (\bibinfo {year}
  {2018})}\BibitemShut {NoStop}%
\bibitem [{\citenamefont {Abrams}\ \emph {et~al.}(2014)\citenamefont {Abrams},
  \citenamefont {Trusheim}, \citenamefont {Englund}, \citenamefont {Shattuck},\
  and\ \citenamefont {Meriles}}]{abrams2014dynamic}%
  \BibitemOpen
  \bibfield  {author} {\bibinfo {author} {\bibfnamefont {D.}~\bibnamefont
  {Abrams}}, \bibinfo {author} {\bibfnamefont {M.~E.}\ \bibnamefont
  {Trusheim}}, \bibinfo {author} {\bibfnamefont {D.~R.}\ \bibnamefont
  {Englund}}, \bibinfo {author} {\bibfnamefont {M.~D.}\ \bibnamefont
  {Shattuck}}, \ and\ \bibinfo {author} {\bibfnamefont {C.~A.}\ \bibnamefont
  {Meriles}},\ }\href@noop {} {\bibfield  {journal} {\bibinfo  {journal} {Nano
  letters}\ }\textbf {\bibinfo {volume} {14}},\ \bibinfo {pages} {2471}
  (\bibinfo {year} {2014})}\BibitemShut {NoStop}%
\bibitem [{\citenamefont {Scheuer}\ \emph {et~al.}(2016)\citenamefont
  {Scheuer}, \citenamefont {Schwartz}, \citenamefont {Chen}, \citenamefont
  {Schulze-S{\"u}nninghausen}, \citenamefont {Carl}, \citenamefont {H{\"o}fer},
  \citenamefont {Retzker}, \citenamefont {Sumiya}, \citenamefont {Isoya},
  \citenamefont {Luy} \emph {et~al.}}]{scheuer2016optically}%
  \BibitemOpen
  \bibfield  {author} {\bibinfo {author} {\bibfnamefont {J.}~\bibnamefont
  {Scheuer}}, \bibinfo {author} {\bibfnamefont {I.}~\bibnamefont {Schwartz}},
  \bibinfo {author} {\bibfnamefont {Q.}~\bibnamefont {Chen}}, \bibinfo {author}
  {\bibfnamefont {D.}~\bibnamefont {Schulze-S{\"u}nninghausen}}, \bibinfo
  {author} {\bibfnamefont {P.}~\bibnamefont {Carl}}, \bibinfo {author}
  {\bibfnamefont {P.}~\bibnamefont {H{\"o}fer}}, \bibinfo {author}
  {\bibfnamefont {A.}~\bibnamefont {Retzker}}, \bibinfo {author} {\bibfnamefont
  {H.}~\bibnamefont {Sumiya}}, \bibinfo {author} {\bibfnamefont
  {J.}~\bibnamefont {Isoya}}, \bibinfo {author} {\bibfnamefont
  {B.}~\bibnamefont {Luy}},  \emph {et~al.},\ }\href@noop {} {\bibfield
  {journal} {\bibinfo  {journal} {New Journal of Physics}\ }\textbf {\bibinfo
  {volume} {18}},\ \bibinfo {pages} {013040} (\bibinfo {year}
  {2016})}\BibitemShut {NoStop}%
\bibitem [{\citenamefont {London}\ \emph {et~al.}(2013)\citenamefont {London},
  \citenamefont {Scheuer}, \citenamefont {Cai}, \citenamefont {Schwarz},
  \citenamefont {Retzker}, \citenamefont {Plenio}, \citenamefont {Katagiri},
  \citenamefont {Teraji}, \citenamefont {Koizumi}, \citenamefont {Isoya} \emph
  {et~al.}}]{london2013detecting}%
  \BibitemOpen
  \bibfield  {author} {\bibinfo {author} {\bibfnamefont {P.}~\bibnamefont
  {London}}, \bibinfo {author} {\bibfnamefont {J.}~\bibnamefont {Scheuer}},
  \bibinfo {author} {\bibfnamefont {J.-M.}\ \bibnamefont {Cai}}, \bibinfo
  {author} {\bibfnamefont {I.}~\bibnamefont {Schwarz}}, \bibinfo {author}
  {\bibfnamefont {A.}~\bibnamefont {Retzker}}, \bibinfo {author} {\bibfnamefont
  {M.~B.}\ \bibnamefont {Plenio}}, \bibinfo {author} {\bibfnamefont
  {M.}~\bibnamefont {Katagiri}}, \bibinfo {author} {\bibfnamefont
  {T.}~\bibnamefont {Teraji}}, \bibinfo {author} {\bibfnamefont
  {S.}~\bibnamefont {Koizumi}}, \bibinfo {author} {\bibfnamefont
  {J.}~\bibnamefont {Isoya}},  \emph {et~al.},\ }\href@noop {} {\bibfield
  {journal} {\bibinfo  {journal} {Physical review letters}\ }\textbf {\bibinfo
  {volume} {111}},\ \bibinfo {pages} {067601} (\bibinfo {year}
  {2013})}\BibitemShut {NoStop}%
\end{thebibliography}%
	        
\end{document}


\title[Short Title]{Supplementary information of Confined nano-NMR spectroscopy using NV centers}
\author{D. Cohen}
\email{email: daniel.cohen7@mail.huji.ac.il}
\affiliation{Racah Institute of Physics, The Hebrew University of Jerusalem, Jerusalem 
	91904, Givat Ram, Israel}
\author{R. Nigmatullin}
\affiliation{Center for Engineered Quantum Systems, Dept. of Physics \& Astronomy, Macquarie University, 2109 NSW, Australia}
\author{M. Eldar}
\author{A. Retzker}
\affiliation{Racah Institute of Physics, The Hebrew University of Jerusalem, Jerusalem 
	91904, Givat Ram, Israel}

	\maketitle
	
	\section{Decomposing the dipole-dipole interaction into angular momentum representation}
	\label{Appendix: DD}
	We start with the dipole-dipole interaction Hamiltonian:
	\beq
	H_{DD}/(\gamma_e J)=-\left|\bar{r}\right|^{-3}\left(3\left(\bar{S}\cdot\hat{r}\right)\left(\bar{I}\cdot\hat{r}\right)-\bar{S}\cdot\bar{I}\right)
	\eeq
	where we denote $J=\frac{\mu_0 \gamma_{N}}{4\pi}$ for short.
	We write the inner product explicitly:
	\begin{align}
	H_{DD}/J=-&\left|\bar{r}\right|^{-3}\left(3\left(S_x\sin(\theta)\cos(\phi)+S_y\sin(\theta)\sin(\phi)+S_z\cos(\theta)\right)\times\right.\\
	\nonumber&\left.\left(I_x\sin(\theta)\cos(\phi)+I_y\sin(\theta)\sin(\phi)+I_z\cos(\theta)\right)-S_xI_x-S_yI_y-S_zI_z\right)\\
	\end{align}
	The product of the two z components will yield (up to a minus sign):
	\beq
	A=-\left|\bar{r}\right|^{-3}\left(3\cos^2(\theta)-1\right)S_zI_z=4\sqrt{\frac{\pi}{5}}\left|\bar{r}\right|^{-3}Y_2^0(\Omega)S_zI_z
	\eeq
	Where $Y_l^m(\Omega)$ is a spherical harmonic. This term corresponds to $m=0$.
	The product of the other same axis terms will yield:
	\begin{align}
	-&\left|\bar{r}\right|^{-3}\left[\left(3\sin^2(\theta)\cos^2(\phi)-1\right)S_xI_x+\left(3\sin^2(\theta)\sin^2(\phi)-1\right)S_yI_y\right]\\
	&=-\frac{1}{4}\left|\bar{r}\right|^{-3}\left[\left(3\sin^2(\theta)\cos^2(\phi)-1\right)(S_++S_-)(I_++I_-)-\left(3\sin^2(\theta)\sin^2(\phi)-1\right)(S_+-S_-)(I_+-I_-)\right]\\
	&=-\frac{1}{4}\left|\bar{r}\right|^{-3}\left[\left(S_+I_-+S_-I_+\right)\left(3\cos^2(\theta)-1\right)+3\left(S_+I_++S_-I_-\right)\sin^2(\theta)\cos(2\phi)\right]
	\end{align}
	We denote:
	\beq
	B=-\frac{1}{4} \left(4\sqrt{\frac{\pi}{5}}\right)\left|\bar{r}\right|^{-3}\left(S_+I_-+S_-I_+\right)Y_2^0(\Omega)
	\eeq
	Which is the second $m=0$ term.
	The mixed XY terms will have similar contributions as the rest of the previous product:
	\begin{align}
	-&3\left|\bar{r}\right|^{-3}\left(S_xI_y+S_yI_x\right)\sin^2(\theta)\sin(\phi)\cos(\phi)\\
	&=\frac{3}{4}\left|\bar{r}\right|^{-3} i \left[S_+I_+-S_-I_-\right]\sin^2(\theta)\sin(2\phi)
	\end{align}
	Summing both contributions we can define:
	\begin{align}
	E&=-\frac{3}{4}\left(4\sqrt{(\frac{2\pi}{15})}\right)Y_2^{-2}(\Omega)S_+I_+\\
	F&=-\frac{3}{4}\left(4\sqrt{(\frac{2\pi}{15})}\right)Y_2^{2}(\Omega)S_-I_-\\
	\end{align}
	These elements correspond to $m=\pm2$.
	The remaining elements are:
	\begin{align}
	-&3\left|\bar{r}\right|^{-3}\left[S_z\left(I_x\cos(\phi)+I_y\sin(\phi)\right)+I_z\left(S_x\cos(\phi)+S_y\sin(\phi)\right)\right]\cos(\theta)\sin(\theta)\\
	-&\frac{3}{2}\left|\bar{r}\right|^{-3}\left[\left(S_zI_++I_zS_+\right)e^{-i\phi}+\left(S_zI_-+I_zS_-\right)e^{i\phi}\right]\cos(\theta)\sin(\theta)
	\end{align}
	Which leads us to our final definitions of the $m=\pm1$ contributions:
	\begin{align}
	C=-\frac{3}{2}\left|\bar{r}\right|^{-3}\left(2\sqrt{(\frac{2\pi}{15})}\right)\left(S_zI_++I_zS_+\right)Y_2^{-1}(\Omega)\\
	D=\frac{3}{2}\left|\bar{r}\right|^{-3}\left(2\sqrt{(\frac{2\pi}{15})}\right)\left(S_zI_-+I_zS_-\right)Y_2^{1}(\Omega)
	\end{align}
	We define the prefactors
	\begin{align}
	\label{SH_coeff}
	&\zeta_0=-J,\ \tilde{\zeta}_0=-4\sqrt{\frac{\pi}{5}}J\\\nonumber
	&\zeta_1=\frac{3}{2}J,\ \tilde{\zeta}_1=\frac{3}{2}\left(2\sqrt{\frac{2\pi}{15}}\right)J\\\nonumber
	&\zeta_2=-\frac{3}{4}J,\ \tilde{\zeta}_2=-\frac{3}{4}\left(4\sqrt{\frac{2\pi}{15}}\right)J\\\nonumber
	&\zeta_{-m}= \zeta_m,\ 	\tilde{\zeta}_{-m}=\left(-1\right)^m \tilde{\zeta}_m
	\end{align}
	
	\section{Measurement protocol}
	We explore two measurement schemes - correlation spectroscopy \cite{Corr_spect} and a Qdyne type measurement \cite{qdyne}. In the following we provide an outline of the two.  
	We consider the NV center as a two level system (spin $\frac{1}{2}$) with an energy gap of $\gamma_e B_{ext}$, where $B_{ext}$ is the externally applied magnetic field \cite{Review_NV}.
	We denote $S_i\ (I_i)$ as the spin operator of the NV (nuclear spin) at the $i$ direction, $S_{\pm}\ (I_\pm)$ as the NV's (nuclear spin's) raising/lowering operators and $\theta_i$ as the angle between the NV's magnetization axis to the vector connecting the NV to the $i$-th nucleus.
	The free Hamiltonian of system in the presence of an external magnetic field is
	\beq\label{free1}
	H_0 =\frac{\omega_0}{2}S_z+\frac{\omega_N}{2}\sum_{i=1}^N I_z^i.
	\eeq
	The NV interacts with the ensemble via the dipolar interaction, which is of the general form
	\beq\label{dipolar1}
	H_{DD}=\sum_{i=1}^N\sum_{j,k=1}^3 g_{i,j,k}\left(\bar{r}_i\right)S_j I_k^i.
	\eeq
	We assume that $\omega_0$ is the largest frequency in the system. Therefore, in the interaction picture with respect to the \eqref{free1}, Eq. \eqref{dipolar1} can be written as
	\begin{align}
	&H_{DD}^I\approx S_z\sum_{j=1}^N\label{dipolar2}\left[ \tilde{g}_{+}^j\left(\bar{r}_j\right)I_+^j e^{i\omega_N t}+\tilde{g}_{-}^j\left(\bar{r}_j\right)I_-^j e^{-i\omega_N t}+\tilde{g}_{z}^j\left(\bar{r}_j\right) I_z\right],
	\end{align}
	where took the rotating wave approximation. 
	Driving the NV center with $N_p$ $\pi-$pulses applied at a frequency of $\omega_p$, transforms \eqref{dipolar2} to the effective Hamiltonian
	\beq\label{dipolar3}
	\tilde{H}_{DD}^I = w\left(t\right) H_{DD}^I,
	\eeq
	where
	\beq\label{window}
	w\left(t\right)=\sum_{n=0}^{N_{p}-1}\left(-1\right)^{n}\Theta\left(t-n\tau_{p}\right)\Theta\left(t-\left(n+1\right)\tau_{p}\right)
	\eeq
	and $\tau_p=\frac{\pi}{\omega_p}$. Assuming $N_p$ is odd, the Fourier series of \eqref{window} is
	\beq\label{window2}
	h\left(t\right)=-\frac{2i}{\pi}\sum_{n=-\infty}^\infty\frac{1}{2n+1}e^{i(2n+1)\omega_pt}.
	\eeq
	Hence, if $\omega_p$ is close to $\omega_N$, such that $\left(\omega_N-\omega_p\right) t \equiv \delta t \ll 1$, applying the rotating wave approximation to \eqref{dipolar3} leads to
	\beq\label{dipolar4}
	\tilde{H}_{DD}\approx-\frac{2}{\pi}\sum_{i=1}^N g^i_{\pm}S_z\left[  I_x^i\sin\left(\delta t-\varphi_i\right) + I_y^i\cos\left(\delta t-\varphi_i\right)\right],
	\eeq
	where we used the explicit spatial forms $\tilde{g}_{\pm}^j=-\frac{3}{2}Jr_j^{-3}\cos\theta_j\sin\theta_je^{\left(-1\right)^ki\varphi_j}$ and denoted $g_{\pm}^j= -J\frac{3}{2}r_j^{-3}\cos\theta_j\sin\theta_j$ \cite{Laura_arxiv}.
	
	In correlation spectroscopy the NV interacts with nuclear spins for a period of time $\tau$ via \eqref{dipolar4} at the end of which the NV state is written onto a memory qubit. After a waiting time $t$ the experiment is carried out again for the same period time $\tau$. In the weak back-action limit, namely when $N g \tau \ll 1$ for a characteristic interaction strength $g$, the operators in \eqref{dipolar4} can be approximated to their classical values $\left<I_x^i\right>=\cos\left(\phi_i\right),\ \left<I_y^i\right>=\sin\left(\phi_i\right)$, where $\phi$ is a uniform random variable on $\left[0,2\pi\right]$. Eq. \eqref{dipolar4} can be written accordingly  
	\begin{align}\label{dipolar5}
	\tilde{H}_{DD}&\approx-\frac{2}{\pi}\sum_{i=1}^N g^i_{\pm}S_z\left[  \cos\left(\phi_i\right)\sin\left(\delta t-\varphi_i\right) + \sin\left(\phi_i\right)\cos\left(\delta t-\varphi_i\right)\right]\\\nonumber
	&=-\frac{2}{\pi}\sum_{i=1}^N g^i_{\pm}S_z\sin\left(\delta t-\varphi_i+\phi_i\right).
	\end{align}
	
	We now turn to the explicit derivation of the protocol.
	The NV is initialized to the state
	\beq\label{corr_spect1}
	\ket{\psi_0}=\frac{1}{\sqrt{2}}\left(\ket{\uparrow}+\ket{\downarrow}\right).
	\eeq
	The state \eqref{corr_spect1} is propagated with \eqref{dipolar5} for a period of time $\tau$. Assuming $\delta\tau\ll1$ this results in
	\beq\label{corr_spect2}
	\ket{\psi_\tau}=\frac{1}{\sqrt{2}}\left(e^{\frac{4i\tau}{\pi}\sum_{i=1}^N g^i_{\pm}\left(0\right)\sin\left(-\varphi_i+\phi_i\right)}\ket{\uparrow}+\ket{\downarrow}\right).
	\eeq
	Note, that $g_\pm^i$ depends on time through the diffusion process. 
	After a waiting time $t$, the state \eqref{corr_spect2} evolves again under \eqref{dipolar5} yielding 
	\beq\label{corr_spect3}
	\ket{\psi_{t+2\tau}}=\frac{1}{\sqrt{2}}\left(e^{\frac{4i\tau}{\pi}\sum_{i=1}^N \left[g^i_{\pm}\left(0\right)\sin\left(-\varphi_i+\phi_i\right)+g^i_{\pm}\left(t\right)\sin\left(\delta t-\varphi_i+\phi_i\right)\right]}\ket{\uparrow}+\ket{\downarrow}\right).
	\eeq
	The probability of measuring $\ket{\psi_0}$ after a time $t+2\tau$ is then
	\beq\label{corr_spect4}
	P_{\ket{\psi_0}}=\frac{1}{2}\left\{1+\cos\left[\frac{4\tau}{\pi}\sum_{i=1}^N \left[g^i_{\pm}\left(0\right)\sin\left(-\varphi_i+\phi_i\right)+g^i_{\pm}\left(t\right)\sin\left(\delta t-\varphi_i+\phi_i\right)\right]\right]\right\}.
	\eeq
	In the weak back-action limit $N g \tau\ll1$ Eq. \eqref{corr_spect4} can be simplified further 
	\beq\label{corr_spect6}
	P_{\ket{\psi_0}}\approx 1-4\left(\frac{\tau}{\pi}\right)^2\sum_{i=1}^N \left[g^i_{\pm}\left(0\right)\sin\left(-\varphi_i+\phi_i\right)+g^i_{\pm}\left(t\right)\sin\left(\delta t-\varphi_i+\phi_i\right)\right]^2.
	\eeq
	The averaging of $\phi_i$ in \eqref{corr_spect6} will result in 
	\beq\label{corr_spect7}
	P_{\ket{\psi_0}}\approx 1-4\left(\frac{\tau}{\pi}\right)^2\left[\sum_{i=1}^N \left(g^i_{\pm}\right)^2+\cos\left(\delta t\right)\sum_{i=1}^N g^i_{\pm}\left(0\right)g^i_{\pm}\left(t\right)\right].
	\eeq
	Averaging \eqref{corr_spect7} over realizations yields
	\beq
		P_{\ket{\psi_0}}\approx 1-4\left(\frac{\gamma_e B_{rms}\tau}{\pi}\right)^2\left[1+\cos\left(\delta t\right)\frac{C\left(t\right)}{\gamma_e^2 B_{rms}^2}\right],
	\eeq 	
	where $C\left(t\right)$ is the correlation function at time $t$ and $C\left(0\right)\equiv\gamma_e^2B_{rms}^2$.
	
	\section{The mean field calculations}
	\label{signal_finite}
	Assuming the nuclei have finite polarization $p$. 
	The mean field of an NV whose magnetization axis coincides with the $\hat{z}$ axis is \cite{Laura_arxiv}
	\beq\label{mean_field}
	\left<B\right>^0=p\tilde{\zeta}_0\int\frac{d^3r}{r^3}Y_2^{(0}\left(\Omega\right),\equiv p J I_1^{(0)}.
	\eeq
	Given the that the nuclei are confined to a cylinder of height $L$ and radius $R$ the integral \eqref{mean_field} can be calculated by  
	\begin{align}
	\label{Cylindrical}
	I_1^{\left(0\right)}&=-\int\limits_0^{2\pi}d\phi\int\limits_0^{R} d\rho\int\limits_d^{d+L} dz \frac{\rho}{\left(\rho^2+z^2\right)^{3/2}}\left(\frac{3z^2}{\rho^2+z^2}-1\right)=-2\pi \int\limits_0^{R} d\rho\int\limits_d^{L+d} dz \frac{\rho}{\left(\rho^2+z^2\right)^{3/2}}\left(\frac{3z^2}{\rho^2+z^2}-1\right)\\\nonumber
	&=-2\pi \int\limits_d^{L+d} dz \frac{R^2}{\left(R^2+z^2\right)^{3/2}}=-2\pi\left(\frac{L+d}{\sqrt{\left(L+d\right)^2+R^2}}-\frac{d}{\sqrt{d^2+R^2}}\right).
	\end{align}
	
	The same calculation carried out for a hemispherical volume of radius $R$ yields
	\begin{align}
	\label{Hemispherical}
	I_1^{\left(0\right)}&=-\int\limits_0^{2\pi}d\phi\int\limits_d^{d+R} dz \int\limits_{0}^{\sqrt{R^2-(z-d)^2}} \frac{\rho d\rho}{\left(\rho^2+z^2\right)^{3/2}}\left(3\frac{z^2}{\rho^2+z^2}-1\right)=-2\pi\int\limits_d^{d+R} dz \int\limits_{0}^{\sqrt{R^2-(z-d)^2}} \frac{\rho d\rho}{\left(\rho^2+z^2\right)^{3/2}}\left(3\frac{z^2}{\rho^2+z^2}-1\right)\\\nonumber
	&=-2\pi\int\limits_{d}^{R+d}dz\frac{R^2-(d-z)^2}{\left(R^2-d (d-2 z)\right){}^{3/2}}=-\frac{2\pi}{3}\left[2-\frac{2d}{\sqrt{d^{2}+R^{2}}}-\frac{R^{2}}{d\sqrt{d^{2}+R^{2}}}-\frac{2R^{3}}{d^{3}}\left(\frac{R}{\sqrt{d^{2}+R^{2}}}-1\right)\right]
	\end{align}
	Finally, we calculate the same integral for a spherical volume of radius $R$
	\begin{align}\label{signal_sphere}
	&I_1^{\left(0\right)}&=-2\pi\int\limits_{d}^{d+2R} dz \int\limits_{0}^{\sqrt{R^2-(z-d-R)^2}} \frac{\rho d\rho}{\left(\rho^2+z^2\right)^{3/2}}\left(3\frac{z^2}{\rho^2+z^2}-1\right)=2\pi \int\limits_{d}^{d+2R} dz \frac{(d-z) (d+2 R-z)}{(2 z (d+R)-d (d+2 R))^{3/2}}
	=-\frac{8\pi R^3}{3 (d+R)^3}.
	\end{align}
	We notice that \eqref{Hemispherical} recovers the half-space calculation of \cite{Laura_arxiv}  in the limit $R_0\rightarrow\infty$ as expect.  Eq. \eqref{signal_sphere} recovers the half-space limit upto a factor of $2$. 
	Yet, Eq. \eqref{Cylindrical} in the limit $R\rightarrow\infty$ goes to zero. This is because the limit $R\rightarrow\infty$ implies $R\gg L,d$ and specifically $L-d\ll R$ which is the 2D limit. Since $\int\limits_{0}^1d\cos\theta Y_2^0\left(\Omega\right)=0$, the weighting of the $r^{-3}$ of the dipolar interaction is crucial for a non-zero signal. But if we take the 2D limit  \begin{align}
	I_1^{\left(0\right)}&=-\int\limits_0^{2\pi}d\phi\int\limits_0^{R} d\rho\int\limits_d^{L+d} dz \frac{\rho}{\left(\rho^2+z^2\right)^{3/2}}\left(\frac{3z^2}{\rho^2+z^2}-1\right)\approx-2\pi L \int\limits_0^{\infty} d\rho \frac{\rho}{\left(\rho^2+d^2\right)^{3/2}}\left(\frac{3d^2}{\rho^2+d^2}-1\right).
	\end{align}
	Since $\rho=d\tan\theta$ we have
	\beq
	I_1^{\left(0\right)}=-2\pi\frac{L}{d} \int\limits_0^{1} d\cos\theta \left(3\cos^2\theta-1\right),
	\eeq
	which is surprisingly the result of constant distance.
	\subsection{The diffusion propagator in a cylinder}
In the following we calculate the diffusion propagator in a cylindrical geometry required for the calculation of the correlation function. 

The propagator is the solution to the equation  
	
	\begin{equation}\label{cylinder_prop1}
	\frac{\partial P}{\partial t}-D\left(\frac{1}{\rho}\frac{\partial}{\partial\rho}\left(\rho\frac{\partial P}{\partial\rho}\right)+\frac{1}{\rho^{2}}\frac{\partial^{2}P}{\partial\phi^{2}}+\frac{\partial^{2}P}{\partial z^{2}}\right)=0
	\end{equation}
	with the boundary conditions 
	\begin{equation}\label{cylinder_prop2}
	\left(\frac{\partial P}{\partial\rho}\right)_{\rho=R}=0,\ \ \left(\frac{\partial P}{\partial z}\right)_{z=d}=0,\ \ \left(\frac{\partial P}{\partial z}\right)_{z={L+d}}=0,\ P\left(\phi+2\pi\right)=P\left(\phi\right)
	\end{equation}
	and the initial conditions 
	\beq\label{initial_cond_cylinder_prop}
	P\left(\rho,\phi,z,t=0\right)=\delta\left(\rho-\rho_{0}\right)\delta\left(z-z_{0}\right)\frac{\delta\left(\phi-\phi_{0}\right)}{\rho}
	\eeq
	We guess a solution to Eq. \eqref{cylinder_prop1} of the form:
	
	\begin{equation}\label{cylinder_prop3}
	P=T\left(t\right)f\left(\rho,\phi,z\right)
	\end{equation}
	Substituting \eqref{cylinder_prop3} into \eqref{cylinder_prop1} we arrive at
	\begin{equation}\label{cylinder_prop4}
	\frac{\dot{T}}{T}=\frac{D}{f}\left(\frac{1}{\rho}\frac{\partial}{\partial\rho}\left(\rho\frac{\partial f}{\partial\rho}\right)+\frac{1}{\rho^{2}}\frac{\partial^{2}f}{\partial\phi^{2}}+\frac{\partial^{2}f}{\partial z^{2}}\right).
	\end{equation}
	
	Since both sides of Eq. \eqref{cylinder_prop4} depend on different independent variables, we deduce that 
	
	\begin{equation}\label{cylinder_prop5}
	\frac{\dot{T}}{T}=-C_{1}=\frac{D}{f}\left(\frac{1}{\rho}\frac{\partial}{\partial\rho}\left(\rho\frac{\partial f}{\partial\rho}\right)+\frac{1}{\rho^{2}}\frac{\partial^{2}f}{\partial\phi^{2}}+\frac{\partial^{2}f}{\partial z^{2}}\right),
	\end{equation}
	where $C_1$ is a constant.
	The solution to the temporal equation of \eqref{cylinder_prop5} is straight forward
	
	\begin{equation}\label{cylinder_prop_time}
	T=A_{T}\exp\left(-C_{1}t\right).
	\end{equation}
	
	We are left with the right-hand side of Eq. \eqref{cylinder_prop5}  equation
	
	\begin{equation}\label{cylinder_prop6}
	-\frac{C_{1}}{D}f=\frac{1}{\rho}\frac{\partial}{\partial\rho}\left(\rho\frac{\partial f}{\partial\rho}\right)+\frac{1}{\rho^{2}}\frac{\partial^{2}f}{\partial\phi^{2}}+\frac{\partial^{2}f}{\partial z^{2}}
	\end{equation}
	
	We guess a separated solution of the form 
	
	\begin{equation}\label{cylinder_prop7}
	f\left(\rho,\phi,z\right)=Z\left(z\right)g\left(\rho,\phi\right).
	\end{equation}
	
Substituting \eqref{cylinder_prop7} into \eqref{cylinder_prop6} yields
	
	\begin{equation}\label{cylinder_prop8}
	-\frac{1}{Z}\frac{\partial^{2}Z}{\partial z^{2}}=\frac{1}{g}\left(\frac{1}{\rho}\frac{\partial}{\partial\rho}\left(\rho\frac{\partial g}{\partial\rho}\right)+\frac{1}{\rho^{2}}\frac{\partial^{2}g}{\partial\phi^{2}}+\frac{C_{1}}{D}g\right).
	\end{equation}
	
Using the same argument as before, we deduce that
	
	\begin{equation}\label{cylinder_prop9}
	-\frac{1}{Z}\frac{\partial^{2}Z}{\partial z^{2}}=-C_{2}=\frac{1}{g}\left(\frac{1}{\rho}\frac{\partial}{\partial\rho}\left(\rho\frac{\partial g}{\partial\rho}\right)+\frac{1}{\rho^{2}}\frac{\partial^{2}g}{\partial\phi^{2}}+\frac{C_{1}}{D}g\right),
	\end{equation}
	where $C_2$ is constant.
	
	The solution to the part of equation \eqref{cylinder_prop9} on  $Z$ together with the boundary conditions \eqref{cylinder_prop2} leads to 
	
	\begin{equation}\label{cylinder_prop10}
	C_{2}\left(k\right)=-\left(\frac{\pi k}{L}\right)^{2},\ k\in\mathbb{\mathbb{N}_0}
	\end{equation}
and
	\begin{equation}\label{cylinder_propr_Z}
	Z_{k}=A_z^k\cos\left[\frac{\pi k\left(z-d\right)}{L}\right].
	\end{equation}
	
	We continue solving Eq. \eqref{cylinder_prop9} for $g$ by guessing
	
	\begin{equation}\label{cylinder_prop11}
	g_{n}\left(\rho,\phi\right)=R_{n}\left(\rho\right)e^{in\phi},
	\end{equation}
	where $n\in\mathbb{Z}$ is required in order to satisfy the periodic boundary condition \eqref{cylinder_prop2}.
Substituting Eq. \eqref{cylinder_prop11} into \eqref{cylinder_prop9} we arrive at

	\begin{equation}\label{cylinder_prop12}
	0=\rho^{2}\frac{\partial^{2}R_{n}}{\partial\rho^{2}}+\rho\frac{\partial R_{n}}{\partial\rho}+\left[\left(\frac{C_{1}}{D}+C_{2}\right)\rho^{2}-n^{2}\right]R_{n}
	\end{equation}
	
	We define
	
	\begin{equation}\label{cylinder_prop13}
	\tilde{\rho}=\alpha\rho,
	\end{equation}
	
	such that
	
	\begin{equation}\label{cylinder_prop14}
	\frac{\partial R_{n}}{\partial\rho}=\frac{\partial R_{n}}{\partial\tilde{\rho}}\frac{\partial\tilde{\rho}}{\partial\rho}=\alpha\frac{\partial R_{n}}{\partial\tilde{\rho}}
	\end{equation}
	
 Rewriting \eqref{cylinder_prop12} in terms of \eqref{cylinder_prop13} leads to
	
	\beq\label{cylinder_prop15}
	0=\tilde{\rho}^{2}\frac{\partial^{2}R_{n}}{\partial\tilde{\rho}^{2}}+\tilde{\rho}\frac{\partial R_{n}}{\partial\tilde{\rho}}+\left[\tilde{\rho}^{2}-n^{2}\right]R_{n},
	\eeq
	
	where we chose $\alpha=\sqrt{\left(C_{2}+\frac{C_{1}}{D}\right)}$. This is the
	Bessel equation with solutions
	
	\begin{equation}
	R_{n}=A_{n}J_{n}\left(\tilde{\rho}\right)+B_{n}Y_{n}\left(\tilde{\rho}\right)=A_{n}J_{n}\left(\sqrt{\left(\frac{C_{1}}{D}+C_{2}\right)}\rho\right)+B_{n}Y_{n}\left(\sqrt{\left(\frac{C_{1}}{D}+C_{2}\right)}\rho\right).
	\end{equation}
	
	The propagator should be regular at $\rho=0$, so $\forall n\ B_{n}=0$,
	reducing the solution to
	
	\begin{equation}\label{cylinder_prop16}
	R_{n}=A_{n}J_{n}\left(\sqrt{\left(\frac{C_{1}}{D}+C_{2}\right)}\rho\right)
	\end{equation}
	
	Imposing the boundary condition \eqref{cylinder_prop2} on $\rho$ leads to 
	
	\begin{equation}\label{cylinder_prop17}
	\left(\frac{\partial R_{n}}{\partial\rho}\right)_{\rho=R}=0=\sqrt{\left(\frac{C_{1}}{D}+C_{2}\right)}A_{n}\left(\frac{\partial}{\partial\tilde{\rho}}J_{n}\left(\tilde{\rho}\right)\right)_{\tilde{\rho}=\sqrt{\left(\frac{C_{1}}{D}+C_{2}\right)}R}
	\end{equation}
	
	Denoting by $\nu_{m,n}$ the m'th zero of the derivative of the n'th
	Bessel function of the first kind, Eq. \eqref{cylinder_prop17} requires
	
	\begin{equation}\label{cylinder_prop18}
	\sqrt{\left(\frac{C_{1}\left(n,m,k\right)}{D}+C_{2}\left(k\right)\right)}R=\nu_{m,n}.
	\end{equation}
The radial dependence of the propagator is therefore
	\begin{equation}\label{cylinder_prop_radial}
	R_{n,m}=A_{n}J_{n}\left(\nu_{m,n}\frac{\rho}{R}\right).
	\end{equation}
	Eq. \eqref{cylinder_prop18} together with Eq. \eqref{cylinder_prop10} determines the value of $C_1$,
	\begin{equation}\label{cylinder_prop19}
	C_{1}\left(n,m,k\right)=D\left[\left(\frac{\nu_{m,n}}{R}\right)^{2}+\left(\frac{\pi k}{L}\right)^{2}\right]
	\end{equation}
	We note that for $n=0=C_{1}=0=C_{2}=0$ the solution Eq. \eqref{cylinder_prop12} is a
	constant that we need to include in the final solution.
	
	Summing up Eqs. \eqref{cylinder_prop_time}, \eqref{cylinder_propr_Z} and \eqref{cylinder_prop_radial} the general solution to Eq. \eqref{cylinder_prop1} with the boundary conditions \eqref{cylinder_prop2} is 
	\begin{equation}\label{Prop_cylinder_general}
	P=C+\sum_{m=1}^{\infty}\sum_{k=0}^{\infty}\sum_{n=-\infty}^{\infty}A_{m,k,n}\cos\left[\pi k\left(\frac{z-d}{L}\right)\right]J_{n}\left(\nu_{m,n}\frac{\rho}{R}\right)e^{in\phi}\exp\left[-D\left[\left(\frac{\nu_{m,n}}{R}\right)^{2}+\left(\frac{\pi k}{L}\right)^{2}\right]t\right].
	\end{equation}
	The initial condition \eqref{initial_cond_cylinder_prop} will set the coefficients $A_{m,k,n}:$ 
	\begin{equation}\label{cylinder_prop20}
	 C+\sum_{m=1}^{\infty}\sum_{k=0}^{\infty}\sum_{n=-\infty}^{\infty}A_{m,k,n}\cos\left[\pi k\left(\frac{z-d}{L}\right)\right]J_{n}\left(\nu_{m,n}\frac{\rho}{R}\right)e^{in\phi}=\delta\left(\rho-\rho_{0}\right)\delta\left(z-z_{0}\right)\frac{\delta\left(\phi-\phi_{0}\right)}{\rho}.
	\end{equation}
	The sum over k is independent, hence, we can solve the $z$ dependency of \eqref{cylinder_prop20} separately:
	\begin{equation}\label{cylinder_prop21}
	\sum_{k=0}^{\infty}A_{k}\cos\left[\pi k\left(\frac{z-d}{L}\right)\right]=\delta\left(z-z_{0}\right).
	\end{equation}
	We  define
	\begin{align}\label{cylinder_prop22}
	&\tilde{z}=\pi\left(\frac{z-d}{L}\right),\ z=d+\frac{\tilde{z}}{\pi}L\\\nonumber
	&\tilde{z}_{0}=\pi\left(\frac{z_{0}-d}{L}\right),z_{0}=d+\frac{\tilde{z}_{0}}{\pi}L.
	\end{align}
	Substituting Eq. \eqref{cylinder_prop22} into \eqref{cylinder_prop21} and using the delta function properties, 
	\begin{equation}\label{cylinder_prop23}
	\sum_{k=0}^{\infty}A_{k}\cos\left[k\tilde{z}\right]=\delta\left(\frac{L}{\pi}\left(\tilde{z}-\tilde{z}_{0}\right)\right)=\frac{\pi}{L}\delta\left(\tilde{z}-\tilde{z}_{0}\right).
	\end{equation}
	By multiplying both sides of \eqref{cylinder_prop23} by $\cos\left(s\tilde{z}\right)$ and integrating
	$\tilde{z}$ on $\left[0,\pi\right]$ we arrive at
	\begin{equation}\label{cylinder_prop24}
	A_{s}=\frac{2}{L\left(1+\delta_{s,0}\right)}\cos\left[s\tilde{z}_{0}\right].
	\end{equation}
	Therefore Eq. \eqref{Prop_cylinder_general} simplifies to 
	\begin{align}\label{cylinder_prop25}
	P=\frac{\tilde{C}}{L}+&\sum_{k=0}^{\infty}\frac{2}{L\left(1+\delta_{k,0}\right)}\cos\left[k\pi\left(\frac{z_{0}-d}{L}\right)\right]\cos\left[\pi k\left(\frac{z-d}{L}\right)\right]\times\\\nonumber
	&\sum_{m=1}^{\infty}\sum_{n=-\infty}^{\infty}A_{m,n}J_{n}\left(\nu_{m,n}\frac{\rho}{R}\right)e^{in\phi}\exp\left[-D\left[\left(\frac{\nu_{m,n}}{R}\right)^{2}+\left(\frac{\pi k}{L}\right)^{2}\right]t\right].
	\end{align}
	We continue with the rest of Eq. \eqref{cylinder_prop20}, 
	\begin{equation}\label{cylinder_prop26}
	\frac{\tilde{C}}{L}+\sum_{m=1}^{\infty}\sum_{n=-\infty}^{\infty}A_{m,n}J_{n}\left(\nu_{m,n}\frac{\rho}{R}\right)e^{in\phi}=\delta\left(\rho-\rho_{0}\right)\frac{\delta\left(\phi-\phi_{0}\right)}{\rho}.
	\end{equation}
	We multiply both sides of Eq. \eqref{cylinder_prop26} by $\rho e^{-il\phi}J_{l}\left(\nu_{c,l}\frac{\rho}{R}\right)$
	\begin{equation}\label{cylinder_prop27}
	\frac{\tilde{C}}{L}e^{-il\phi}\rho J_{l}\left(\nu_{c,l}\frac{\rho}{R}\right)+\sum_{m=1}^{\infty}\sum_{n=-\infty}^{\infty}A_{m,n}\rho J_{n}\left(\nu_{m,n}\frac{\rho}{R}\right)J_{l}\left(\nu_{c,l}\frac{\rho}{R}\right)e^{i\left(n-l\right)\phi}=\delta\left(\rho-\rho_{0}\right)\delta\left(\phi-\phi_{0}\right)e^{-il\phi}J_{l}\left(\nu_{c,l}\frac{\rho}{R}\right)
	\end{equation}
	and integrate both sides. The integration of the right hand-side of \eqref{cylinder_prop27} yields
	\begin{equation}\label{cylinder_prop28}
	\int\limits _{0}^{R}d\rho\int\limits _{0}^{2\pi}d\phi\delta\left(\rho-\rho_{0}\right)\delta\left(\phi-\phi_{0}\right)e^{-il\phi}J_{l}\left(\nu_{c,l}\frac{\rho}{R}\right)=e^{-il\phi_{0}}J_{l}\left(\nu_{c,l}\frac{\rho_{0}}{R}\right).
	\end{equation}
	The integration on the left-hand side of \eqref{cylinder_prop27} is equal to
	\begin{align}\label{cylinder_prop29}
	&\sum_{m=1}^{\infty}\sum_{n=-\infty}^{\infty}A_{m,n}\int\limits _{0}^{R}d\rho\rho J_{n}\left(\nu_{m,n}\frac{\rho}{R}\right)J_{l}\left(\nu_{c,l}\frac{\rho}{R}\right)\int\limits _{0}^{2\pi}d\phi e^{i\left(n-l\right)\phi}\\\nonumber
	&=2\pi\sum_{m=1}^{\infty}A_{m,l}\int\limits _{0}^{R}d\rho\rho J_{l}\left(\nu_{m,l}\frac{\rho}{R}\right)J_{l}\left(\nu_{c,l}\frac{\rho}{R}\right)=\pi R^{2}A_{c,l}J_{l}^{2}\left(\nu_{c,l}\right)\left(1-\left(\frac{l}{\nu_{c,l}}\right)^{2}\right),
	\end{align}
	where we used
	\begin{equation}
	\int\limits _{0}^{R}d\rho\rho J_{l}\left(\nu_{m,l}\frac{\rho}{R}\right)J_{l}\left(\nu_{c,l}\frac{\rho}{R}\right)\underbrace{=}_{c\neq m}\frac{\nu_{c,l}J_{l}(\nu_{m,l})J'(\nu_{c,l})-\nu_{m,l}J_{l}(\nu_{c,l})J'_{l}(\nu_{m,l}))}{\left(\nu_{m,l}\right)^{2}-\left(\nu_{c,l}\right)^{2}}.
	\end{equation}
	
	Substituting Eqs. \eqref{cylinder_prop28} and \eqref{cylinder_prop29} into \eqref{cylinder_prop27} yields:
	
	\begin{equation}\label{cylinder_prop30}
	A_{c,l}=\frac{1}{\pi R^{2}J_{l}^{2}\left(\nu_{c,l}\right)\left(1-\left(\frac{l}{\nu_{c,l}}\right)^{2}\right)}e^{-il\phi_{0}}J_{l}\left(\nu_{c,l}\frac{\rho_{0}}{R}\right).
	\end{equation}
	
	To determine the value of $\tilde{C}$ we integrate both sides
	of \eqref{cylinder_prop26} over the entire volume
	
	\begin{align}\label{cylinder_prop31}
	&\tilde{C}\int\limits _{0}^{2\pi}d\phi\int\limits _{0}^{R}d\rho\rho+\sum_{m=1}^{\infty}\sum_{n=-\infty}^{\infty}A_{m,n}\int\limits _{0}^{2\pi}d\phi\int\limits _{0}^{R}d\rho J_{n}\left(\nu_{m,n}\frac{\rho}{R}\right)e^{in\phi}=1,\\\nonumber
	&\tilde{C}\pi R^{2}+2\pi\sum_{m=1}^{\infty}A_{m,0}\int\limits _{0}^{R}d\rho\rho J_{0}\left(\nu_{m,0}\frac{\rho}{R}\right)=1,\\\nonumber
	&\tilde{C}\pi R^{2}+2\pi\sum_{m=1}^{\infty}\frac{A_{m,0}}{\nu_{m,0}}J_{1}\left(\nu_{m,0}\right)=1.
	\end{align}
	However, $\frac{d}{dx}J_{0}=-J_{1}\left(x\right)$ so the entire series
	in Eq. \eqref{cylinder_prop31} is zero, and
	\beq\label{cylinder_prop32}
	\tilde{C}=\frac{1}{\pi R^{2}}.
	\eeq
	
	The Green's function is therefore:
	\begin{align}\label{Propagator_cylinder}
	P=\frac{1}{V}+\frac{1}{V}\sum_{k=0}^{\infty}\sum_{m=1}^{\infty}\sum_{n=-\infty}^{\infty}&\frac{2}{J_{n}^{2}\left(\nu_{m,n}\right)\left(1-\left(\frac{n}{\nu_{m,n}}\right)^{2}\right)\left(1+\delta_{k,0}\right)}\cos\left[k\pi\left(\frac{z_{0}-d}{L}\right)\right]\cos\left[\pi k\left(\frac{z-d}{L}\right)\right]\times\\\nonumber
	&J_{n}\left(\nu_{m,n}\frac{\rho_{0}}{R}\right)J_{n}\left(\nu_{m,n}\frac{\rho}{R}\right)e^{in\left(\phi-\phi_{0}\right)}\exp\left[-D\left[\left(\frac{\nu_{m,n}}{R}\right)^{2}+\left(\frac{\pi k}{L}\right)^{2}\right]t\right],
	\end{align}
	where $V=\pi R^2 L$ is the cylinder's volume.
	\subsection{The diffusion propagator in a hemisphere/sphere}
	
We now repeat the previous section for a hemispherical geometry. We are looking for a solution to the equation
	\begin{equation}\label{sphere_prop1}
	\frac{\partial P}{\partial t}-D\left(\frac{1}{r^{2}}\frac{\partial}{\partial r}\left(r^{2}\frac{\partial P}{\partial r}\right)+\frac{1}{r^{2}sin\theta}\frac{\partial}{\partial\theta}\left(sin\theta\frac{\partial P}{\partial\theta}\right)+\frac{1}{r^{2}sin^{2}\theta}\frac{\partial^{2}P}{\partial\phi^{2}}\right)=0
	\end{equation}
	with the boundary conditions:
	\begin{equation}\label{sphere_prop2}
	\left[\frac{\partial P}{\partial r}\right]_{r=R}=0,\ \ \left[\frac{\partial P}{\partial z}\right]_{z=0}=0,\ \ P\left(\phi+2\pi\right)=P\left(\phi\right)
	\end{equation}
	and initial condition
	\beq\label{sphere_prop3}
	P\left(\bar{r},t=0\right)=\frac{\delta\left(r-r_{0}\right)\delta\left(\theta-\theta_{0}\right)\delta\left(\phi-\phi_{0}\right)}{r_{0}^{2}sin\theta_{0}}
	\eeq
	Using the method of images, we can impose the boundary condition $\left[\frac{\partial P}{\partial z}\right]_{z=0}=0$
	by symmetrically extending the initial conditions and the boundary conditions.
	
	We guess a solution to Eq. \eqref{sphere_prop1} of the form
	
	\begin{equation}\label{sphere_prop4}
	P=T\left(t\right)f\left(r,\theta,\phi\right).
	\end{equation}
	Substituting \eqref{sphere_prop4} into \eqref{sphere_prop1} leads to
	\begin{equation}
	\frac{\dot{T}}{T}=\frac{D}{f}\left(\frac{1}{r^{2}}\frac{\partial}{\partial r}\left(r^{2}\frac{\partial f}{\partial r}\right)+\frac{1}{r^{2}sin\theta}\frac{\partial}{\partial\theta}\left(sin\theta\frac{\partial f}{\partial\theta}\right)+\frac{1}{r^{2}sin^{2}\theta}\frac{\partial^{2}f}{\partial\phi^{2}}\right).
	\end{equation}
	We deduce
	\begin{equation}\label{sphere_prop5}
	\frac{\dot{T}}{T}=-C_{1}=\frac{D}{f}\left(\frac{1}{r^{2}}\frac{\partial}{\partial r}\left(r^{2}\frac{\partial f}{\partial r}\right)+\frac{1}{r^{2}sin\theta}\frac{\partial}{\partial\theta}\left(sin\theta\frac{\partial f}{\partial\theta}\right)+\frac{1}{r^{2}sin^{2}\theta}\frac{\partial^{2}f}{\partial\phi^{2}}\right).
	\end{equation}
	The solution for the temporal part of \eqref{sphere_prop5} is
	\begin{equation}\label{sphere_prop_temp}
	T=A_{T}exp\left(-C_{1}t\right).
	\end{equation}
	We are left with the right-hand side of equation \eqref{sphere_prop5},
	\begin{equation}\label{sphere_prop6}
	-\frac{C_{1}}{D}f=\frac{1}{r^{2}}\frac{\partial}{\partial r}\left(r^{2}\frac{\partial f}{\partial r}\right)+\frac{1}{r^{2}sin\theta}\frac{\partial}{\partial\theta}\left(sin\theta\frac{\partial f}{\partial\theta}\right)+\frac{1}{r^{2}sin^{2}\theta}\frac{\partial^{2}f}{\partial\phi^{2}}.
	\end{equation}
	We guess
	\begin{equation}\label{sphere_prop7}
	f\left(r,\theta,\phi\right)=g\left(r\right)Y\left(\theta,\phi\right)
	\end{equation}
	and substitute Eq. \eqref{sphere_prop7} into Eq. \eqref{sphere_prop6},
	\begin{equation}\label{sphere_prop8}
	\frac{1}{g}\frac{\partial}{\partial r}\left(r^{2}\frac{\partial g}{\partial r}\right)+\frac{C_{1}r^{2}}{D}=-\frac{1}{Y}\left(\frac{1}{sin\theta}\frac{\partial}{\partial\theta}\left(sin\theta\frac{\partial Y}{\partial\theta}\right)+\frac{1}{sin^{2}\theta}\frac{\partial^{2}Y}{\partial\phi^{2}}\right).
	\end{equation}
	We can again assume
	\begin{equation}\label{sphere_prop9}
	\frac{1}{g}\frac{\partial}{\partial r}\left(r^{2}\frac{\partial g}{\partial r}\right)+\frac{C_{1}r^{2}}{D}=C_{2}=-\frac{1}{Y}\left(\frac{1}{sin\theta}\frac{\partial}{\partial\theta}\left(sin\theta\frac{\partial Y}{\partial\theta}\right)+\frac{1}{sin^{2}\theta}\frac{\partial^{2}Y}{\partial\phi^{2}}\right).
	\end{equation}
	The solutions for $Y$ are the spherical harmonics:
	\begin{equation}\label{sphere_prop_angular}
	Y\left(\theta,\phi\right)=Y_{l}^{(m)}\left(\theta,\phi\right),
	\end{equation}
	and
	\begin{equation}\label{sphere_prop10}
	C_{2}=l\left(l+1\right),\ l\in\mathbb{N}_{0}.
	\end{equation}
	The radial part of Eq. \eqref{sphere_prop9} is therefore 
	\begin{equation}\label{sphere_prop11}
	r^{2}\frac{\partial^{2}g}{\partial r^{2}}+2r\frac{\partial g}{\partial r}+\left[\frac{C_{1}}{D}r^{2}-l\left(l+1\right)\right]g=0.
	\end{equation}
	if $C_{1}=0$ Eq. \eqref{sphere_prop11} simplifies to
	\begin{equation}\label{sphere_prop12}
	r^{2}\frac{\partial^{2}g}{\partial r^{2}}+2r\frac{\partial g}{\partial r}-l\left(l+1\right)g=0
	\end{equation}
	 The solution to \eqref{sphere_prop11} that is regular at $r=0$ is
	\beq\label{sphere_prop13}
	g\left(r\right)=d_{l}\cdot r^{l}
	\eeq
	 The boundary conditions \eqref{sphere_prop2} require from the solution \eqref{sphere_prop13}  
	\begin{equation}
	\left[\frac{\partial}{\partial r}\left(r^{l}\right)\right]_{r=R}=0
	\end{equation}
	from which we deduce that the only valid solution is $g=const$.
	
	Returning to \eqref{sphere_prop11}, if $C_{1}\neq0$, we can define:
	\begin{equation}\label{sphere_prop14}
	\tilde{r}=\sqrt{\frac{C_{1}}{D}}r
	\end{equation}
	Rewriting Eq. \eqref{sphere_prop11} in terms of the variable \eqref{sphere_prop14},
	\begin{equation}\label{sphere_prop15}
	\tilde{r}^{2}\frac{\partial^{2}g}{\partial\tilde{r}^{2}}+2\tilde{r}\frac{\partial g}{\partial\tilde{r}}+\left[\tilde{r}^{2}-l\left(l+1\right)\right]g=0
	\end{equation}
	The two linearly independent solutions to this equation are the spherical
	Bessel functions $j_{l}$ and $y_{l}$:
	\begin{equation}\label{sphere_prop16}
	g_{l}=a_{l}j_{l}\left(\sqrt{\frac{C_{1}}{D}}r\right)+b_{l}y_{l}\left(\sqrt{\frac{C_{1}}{D}}r\right)
	\end{equation}
	The propagator should be regular at r=0, so for all $l\in\mathbb{N}_{0}$
	it follows that $b_{l}=0$, reducing the solution to:
	\begin{equation}\label{sphere_prop17}
	g_{l}\left(r\right)=a_{l}j_{l}\left(\sqrt{\frac{C_{1}}{D}}r\right)
	\end{equation}
	The  radial boundary condition \eqref{sphere_prop2}, now reads:
	\begin{equation}\label{sphere_prop18}
	\left[\frac{\partial j_{l}}{\partial r}\right]_{r=R}=0
	\end{equation}
	We denote by $\tilde{\nu}_{k,l}$ the k'th zero of the derivative of the l'th
	spherical Bessel function of the first kind.  Eq. \eqref{sphere_prop18}, therefore, leads to 
	\begin{equation}\label{sphere_prop19}
	\tilde{\nu}_{k,l}=\sqrt{\frac{C_{1}\left(k,l\right)}{D}}R.
	\end{equation}
	Hence,
	\begin{equation}\label{sphere_prop20}
	C_{1}\left(k,l\right)=D\left(\frac{\tilde{\nu}_{k,l}}{R}\right)^{2}=\frac{\tilde{\nu}_{k,l}^{2}}{\tau_{V}}.
	\end{equation}
Note, that we assumed $C_1\neq0$, whereas $\tilde{\nu}_{1,l}=0$ for all $l\neq1$. We therefore change our definition of $\tilde{\nu}_{1,l}$ to the first root that does not equal zero.
Summing up Eqs. \eqref{sphere_prop_temp}, \eqref{sphere_prop_angular}, \eqref{sphere_prop17} and \eqref{sphere_prop20}, the general solution to \eqref{sphere_prop1} with the boundary conditions \eqref{sphere_prop2} is
	\begin{equation}\label{Prop_sphere_general}
	P=A+\sum_{l=0}^{\infty}\sum_{k=1}^{\infty}\sum_{m=-l}^{l}a_{k,l,m}\cdot j_{l}\left(\frac{\tilde{\nu}_{k,l}}{R}r\right)Y_{l}^{(m)}\left(\theta,\phi\right)exp\left(-\frac{\tilde{\nu}_{k,l}^{2}}{\tau_{V}}t\right).
	\end{equation}
	
	We now need to impose the symmetrically extended initial condition
	\begin{equation}\label{sphere_prop21}
	P\left(r,\theta,\phi,t=0\right)=\frac{1}{r^{2}\sin\theta}\delta\left(r-r_{0}\right)\delta\left(\phi-\phi_{0}\right)\left[\delta\left(\theta-\theta_{0}\right)+\delta\left(\theta-\left(\pi-\theta_{0}\right)\right)\right].
	\end{equation}
	
	First, we find the angular contribution by equating \eqref{Prop_sphere_general} at $t=0$ with
	\eqref{sphere_prop21}, multiplying both sides by $Y_{k}^{(s)}\left(\theta,\phi\right)^*$ and
	integrating:
	\begin{align}\label{sphere_prop22}
	\int d\Omega\frac{Y_{k}^{(s)}\left(\theta,\phi\right)^{*}}{\sin\theta}\delta\left(\phi-\phi_{0}\right)&\left[\delta\left(\theta-\theta_{0}\right)+\delta\left(\theta-\left(\pi-\theta_{0}\right)\right)\right]=\int d\Omega Y_{k}^{(s)}\left(\theta,\phi\right)^{*}\left(A+\sum_{l=0}^{\infty}\sum_{m=-l}^{l}a_{,l,m}\cdot Y_{l}^{(m)}\left(\theta,\phi\right)\right),\\\nonumber
	&a_{k,s}=Y_{k}^{(s)}\left(\theta_{0,}\phi_{0}\right)^{*}+Y_{k}^{(s)}\left(\pi-\theta_{0,}\phi_{0}\right)^{*}=\left[1+\left(-1\right)^{k+s}\right]Y_{k}^{(s)}\left(\theta_{0,}\phi_{0}\right)^{*}.
\end{align}
	
	The radial contribution to the coefficients can be found similarly by
	\begin{align}\label{sphere_prop23}
	&\int dr\delta\left(r-r_{0}\right)j_{l}\left(\frac{\tilde{\nu}_{w,l}}{R}r\right)=\sum_{l=0}^{\infty}\sum_{k=1}^{\infty}a_{k,l}\int drr^{2}j_{l}\left(\frac{\tilde{\nu}_{w,l}}{R}r\right)\cdot j_{l}\left(\frac{\tilde{\nu}_{k,l}}{R}r\right),\\
	&j_{l}\left(\frac{\tilde{\nu}_{w,l}}{R}r_{0}\right)=\frac{\pi R}{2}\sum_{l=0}^{\infty}\sum_{k=1}^{\infty}\frac{a_{k,l}}{\sqrt{\tilde{\nu}_{w,l}\tilde{\nu}_{k,l}}}\int dr r J_{l+1/2}\left(\frac{\tilde{\nu}_{w,l}}{R}r\right)\cdot J_{l+1/2}\left(\frac{\tilde{\nu}_{k,l}}{R}r\right)\\\nonumber
	&=\begin{cases}
	0 & w\neq k\\
	\frac{\pi R^{3}}{4\tilde{\nu}_{w,l}}a_{w,l}J_{l+1/2}^{2}\left(\tilde{\nu}_{w,l}\right)\left(1-\left(\frac{l}{\tilde{\nu}_{w,l}}\right)^{2}\right) & w=k
	\end{cases}
	=\begin{cases}
	0 & w\neq k\\
	\frac{R^{3}}{2}a_{w,l}j_{l}^{2}\left(\tilde{\nu}_{w,s}\right)\left(1-\left(\frac{s}{\tilde{\nu}_{w,s}}\right)^{2}\right) & w=k .
	\end{cases}
	\end{align}
	Therefore,
	\begin{equation}\label{sphere_prop24}
	a_{w,l}=\frac{2j_{l}\left(\frac{\tilde{\nu}_{w,l}}{R}r_{0}\right)}{R^{3}j_{l}^{2}\left(\nu_{w,l}\right)\left(1-\left(\frac{l}{\tilde{\nu}_{w,l}}\right)^{2}\right)}.
	\end{equation}
	Substituting Eqs. \eqref{sphere_prop22} and \eqref{sphere_prop24} into \eqref{Prop_sphere_general}
	leads to
	\beq\label{sphere_prop25}
	P=A+\frac{1}{V}\sum_{l=0}^{\infty}\sum_{k=1}^{\infty}\sum_{m=-l}^{l}\frac{4\pi\left[1+\left(-1\right)^{l+m}\right]}{3j_{l}^{2}\left(\tilde{\nu}_{k,l}\right)\left(1-\left(\frac{l}{\tilde{\nu}_{k,l}}\right)^{2}\right)}j_{l}\left(\frac{\tilde{\nu}_{k,l}}{R}r_{0}\right)j_{l}\left(\frac{\tilde{\nu}_{k,l}}{R}r\right)Y_{l}^{(m)}\left(\theta_{0},\phi_{0}\right)^{*}Y_{l}^{(m)}\left(\theta,\phi\right)exp\left(-\frac{\tilde{\nu}_{k,l}^{2}}{\tau_{V}}t\right),
	\eeq
	where $V=\frac{2\pi}{3}R^{3}$. 
	We find the constant $A$ by equating
	\eqref{sphere_prop25} at $t=0$  to \eqref{sphere_prop21} and integrating over the
	entire volume, which results in $A=\frac{1}{V}$. So finally:
	\beq\label{Hmisphere_prop}
	P=\frac{1}{V}+\frac{1}{V}\sum_{l=0}^{\infty}\sum_{k=1}^{\infty}\sum_{m=-l}^{l}\frac{4\pi\left[1+\left(-1\right)^{l+m}\right]}{3j_{l}^{2}\left(\tilde{\nu}_{k,l}\right)\left(1-\left(\frac{l}{\tilde{\nu}_{k,l}}\right)^{2}\right)}j_{l}\left(\frac{\tilde{\nu}_{k,l}}{R}r_{0}\right)j_{l}\left(\frac{\tilde{\nu}_{k,l}}{R}r\right)Y_{l}^{(m)}\left(\theta_{0},\phi_{0}\right)^{*}Y_{l}^{(m)}\left(\theta,\phi\right)exp\left(-\frac{\tilde{\nu}_{k,l}^{2}}{\tau_{V}}t\right).
	\eeq
	The propagator for a full sphere is found in a similar fashion and is equal to:
	\beq\label{Sphere_prop}
	P=\frac{1}{V}+\frac{1}{V}\sum_{l=0}^{\infty}\sum_{k=1}^{\infty}\sum_{m=-l}^{l}\frac{8\pi}{3j_{l}^{2}\left(\tilde{\nu}_{k,l}\right)\left(1-\left(\frac{l}{\tilde{\nu}_{k,l}}\right)^{2}\right)}j_{l}\left(\frac{\tilde{\nu}_{k,l}}{R}r_{0}\right)j_{l}\left(\frac{\tilde{\nu}_{k,l}}{R}r\right)Y_{l}^{(m)}\left(\theta_{0},\phi_{0}\right)^{*}Y_{l}^{(m)}\left(\theta,\phi\right)exp\left(-\frac{\tilde{\nu}_{k,l}^{2}}{\tau_{V}}t\right),
	\eeq
	with $V=\frac{4\pi}{3}R^3$.
\subsection{Correlation calculation}
In the following we provide general expressions for the magnetic field correlation function of a statistically polarized ensemble and then provide numeric results for the integrals.

The correlation is given by
\beq\label{corr_general}
C^{(m)}(t)=\tilde{\zeta}_m^2\int\frac{d^3r}{r^3}\int\frac{d^3r_0}{r_0^3} Y_2^{(m)}\left(\Omega\right)Y_2^{(m)*}\left(\Omega_0\right)P\left(\bar{r},\bar{r}_0,t\right),
\eeq
where $P$ is the appropriate diffusion propagator. While solving \eqref{corr_general} is generally hard, the asymptotic behavior of the correlation has a simplified expression. In short times $t\ll\tau_D$ the propagator approaches a delta function and therefore
\beq\label{Brms_general}
C^{m}\left(0\right)=B_{rms}^2=\tilde{\zeta}_m^2\int\frac{d^3r}{r^6} Y_2^{(m)}\left(\Omega\right)Y_2^{(m)*}\left(\Omega\right).
\eeq
In long times $t\gg\tau_V$, however, the propagator approaches the uniform distribution $V^{-1}$ and therefor
\beq\label{corr_longtimes_general}
C^{(m)}(t\gg\tau_V)\approx\frac{1}{V}\left(\tilde{\zeta}_m\int \frac{d^3r}{r^3} \left|Y_2^{(m)}\left(\Omega\right)\right|\right)^2.
\eeq 
The correlation in long times is therefore proportional to the squared mean field.
In the following we calculate \eqref{Brms_general} and \eqref{corr_longtimes_general} for three different geometries. In general, we expect the correlation to decay exponentially for times $t\ll\tau_D$, which yields the well-known Lorenztian spectrum at high frequencies. In intermediate times, $\tau_D\ll t\ll\tau_V$ the spectrum decays as a power law \cite{microfludics} as a result of diffusion, and at long times $t\gg\tau_V$ the spectrum decays exponentially to a constant value. In the following we offer some numerical results for  $\tilde{C}^{(m)}=\frac{\left[C^{(m)}(t)-C^{(m)} (t\gg\tau_V)\right]}{C^{(m)(0)}}$. which support these general arguments.
\subsubsection{cylinder}
We first solve \eqref{Brms_general} for each value of $m$ individually:
\begin{align}\label{Brms_cylinder0}
C^{(0)}(0)/J^2=&\frac{\pi}{64}\left\{\frac{1}{R^{2}}\left[\frac{23d}{d^{2}+R^{2}}+\frac{24d^{5}}{\left(d^{2}+R^{2}\right)^{3}}+\frac{16LR^{2}\left(3d^{2}+3dL+L^{2}\right)}{d^{3}(d+L)^{3}}-\frac{38d^{3}}{\left(d^{2}+R^{2}\right)^{2}}-\frac{23(d+L)}{(d+L)^{2}+R^{2}}+\frac{38(d+L)^{3}}{\left((d+L)^{2}+R^{2}\right)^{2}}\right.\right.\\\nonumber
&\left.\left.-\frac{24(d+L)^{5}}{\left((d+L)^{2}+R^{2}\right)^{3}}\right]+\frac{1}{R^{3}}\left[105\tan^{-1}\left(\frac{R}{d+L}\right)+96\tan^{-1}\left(\frac{d+L}{R}\right)-105\tan^{-1}\left(\frac{R}{d}\right)-96\tan^{-1}\left(\frac{d}{R}\right)\right]\right\},
\end{align}
\begin{align}\label{Brms_cylinder1}
&C^{(1)}(0)/J^2=\frac{\pi}{256}\left\{ \frac{1}{R^{2}}\left[-\frac{15d}{d^{2}+R^{2}}-\frac{24d^{5}}{\left(d^{2}+R^{2}\right)^{3}}+\frac{16LR^{2}\left(3d^{2}+3dL+L^{2}\right)}{d^{3}(d+L)^{3}}+\frac{54d^{3}}{\left(d^{2}+R^{2}\right)^{2}}+\frac{15(d+L)}{(d+L)^{2}+R^{2}}\right.\right.\\\nonumber
&\left.\left.-\frac{54(d+L)^{3}}{\left((d+L)^{2}+R^{2}\right)^{2}}+\frac{24(d+L)^{5}}{\left((d+L)^{2}+R^{2}\right)^{3}}\right]-\frac{15}{R^{3}}\left[-7\tan^{-1}\left(\frac{R}{d+L}\right)-6\tan^{-1}\left(\frac{d+L}{R}\right)+7\tan^{-1}\left(\frac{R}{d}\right)+6\tan^{-1}\left(\frac{d}{R}\right)\right]\right\} ,
\end{align}
\begin{align}\label{Brms_cylinder2}
C^{(2)}(0)/J^2=&\frac{\pi}{1024}\left\{ \frac{1}{R^{2}}\left[\frac{183d}{d^{2}+R^{2}}+\frac{24d^{5}}{\left(d^{2}+R^{2}\right)^{3}}+\frac{16LR^{2}\left(3d^{2}+3dL+L^{2}\right)}{d^{3}(d+L)^{3}}-\frac{102d^{3}}{\left(d^{2}+R^{2}\right)^{2}}-\frac{183(d+L)}{(d+L)^{2}+R^{2}}\right.\right.\\\nonumber
&\left.\left.+\frac{102(d+L)^{3}}{\left((d+L)^{2}+R^{2}\right)^{2}}-\frac{24(d+L)^{5}}{\left((d+L)^{2}+R^{2}\right)^{3}}\right]+\frac{105}{R^{3}}\left[\tan^{-1}\left(\frac{R}{d+L}\right)-\tan^{-1}\left(\frac{R}{d}\right)\right]\right\} .
\end{align}
 The limit $d\ll R,L$ recovers the half-plane $B_{rms}^2$  of \cite{microfludics} as expected.
 
 We continue by calculating the long times limit \eqref{corr_longtimes_general}.
 
 \begin{align}\label{longtimes_cylinder0}
 &C^{(0)}(t\gg \tau_V)/J^2\approx\frac{4\pi^2}{V} \left(\frac{L+d}{\sqrt{\left(L+d\right)^2+R^2}}-\frac{d}{\sqrt{d^2+R^2}}\right)^2\\\label{longtimes_cylinder1}
  &C^{(1)}(t\gg \tau_V)/J^2\approx\frac{3 \pi ^2}{V} \left[R \left(\frac{1}{\sqrt{(d+L)^2+R^2}}-\frac{1}{\sqrt{d^2+R^2}}\right)-\sinh ^{-1}\left(\frac{R}{d+L}\right)+\sinh ^{-1}\left(\frac{R}{d}\right)\right]\\\label{longtimes_cylinder2}
  &C^{(2)}(t\gg \tau_V)/J^2\approx\frac{3\pi ^2}{4V}  \left[-\frac{\left(d+L\right)}{\sqrt{(d+L)^2+R^2}}+\frac{d}{\sqrt{d^2+R^2}}-2 \sinh ^{-1}\left(\frac{d+L}{R}\right)+2 \log \left(\frac{d+L}{d}\right)+2 \sinh ^{-1}\left(\frac{d}{R}\right)\right]
 \end{align}
 
\begin{figure}
		\subfloat[]{\includegraphics[width=0.32\textwidth]{R200nm_L200nm_exponential_short_q01155.pdf} \label{cylinder200short}}
	\subfloat[]{\includegraphics[width=0.32\textwidth]{R200nm_L200nm_power_inter_q14967.pdf} \label{cylinder200inter}}
	\subfloat[]{\includegraphics[width=0.32\textwidth]{R200nm_L200nm_exponential_long_q000123_tv1500.pdf} \label{cylinder200long}}
	\hfill
	\subfloat[]{\includegraphics[width=0.32\textwidth]{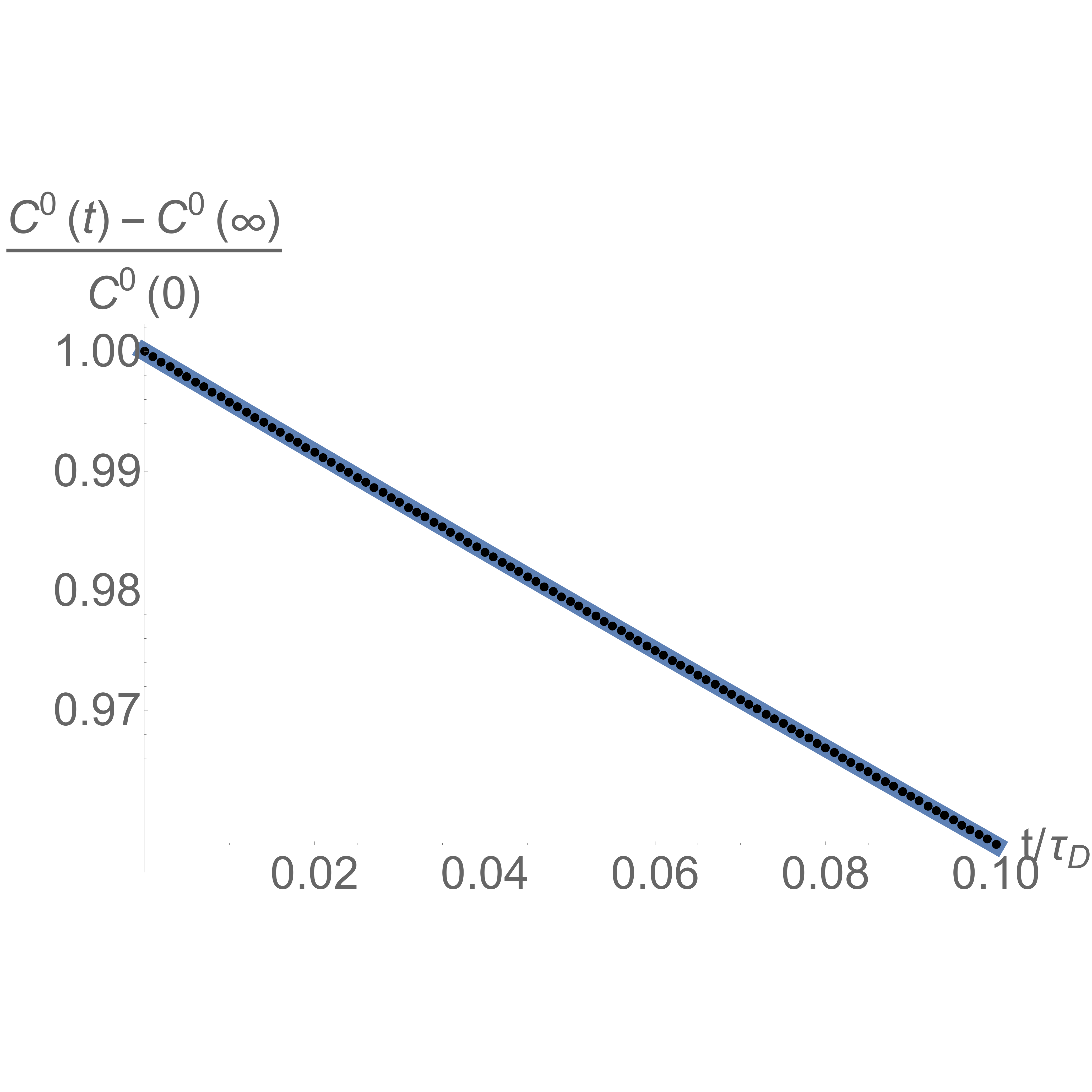} \label{cylinder100short}}
	\subfloat[]{\includegraphics[width=0.32\textwidth]{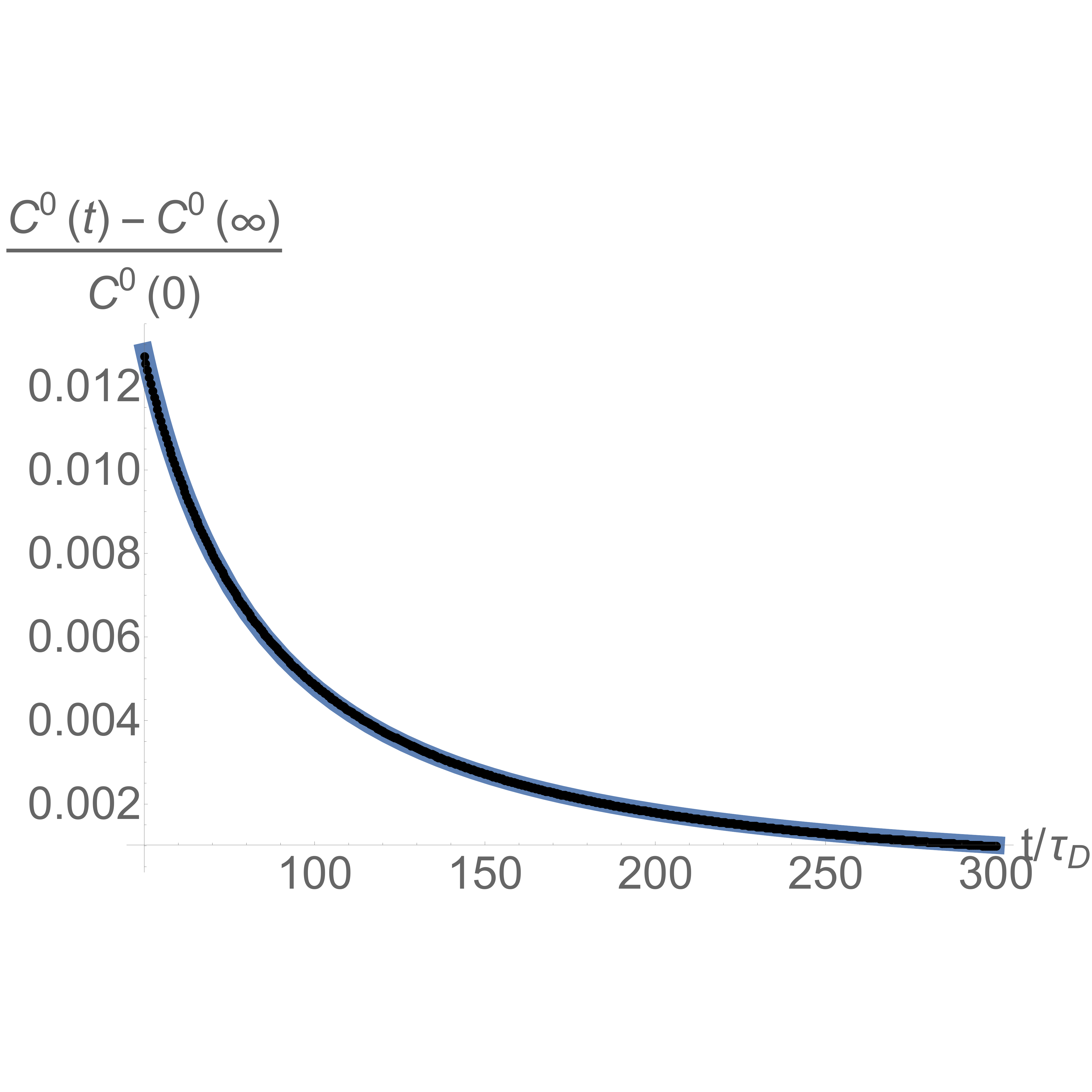} \label{cylinder100inter}}
	\subfloat[]{\includegraphics[width=0.32\textwidth]{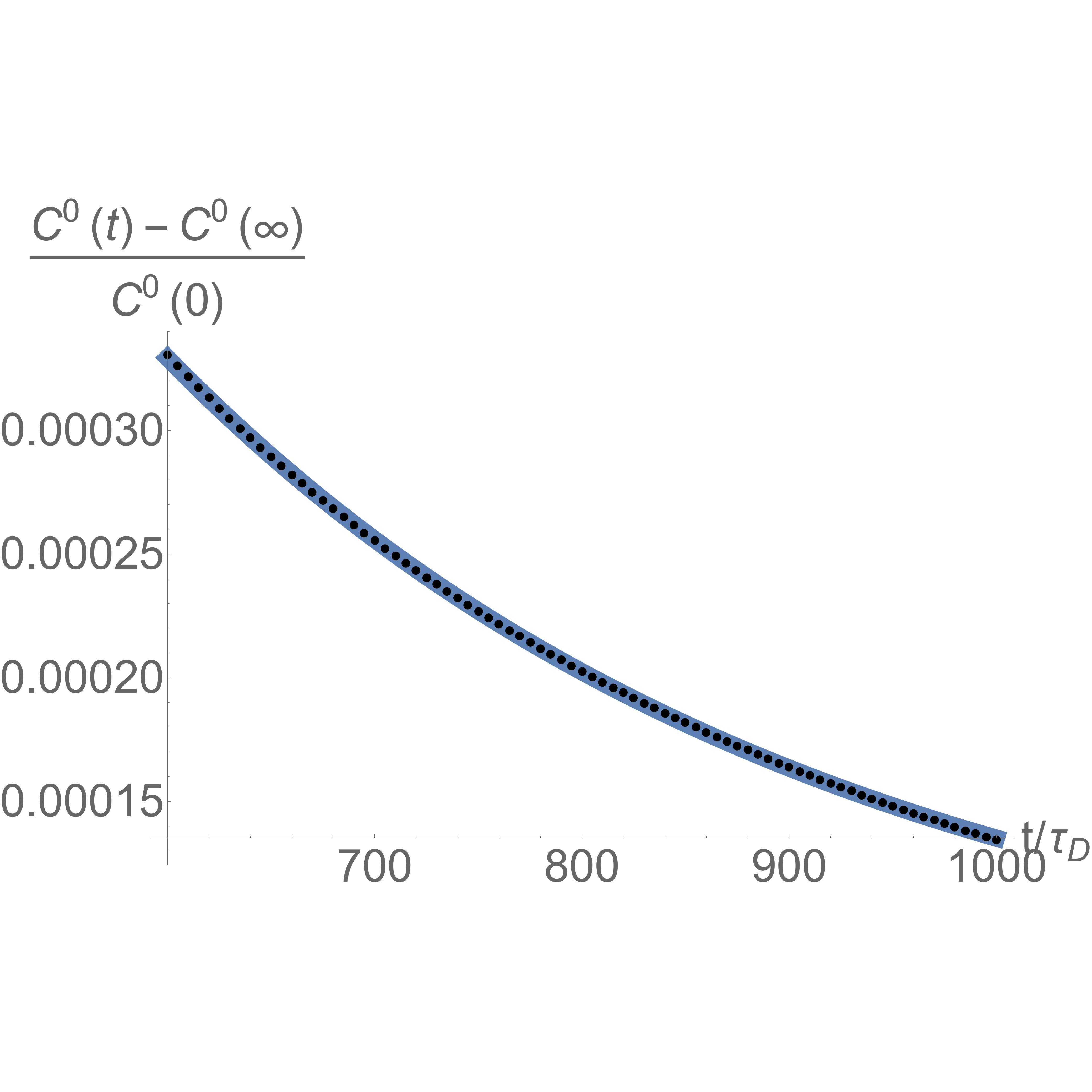} \label{cylinder100long}}
	\hfill
	\subfloat[]{\includegraphics[width=0.32\textwidth]{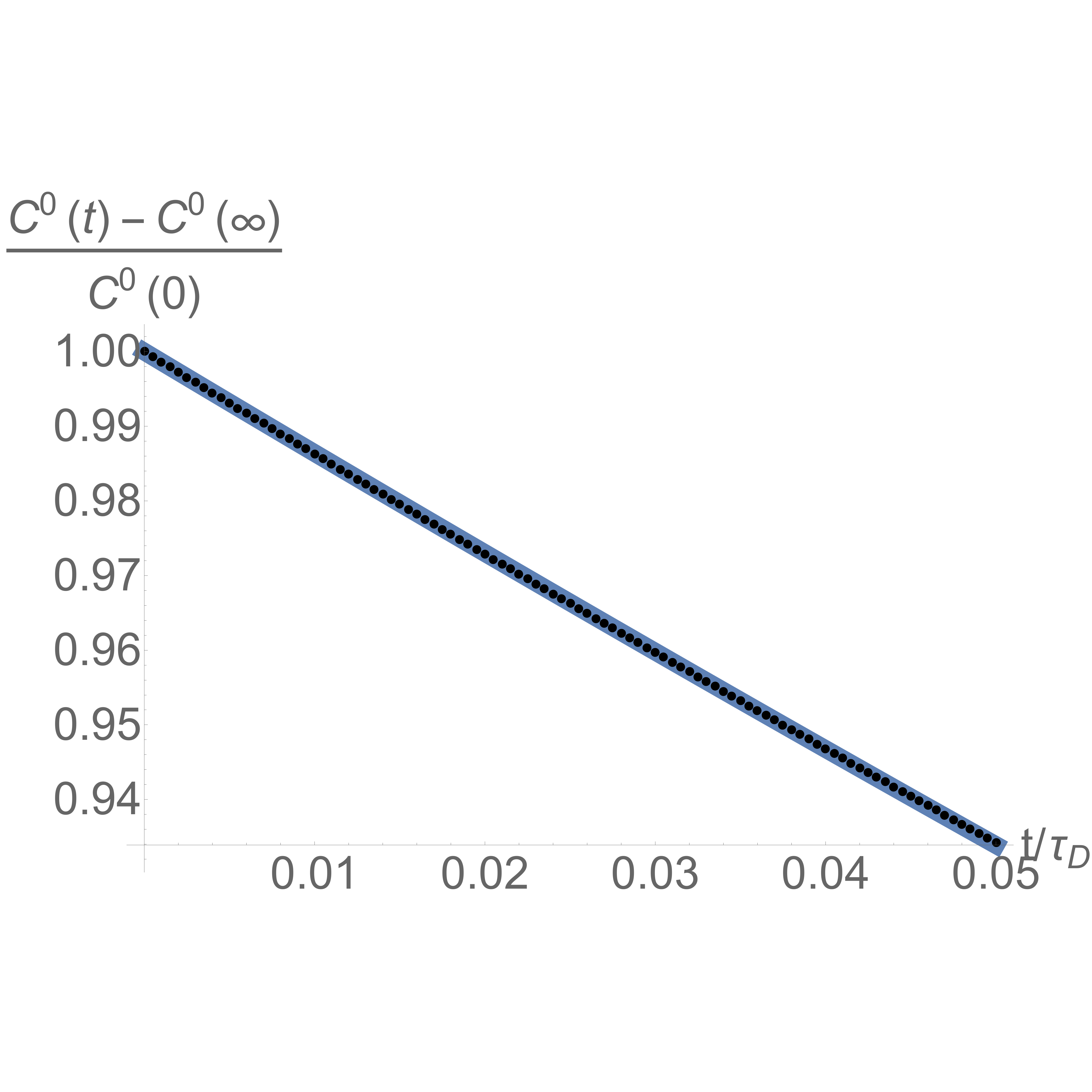} \label{cylinder50short}}
	\subfloat[]{\includegraphics[width=0.32\textwidth]{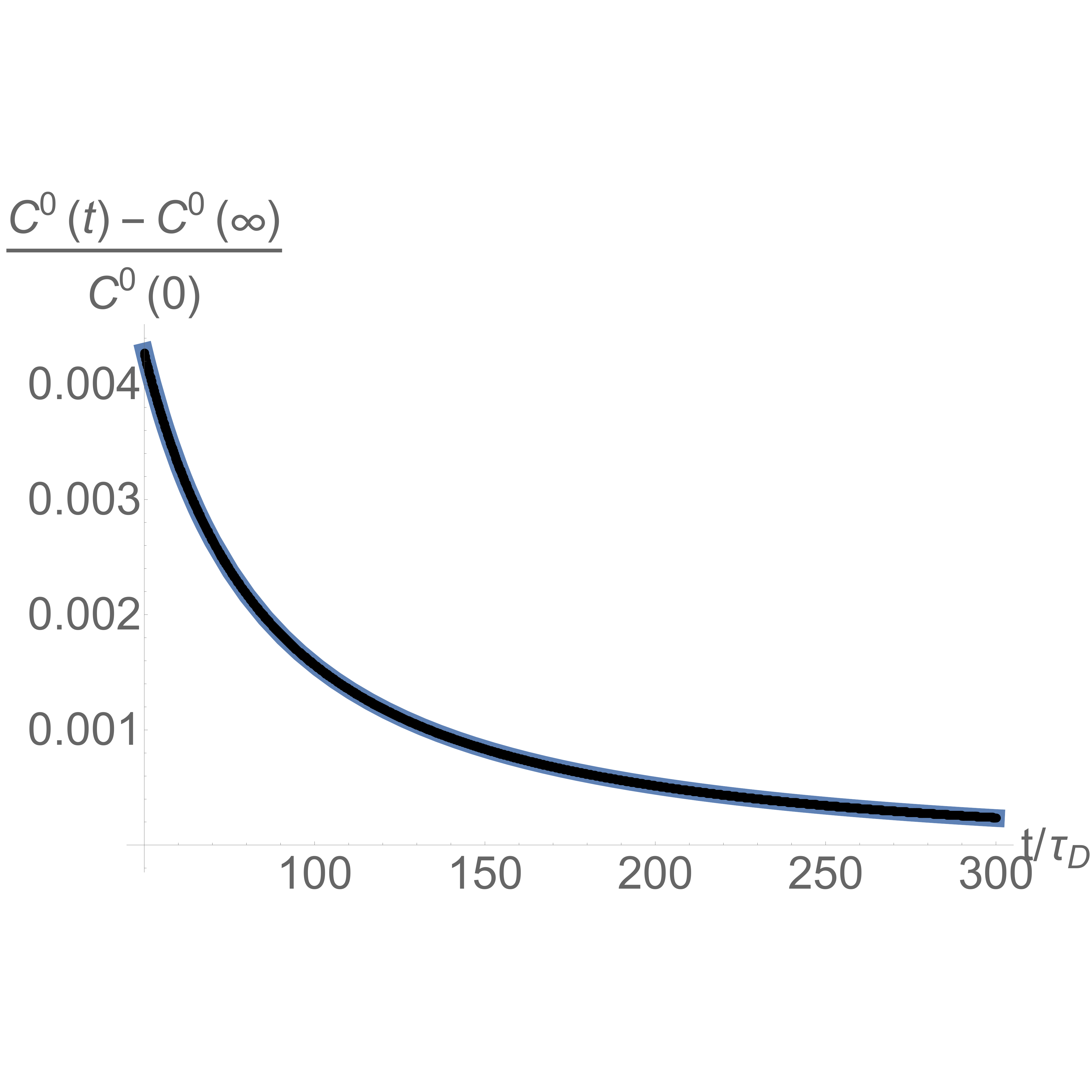} \label{cylinder50inter}}
	\subfloat[]{\includegraphics[width=0.32\textwidth]{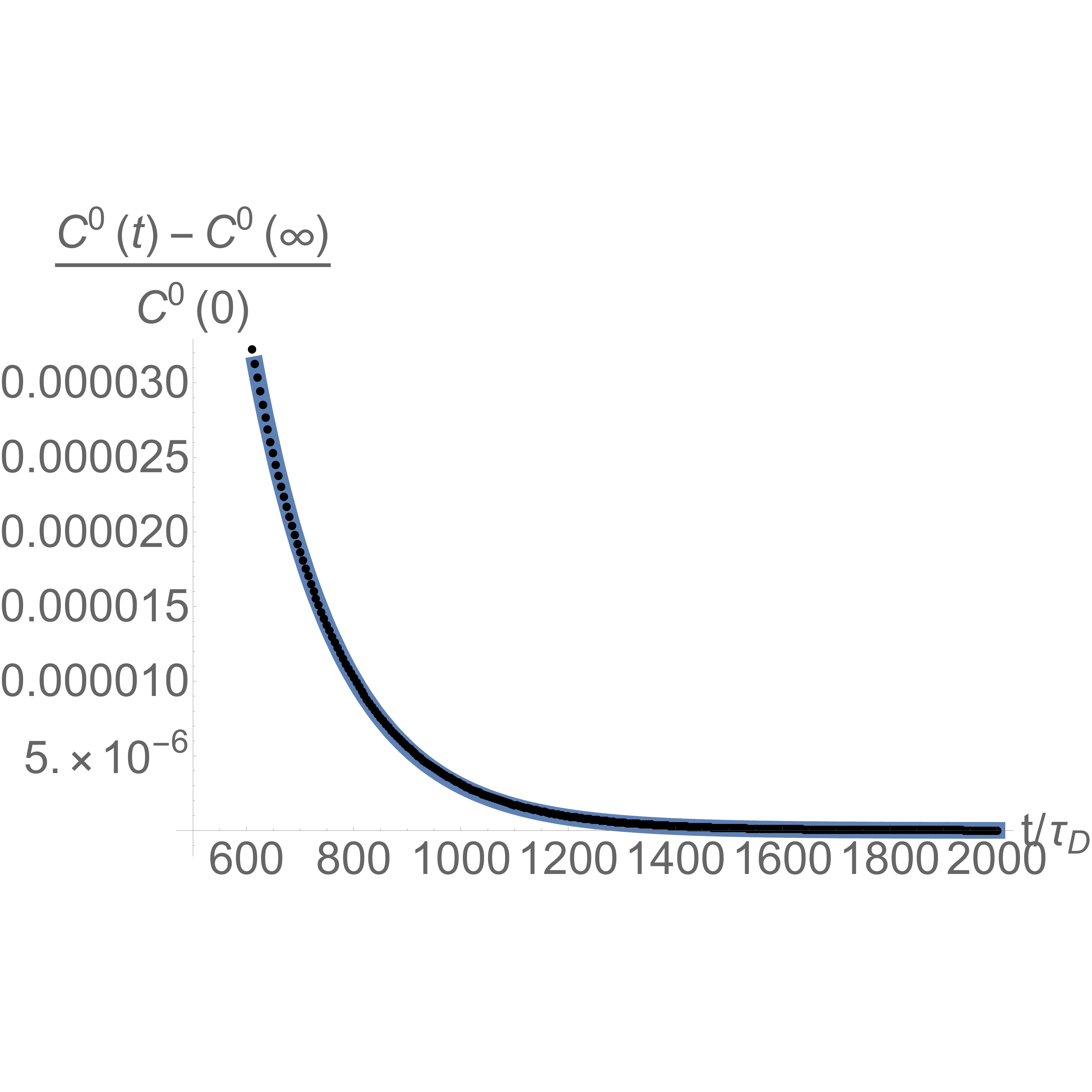} \label{cylinder50long}}
	\caption{The correlation function in a \textit{cylindrical} geometry \eqref{correlation_cylinder_temporal_decay}  (black points) for $m=0$ in the three time regimes - $t\ll\tau_D$ (left column), $\tau_D\ll t\ll \tau_V$ (central column) and $\tau\gg\tau_V$ (right column) . The correlation was calculate for $d=1\ \textrm{nm}, \ D=0.5\ \frac{\textrm{nm}^2}{\mu\textrm{s}}$ and three different cylinder sizes - $R=L=200 \ \textrm{nm}$ (top row), $R=L=100 \ \textrm{nm}$ (central row) and $R=L=50 \ \textrm{nm}$ (bottom row). The short and long times limit of the correlation were fitted to exponential functions and the intermediate times to a power law (blue lines). 
	The decay rate in short times is inversely proportional to the cylinder's volume.
	The power in the intermediate regime fits the expected $-1.5$ scaling predicted in previous works, 
	and the decay rate in long times fits the slowest decaying mode of \eqref{correlation_cylinder_temporal_decay}. \label{corr_cylinder_m0}}
\end{figure}

\begin{figure}
	\subfloat[]{\includegraphics[width=0.32\textwidth]{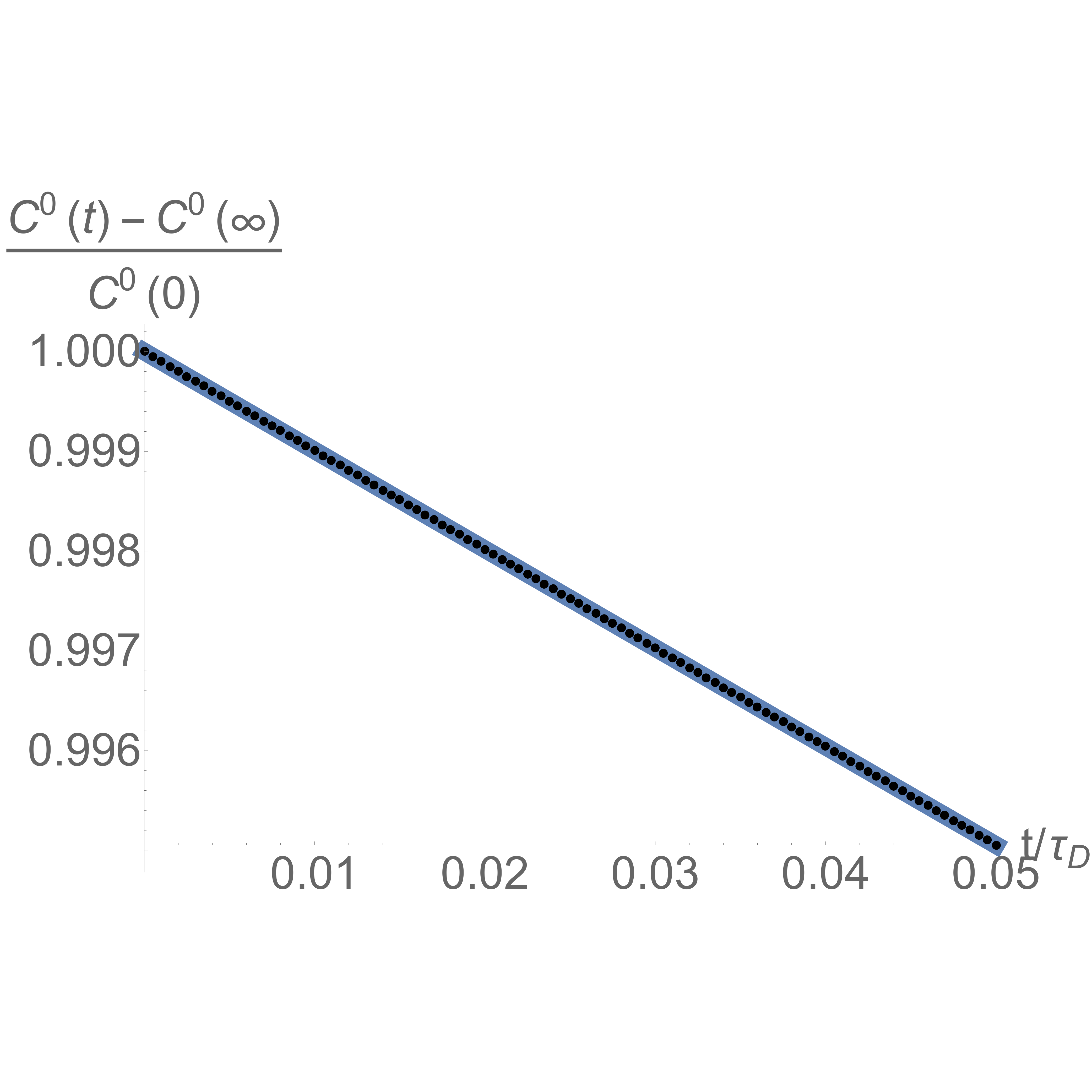} \label{cylinder200shortm1}}
	\subfloat[]{\includegraphics[width=0.32\textwidth]{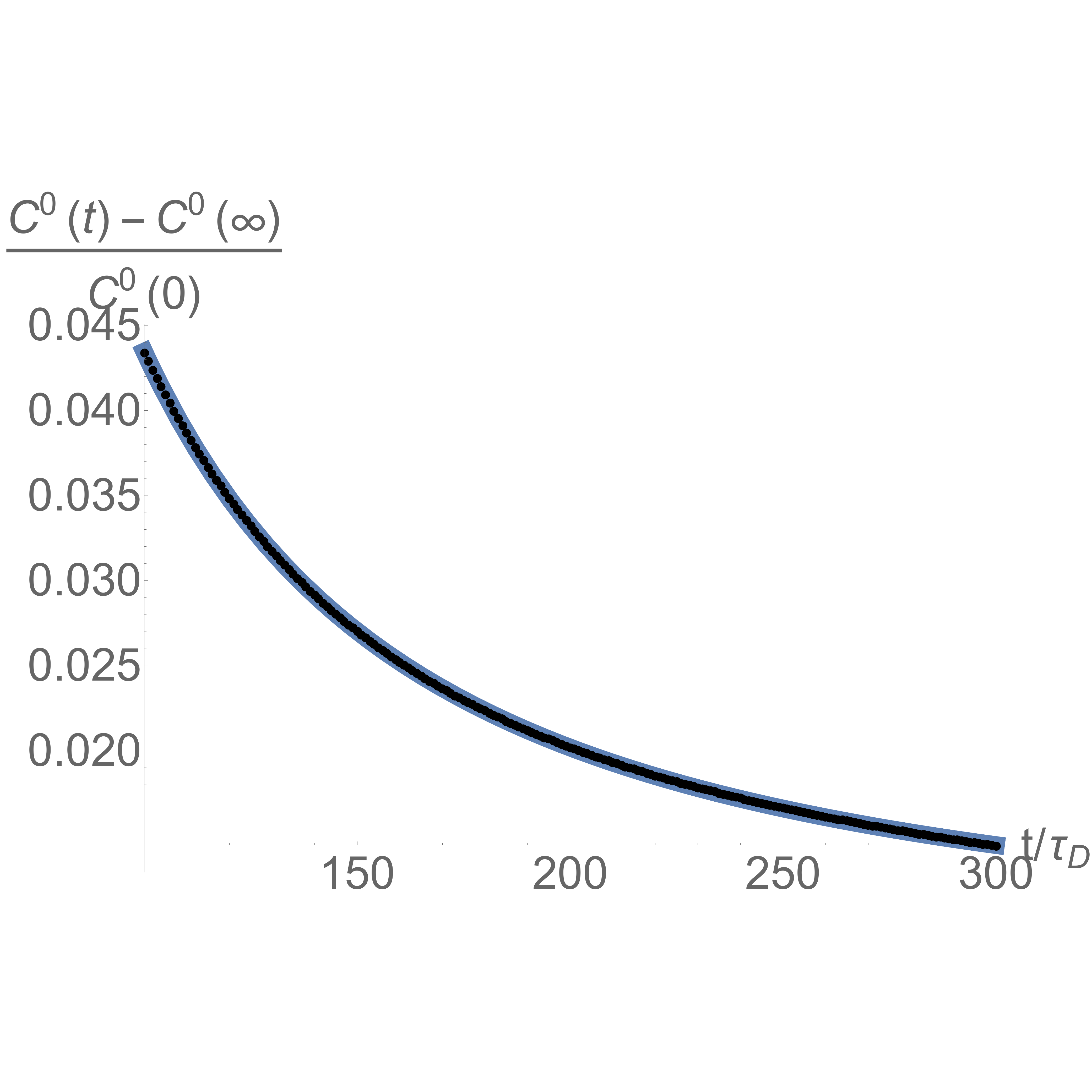} \label{cylinder200interm1}}
	\subfloat[]{\includegraphics[width=0.32\textwidth]{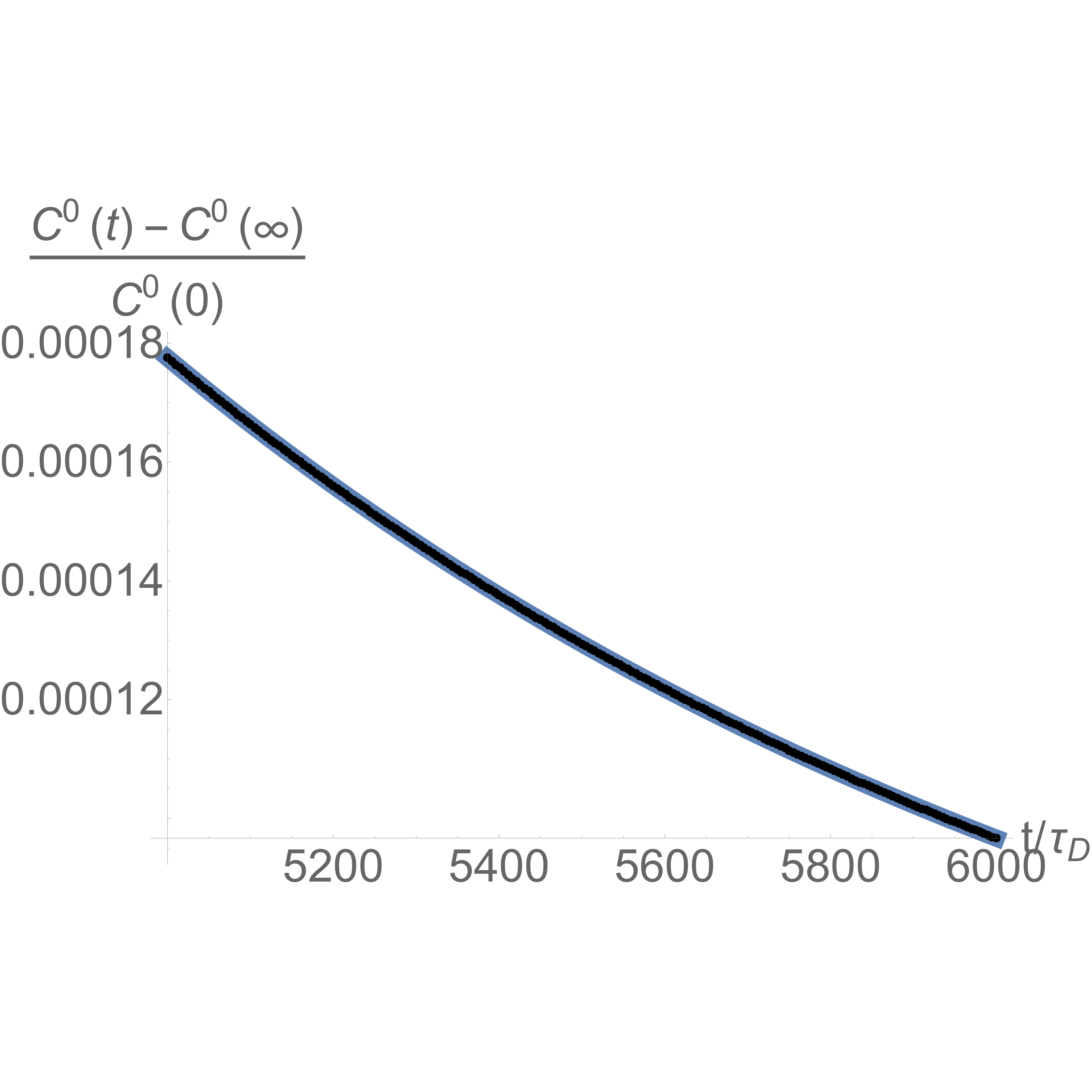} \label{cylinder200longm1}}
	\hfill
	\subfloat[]{\includegraphics[width=0.32\textwidth]{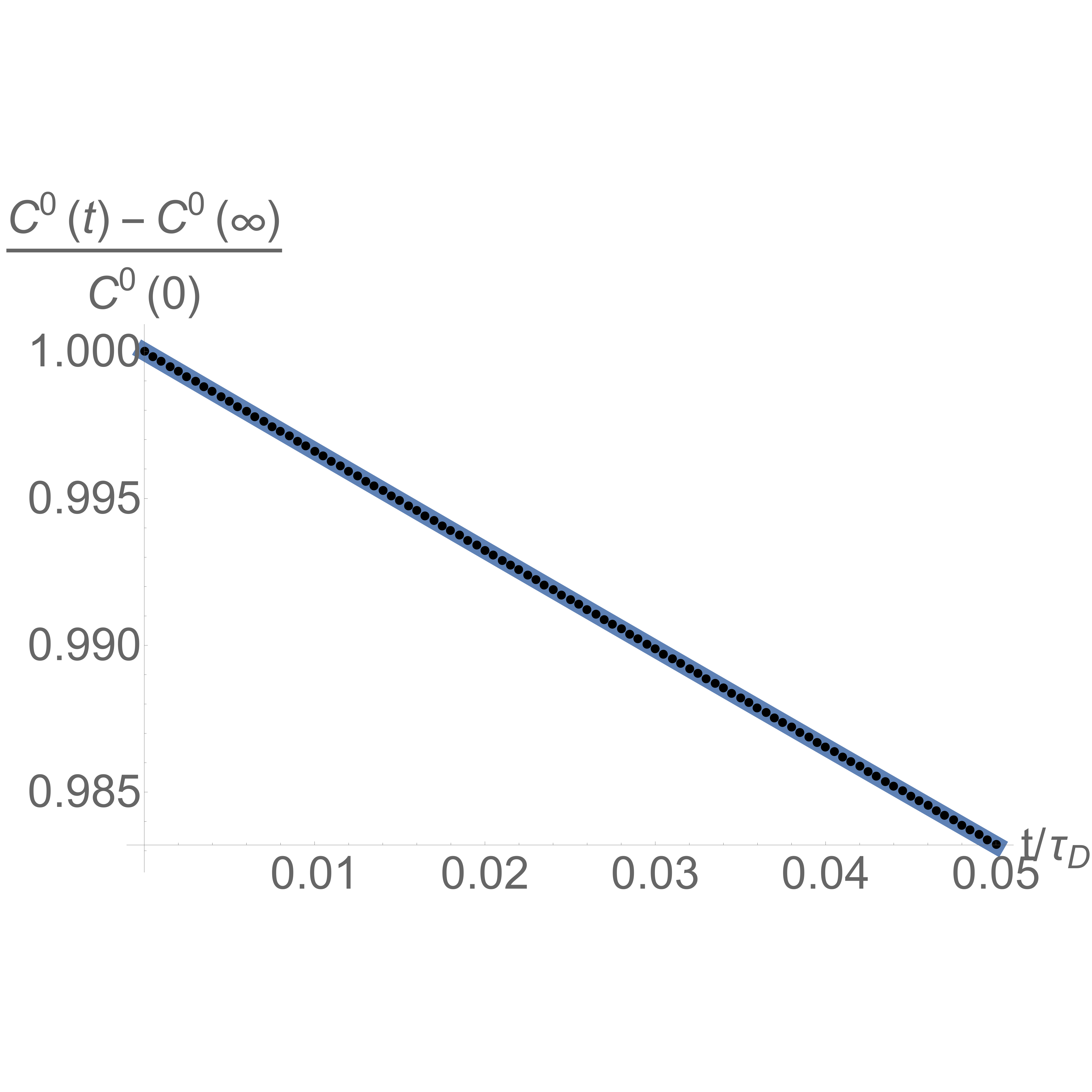} \label{cylinder100shortm1}}
	\subfloat[]{\includegraphics[width=0.32\textwidth]{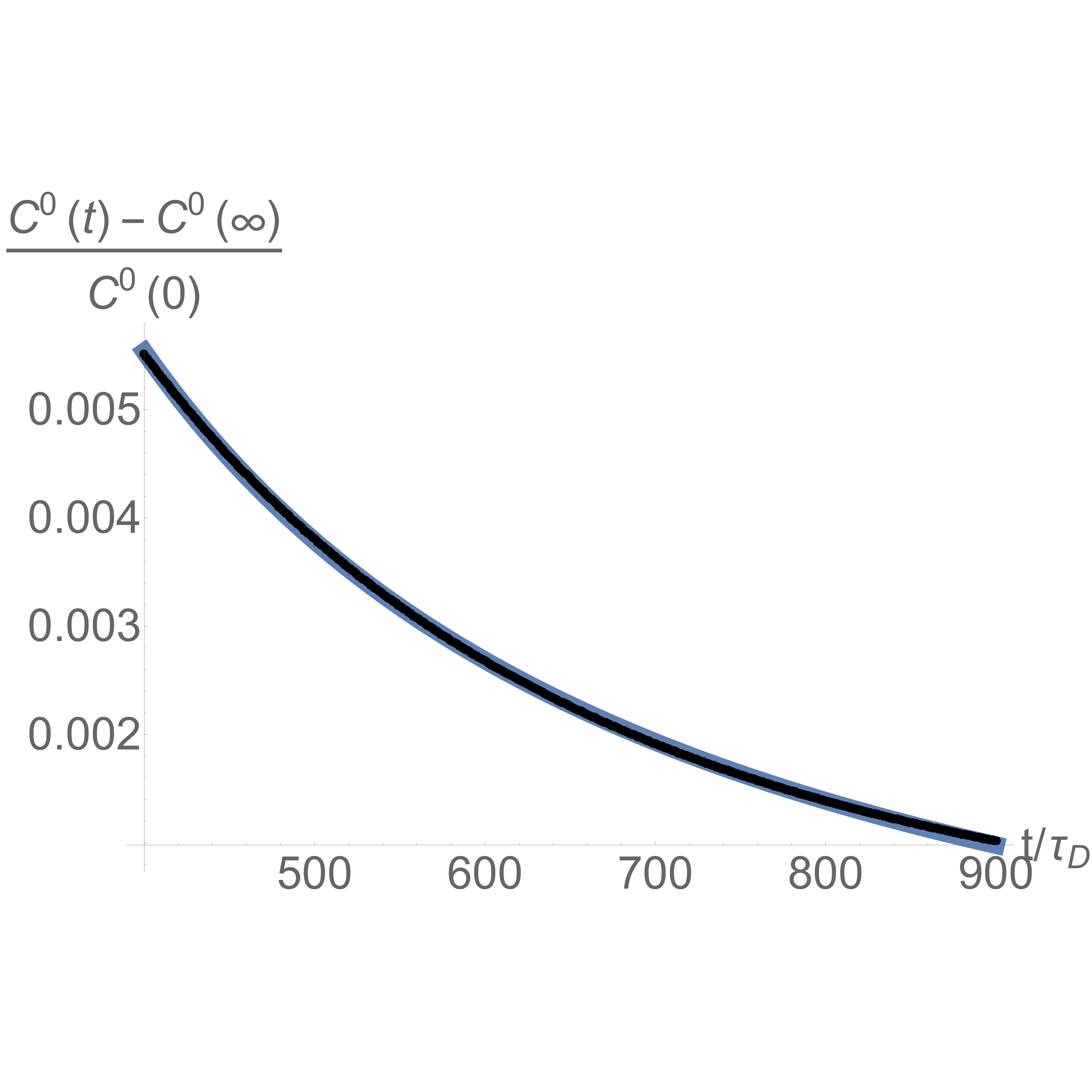} \label{cylinder100interm1}}
	\subfloat[]{\includegraphics[width=0.32\textwidth]{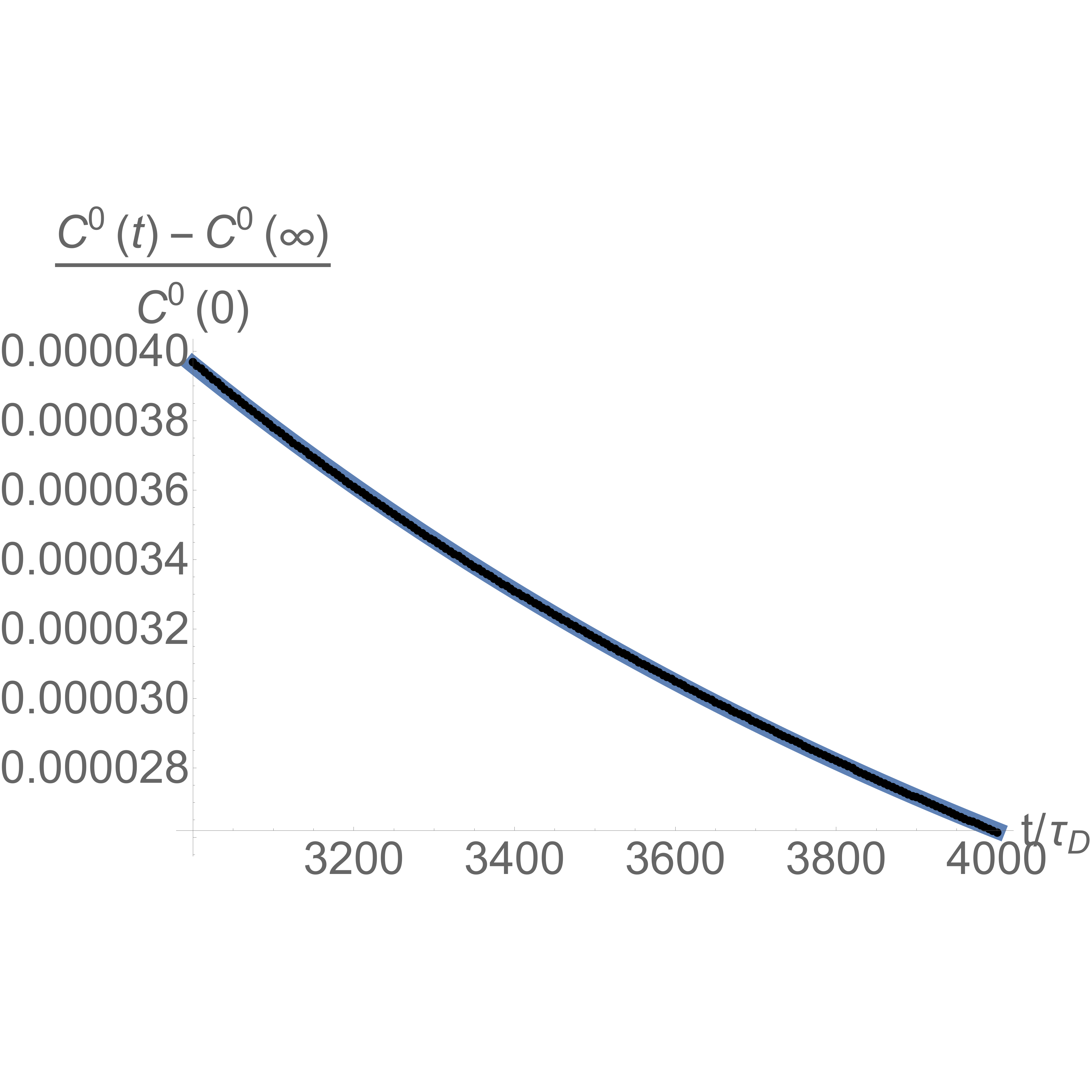} \label{cylinder100longm1}}
	\hfill
	\subfloat[]{\includegraphics[width=0.32\textwidth]{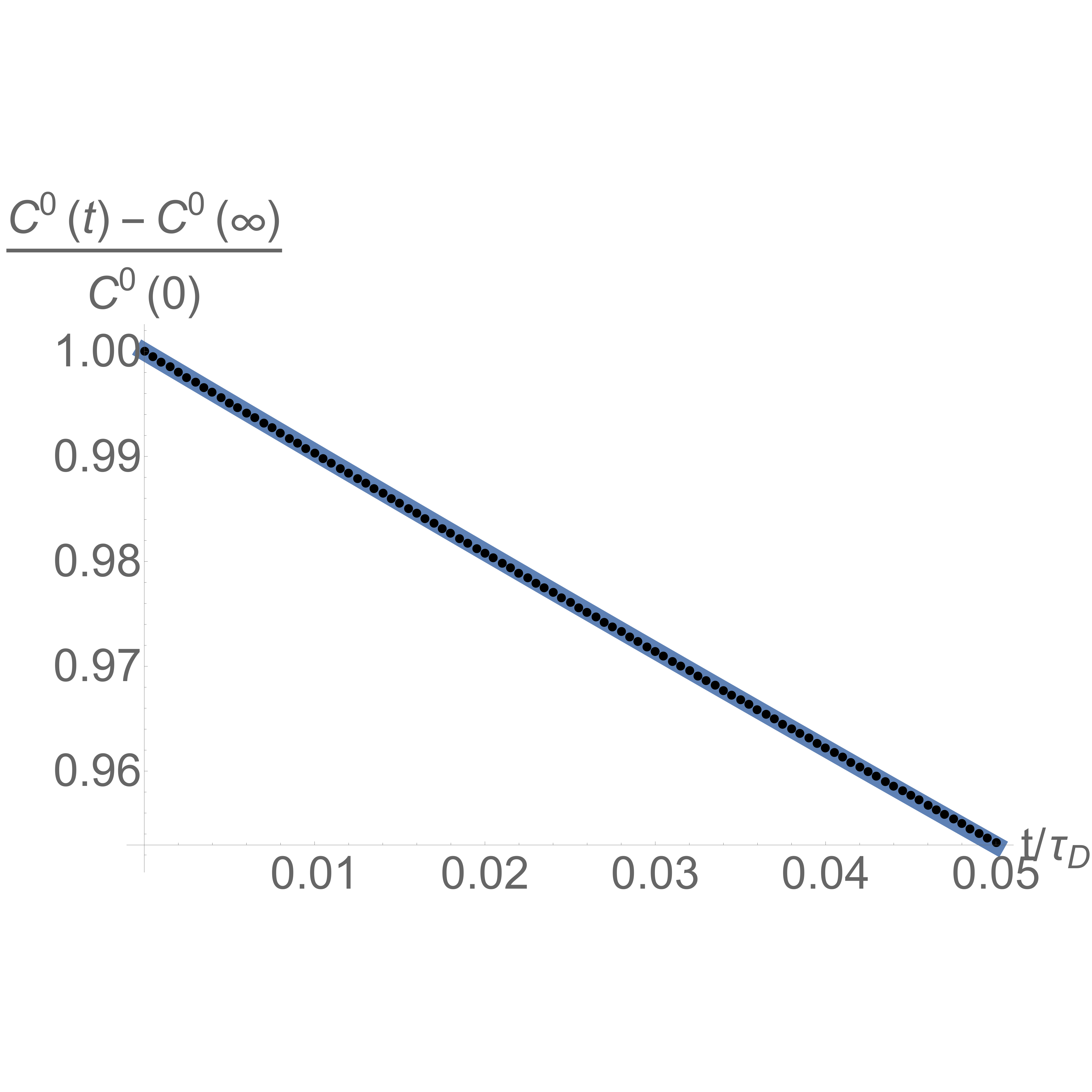} \label{cylinder50shortm1}}
	\subfloat[]{\includegraphics[width=0.32\textwidth]{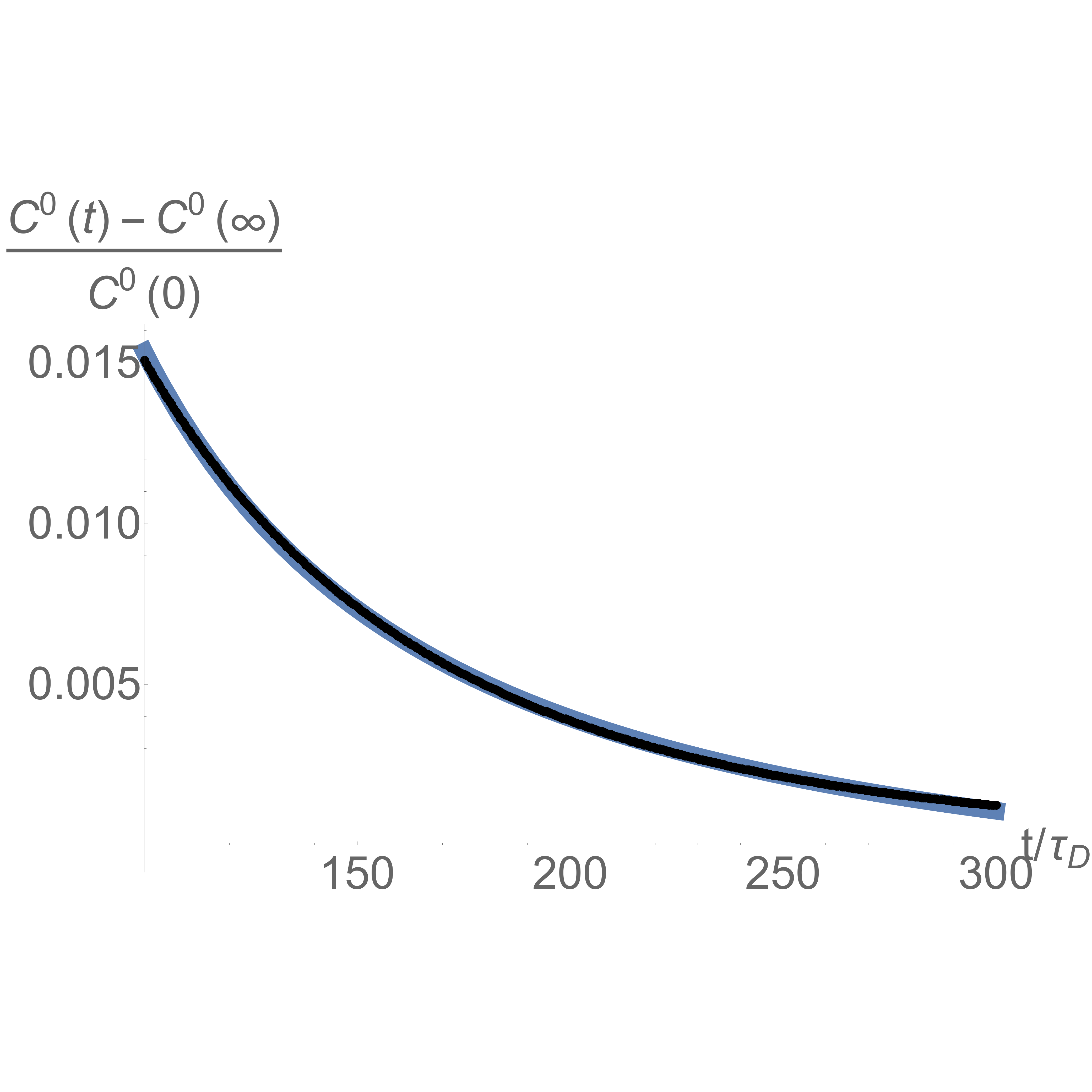} \label{cylinder50interm1}}
	\subfloat[]{\includegraphics[width=0.32\textwidth]{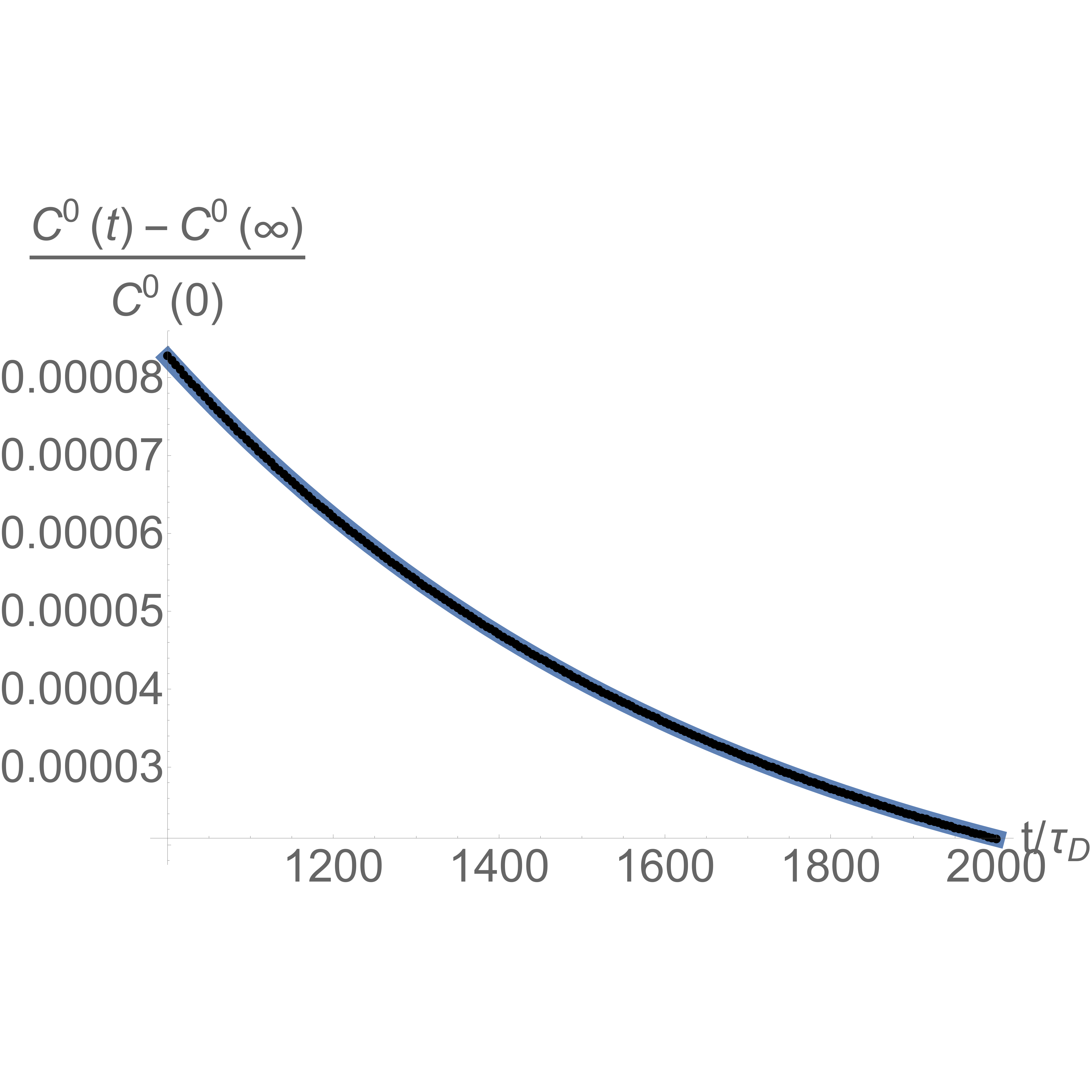} \label{cylinder50longm1}}
	\caption{The correlation function in a \textit{cylindrical} geometry \eqref{correlation_cylinder_temporal_decay}  (black points) for $m=1$ in the three time regimes - $t\ll\tau_D$ (left column), $\tau_D\ll t\ll \tau_V$ (central column) and $\tau\gg\tau_V$ (right column) . The correlation was calculate for $d=1\ \textrm{nm}, \ D=0.5\ \frac{\textrm{nm}^2}{\mu\textrm{s}}$ and three different cylinder sizes - $R=L=200 \ \textrm{nm}$ (top row), $R=L=100 \ \textrm{nm}$ (central row) and $R=L=50 \ \textrm{nm}$ (bottom row). The short and long times limit of the correlation were fitted to exponential functions and the intermediate times to a power law (blue lines). 
		The decay rate in short times is inversely proportional to the cylinder's volume.
		The power in the intermediate regime fits the expected $-1.5$ scaling predicted in previous works, 
		and the decay rate in long times fits the slowest decaying mode of \eqref{correlation_cylinder_temporal_decay}. \label{corr_cylinder_m1}}
\end{figure}

\begin{figure}
		\subfloat[]{\includegraphics[width=0.32\textwidth]{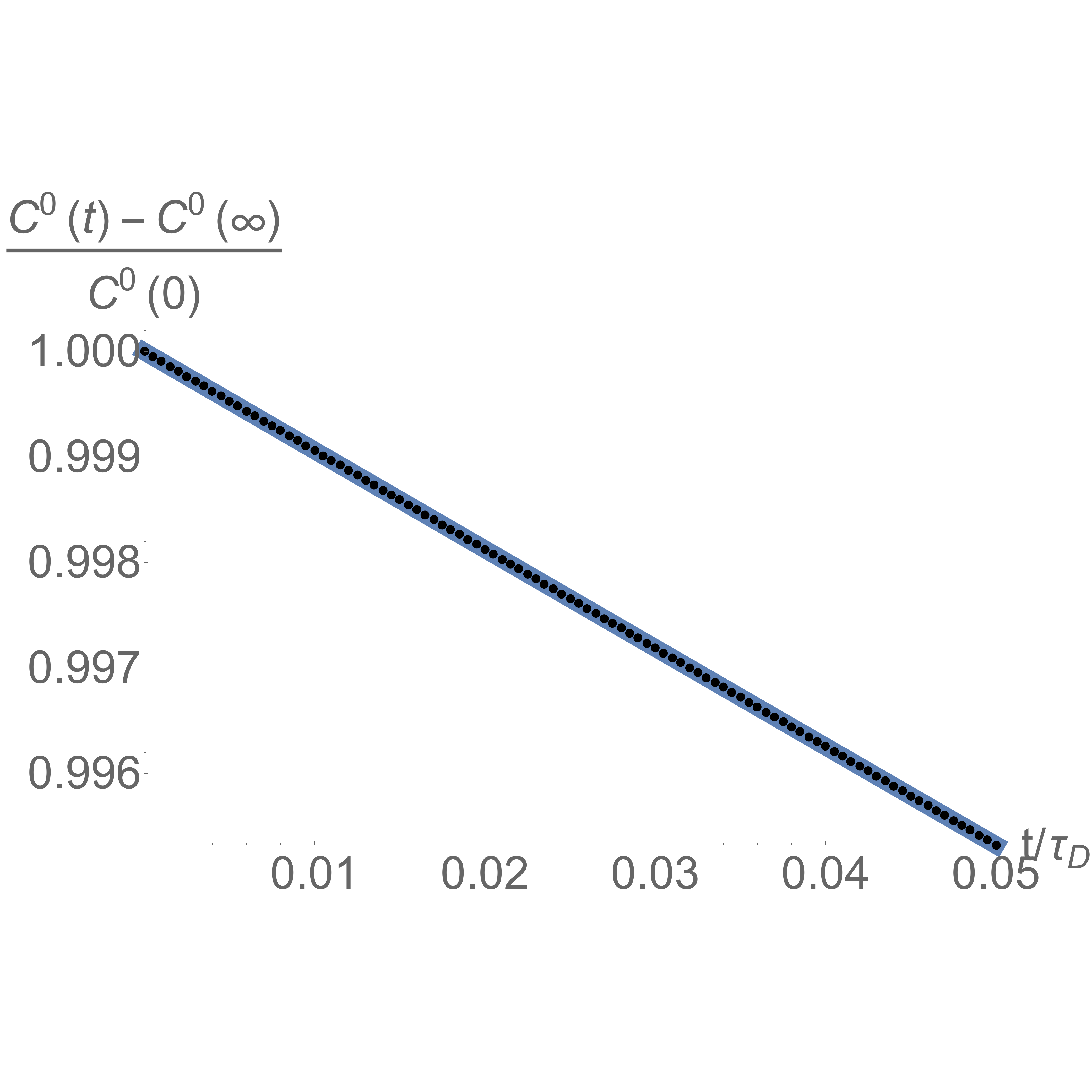} \label{cylinder200shortm2}}
	\subfloat[]{\includegraphics[width=0.32\textwidth]{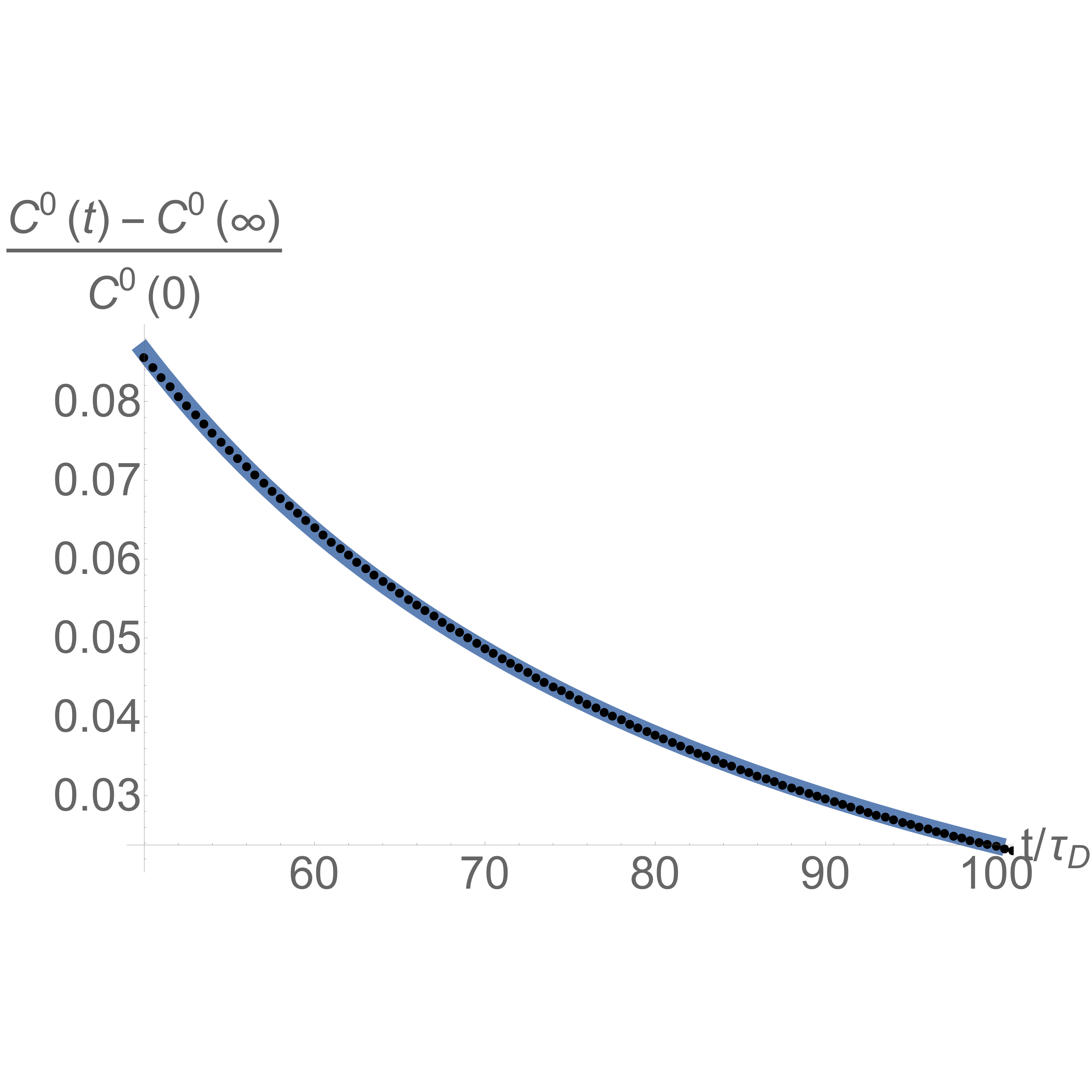} \label{cylinder200interm2}}
	\subfloat[]{\includegraphics[width=0.32\textwidth]{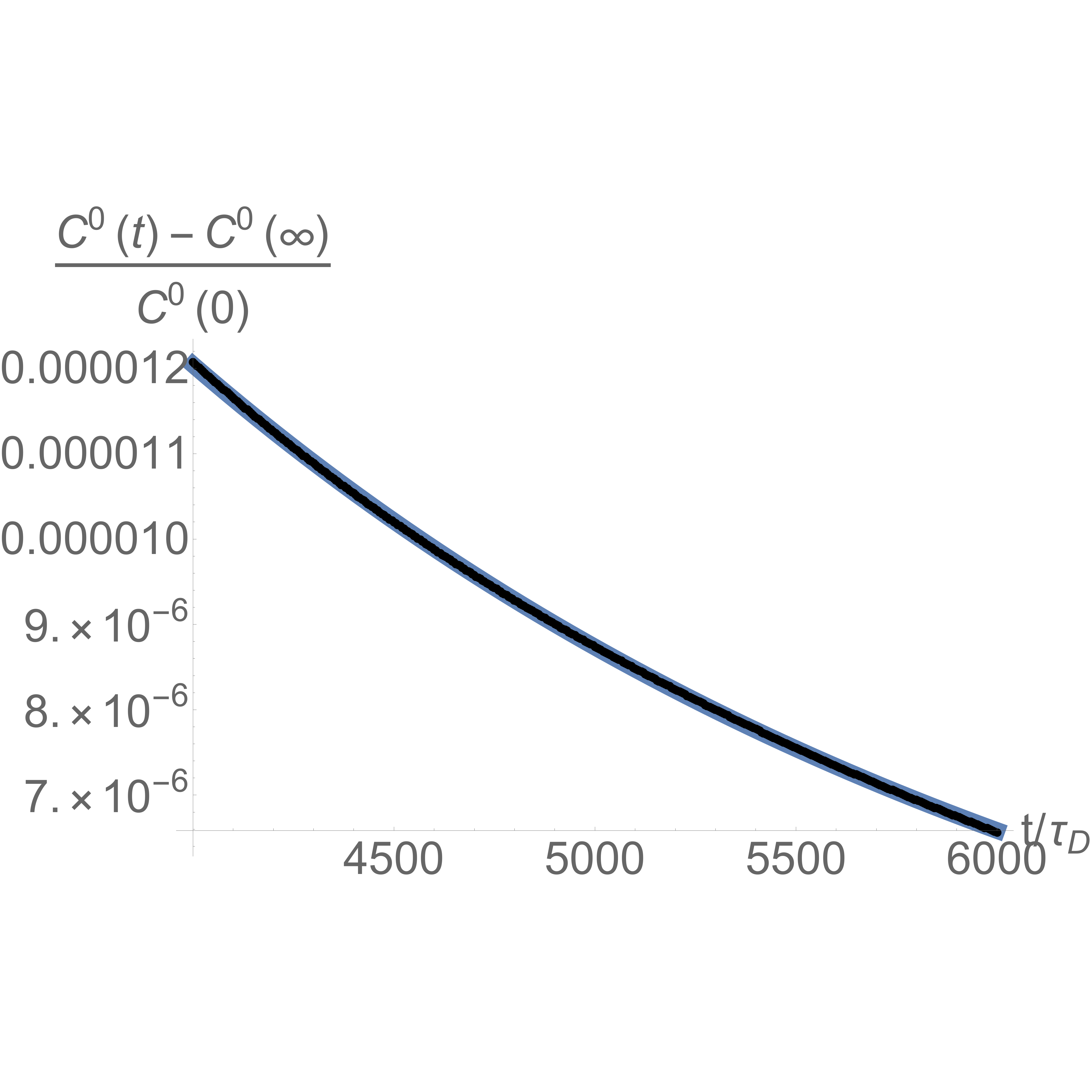} \label{cylinder200longm2}}
	\hfill
	\subfloat[]{\includegraphics[width=0.32\textwidth]{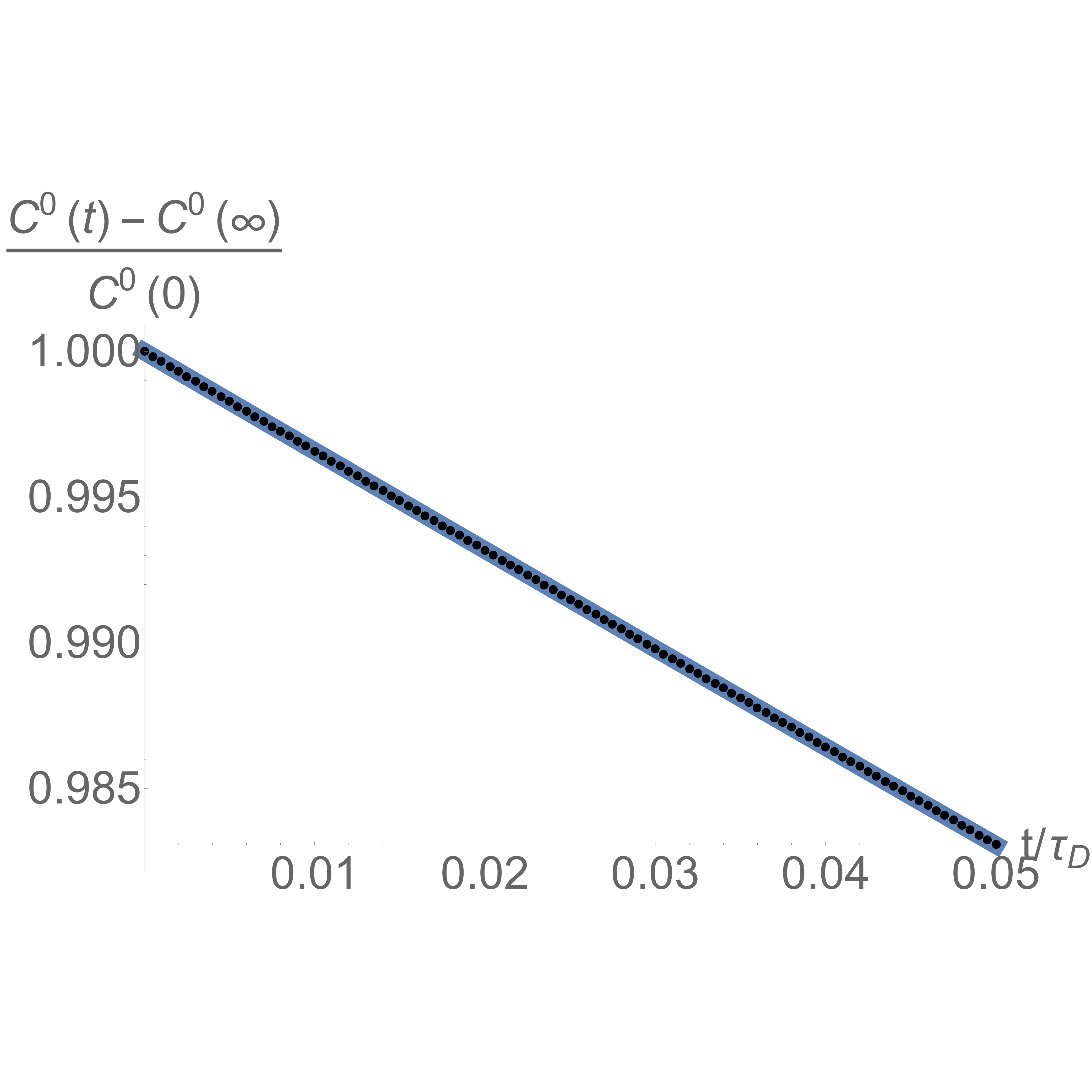} \label{cylinder100shortm2}}
	\subfloat[]{\includegraphics[width=0.32\textwidth]{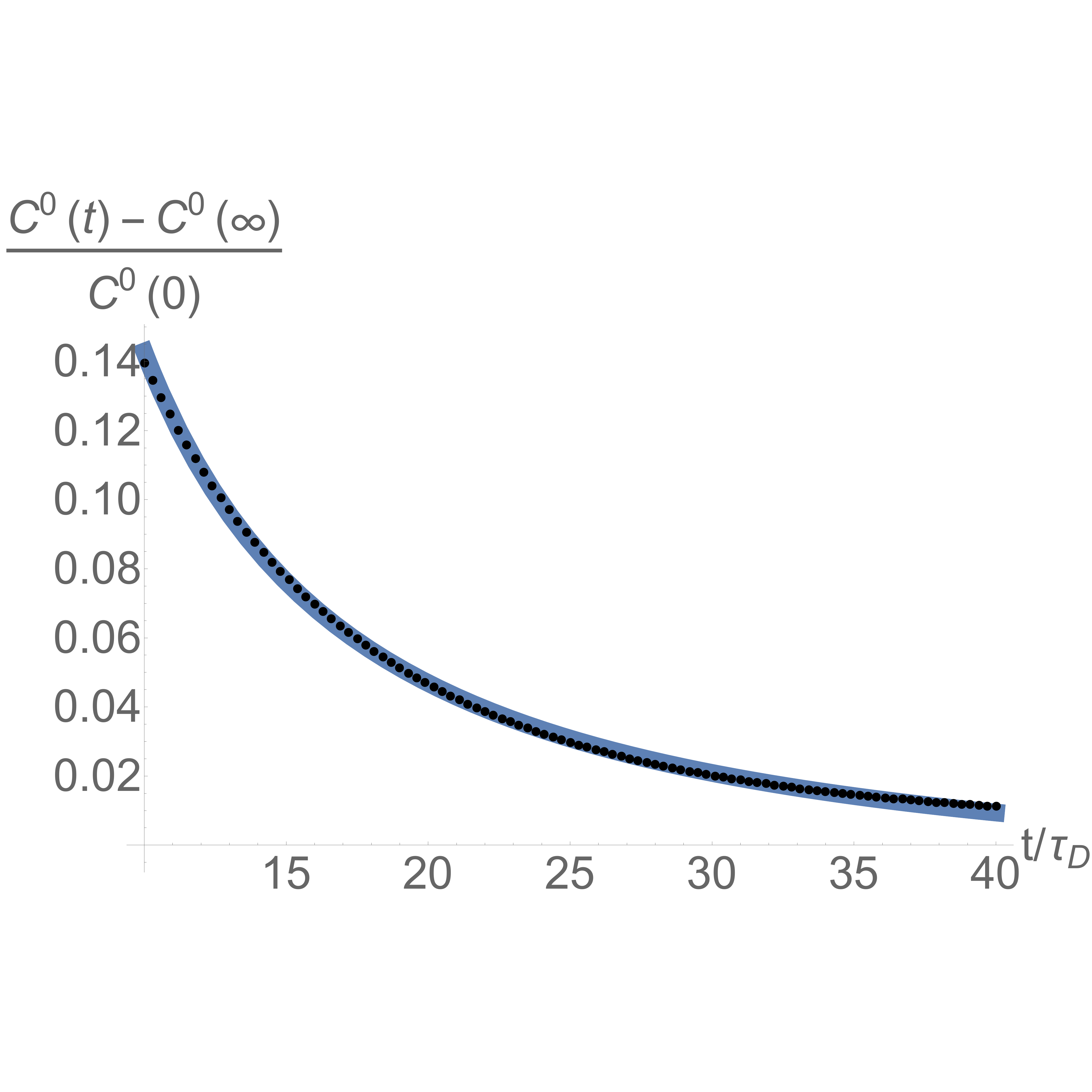} \label{cylinder100interm2}}
	\subfloat[]{\includegraphics[width=0.32\textwidth]{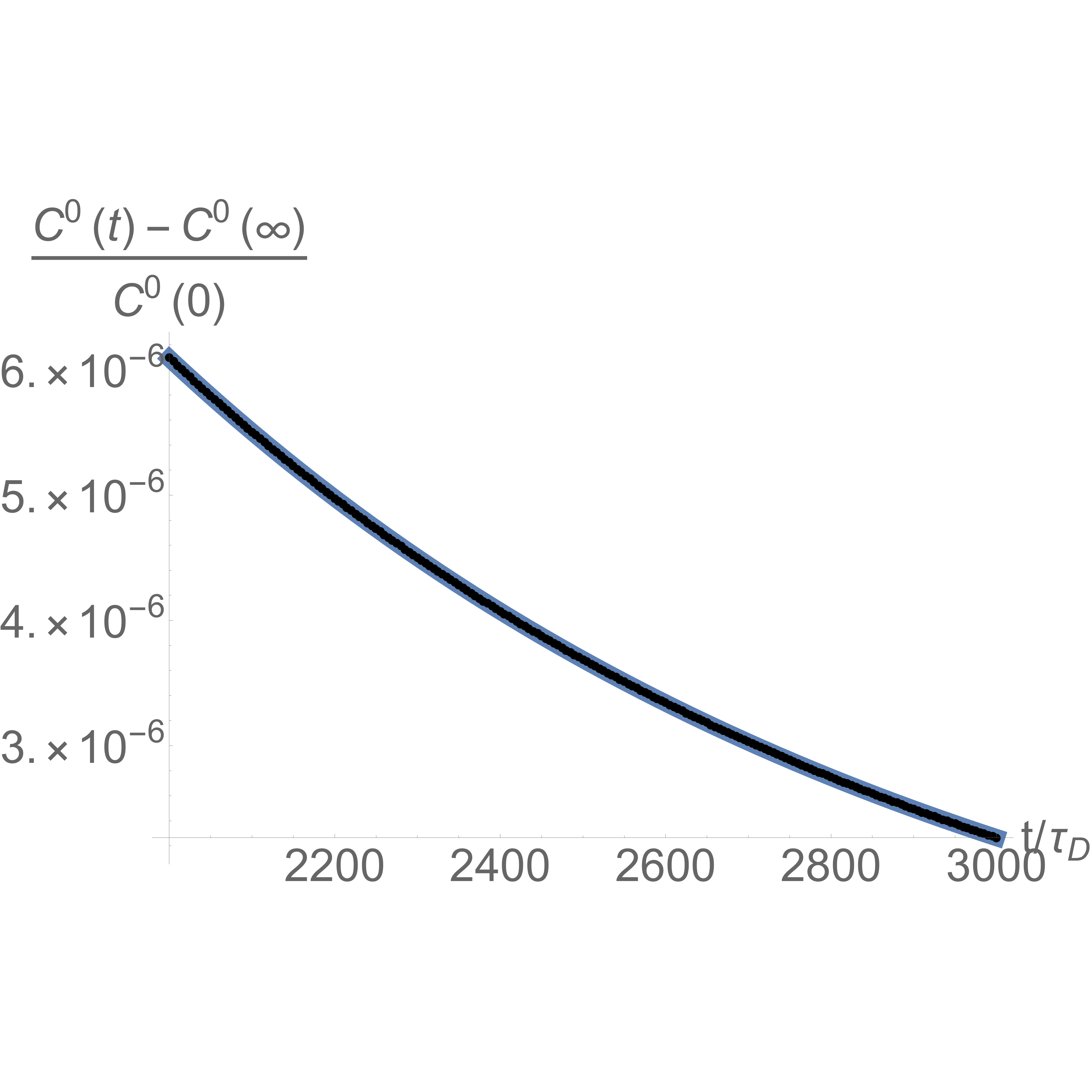} \label{cylinder100longm2}}
	\hfill
	\subfloat[]{\includegraphics[width=0.32\textwidth]{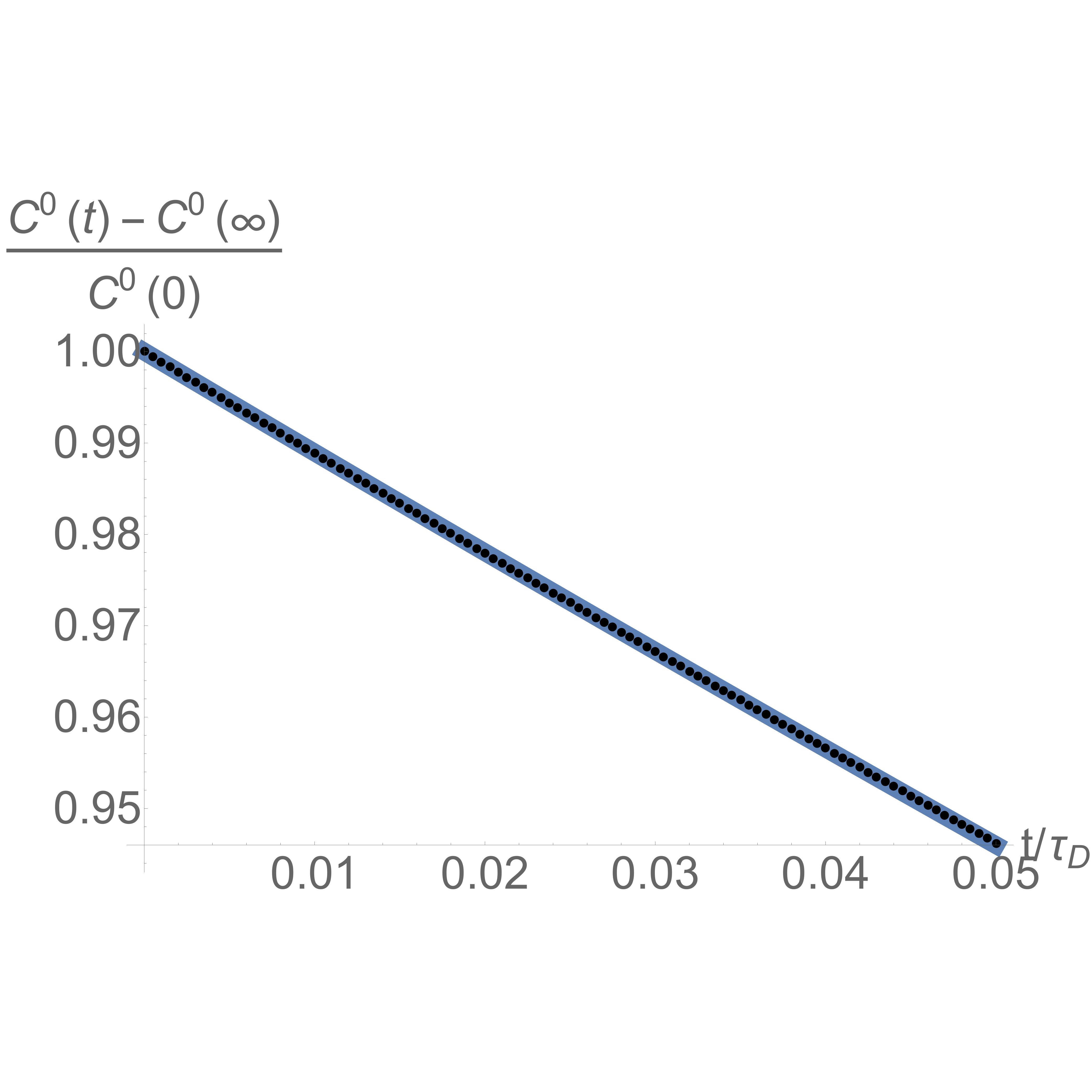} \label{cylinder50shortm2}}
	\subfloat[]{\includegraphics[width=0.32\textwidth]{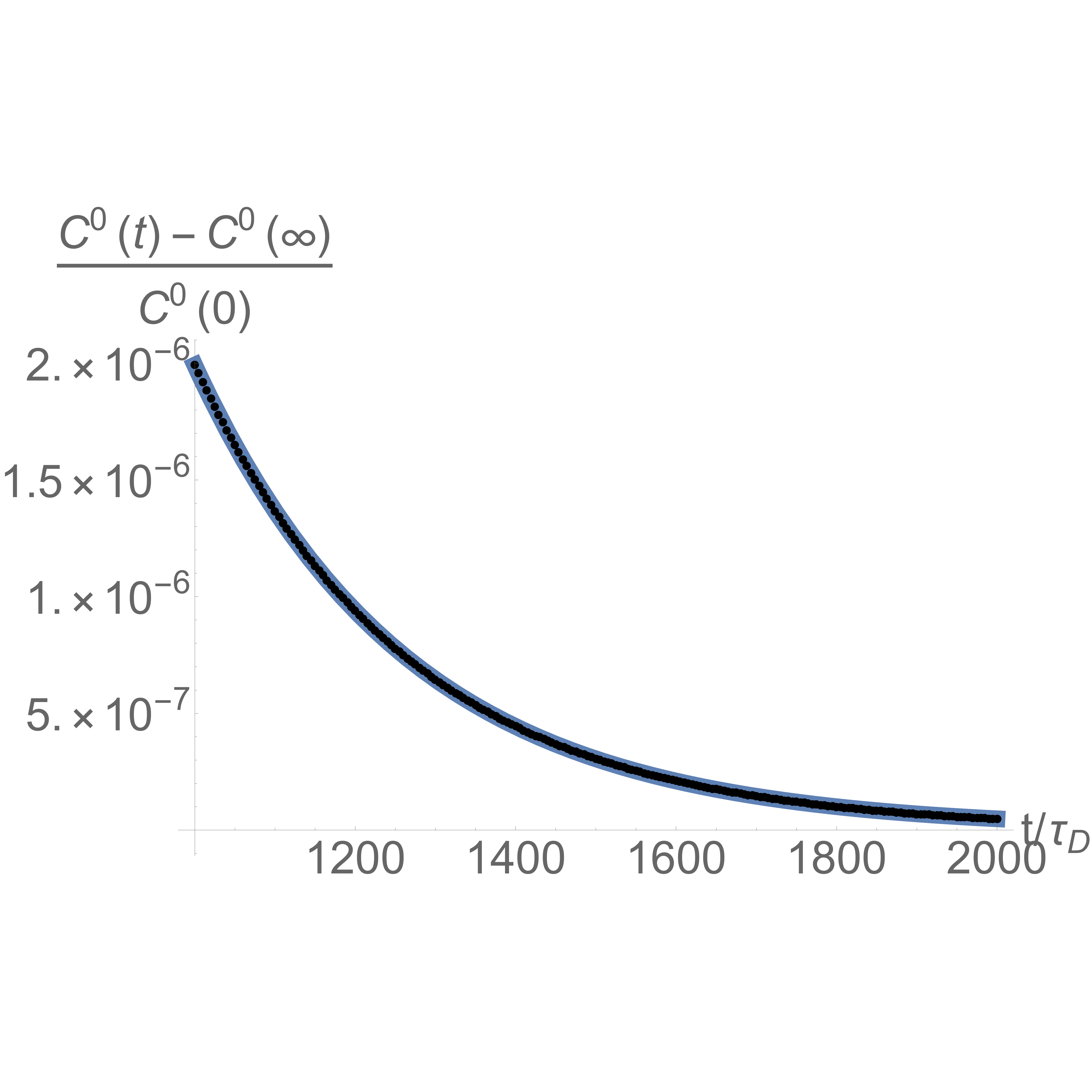} \label{cylinder50longm2}}
	\caption{The correlation function in a \textit{cylindrical} geometry \eqref{correlation_cylinder_temporal_decay}  (black points) for $m=2$ in the three time regimes - $t\ll\tau_D$ (left column), $\tau_D\ll t\ll \tau_V$ (central column) and $\tau\gg\tau_V$ (right column) . The correlation was calculate for $d=1\ \textrm{nm}, \ D=0.5\ \frac{\textrm{nm}^2}{\mu\textrm{s}}$ and three different cylinder sizes - $R=L=200 \ \textrm{nm}$ (top row), $R=L=100 \ \textrm{nm}$ (central row) and $R=L=50 \ \textrm{nm}$ (bottom row). The short and long times limit of the correlation were fitted to exponential functions and the intermediate times to a power law (blue lines). 
	The decay rate in short times is inversely proportional to the cylinder's volume.
	The power in the intermediate regime fits the expected $-1.5$ scaling predicted in previous works, 
	and the decay rate in long times fits the slowest decaying mode of \eqref{correlation_cylinder_temporal_decay}. For $L=R=50\ \textrm{nm}$ the intermediate regime does not exist since the characteristic time of the lowest decaying mode $\sim\tau_D$.
	\label{corr_cylinder_m2}}
\end{figure}

 Finally, we provide numerical results for $\tilde{C}^{(m)}$.
 Substituting the diffusion propagator \eqref{Propagator_cylinder} into \eqref{corr_general} we arrive at
 	\begin{align}\label{correlation_cylinder_temporal_decay}
 \tilde{C}^{(m)}=\frac{1}{V}\sum_{k=1}^{\infty}\sum_{s=1}^{\infty}&\frac{2}{J_{m}^{2}\left(\nu_{s,m}\right)\left(1-\left(\frac{m}{\nu_{s,m}}\right)^{2}\right)}\exp\left[-D\left[\left(\frac{\nu_{s,m}}{R}\right)^{2}+\left(\frac{\pi k}{L}\right)^{2}\right]t\right]\times\\
 &\left[\int\frac{d^3r}{r^3}\cos\left[\pi k\left(\frac{z-d}{L}\right)\right]J_m\left(\nu_{s,m}\frac{\rho}{R}\right)\left|Y_2^{(m)}\left(\Omega\right)\right|\right]^2.
 \end{align}
We evaluated the series \eqref{correlation_cylinder_temporal_decay} by calculating the first terms ($k,s\in[1,25]$). The integral was calculated numerically for $d=\textrm{nm}$ and $L=R=\alpha$ for the values $\alpha_1=200\ \textrm{nm},\ \alpha_2=100\ \textrm{nm},\ \alpha_3=50\ \textrm{nm}$. 
The results where fitted in the regimes $t\ll\tau_D$ and $t\gg\tau_V$ to exponential functions and for  $\tau_D \ll t\ll \tau_V$ to a power law. The results for $m=0$ are presented in  Fig. \ref{corr_cylinder_m0}, for $m=1$ in Fig. \ref{corr_cylinder_m1} and for $m=2$ in Fig. \ref{corr_cylinder_m2}
 \subsubsection{Hemisphere}
 We first solve \eqref{Brms_general} for each value of $m$ individually:
 \begin{align}\label{Brms_hemisphere0}
 &C^{(0)}(0)/J^2=  \frac{\pi}{32d^{5}}\left[\left(d^{2}-9R^{2}\right)\ln\left(\frac{d^{2}+R^{2}}{\left(d+R\right)^{2}}\right)\right.\\\nonumber
 &\left.+2d^{3}\left(-\frac{5d}{d^{2}+R^{2}}-\frac{4d^{2}}{(d+R)^{3}}-\frac{6d^{5}}{\left(d^{2}+R^{2}\right)^{3}}+\frac{2d^{3}}{\left(d^{2}+R^{2}\right)^{2}}+\frac{2d}{(d+R)^{2}}-\frac{2}{d+R}\right)+26d^{2}-18dR\right],
 \end{align}
 \begin{align}
 \label{Brms_hemisphere1}
 &C^{(1)}(0)/J^2=\frac{\pi}{128d^{5}}\left[3\left(d^{2}+3R^{2}\right)\ln\left(\frac{d^{2}+R^{2}}{(d+R)^{2}}\right)\right.\\\nonumber
 &\left.+\frac{2dR}{(d+R)^{3}\left(d^{2}+R^{2}\right)^{3}}\left(3d^{9}+6d^{8}R+19d^{7}R^{2}+33d^{6}R^{3}+81d^{5}R^{4}+75d^{4}R^{5}+75d^{3}R^{6}+45d^{2}R^{7}+22dR^{8}+9R^{9}\right)\right],
 \end{align}
 \begin{align}
 \label{Brms_hemisphere2}
 &C^{(2)}(0)/J^2=\frac{\pi}{512d^{5}}\left[-3\left(5d^{2}+3R^{2}\right)\ln\left(\frac{d^{2}+R^{2}}{\left(d+R\right)^{2}}\right)\right.\\\nonumber
 &\left.+2d\left(-\frac{4d^{4}}{(d+R)^{3}}-\frac{18d^{2}}{d+R}-\frac{6d^{7}}{\left(d^{2}+R^{2}\right)^{3}}+\frac{18d^{5}}{\left(d^{2}+R^{2}\right)^{2}}+3d^{3}\left(\frac{6}{(d+R)^{2}}-\frac{7}{d^{2}+R^{2}}\right)+13d-9R\right)\right]
 \end{align}
 The limit $d\ll R$ recovers the half-plane $B_{rms}^2$  of \cite{microfludics}.
  We continue by calculating the long times limit \eqref{corr_longtimes_general}.
 \begin{align}\label{longtimes_hemisphere0}
 &C^{(0)}(t\gg \tau_V)/J^2\approx\frac{4\pi^2}{9V}\left[2-\frac{2d}{\sqrt{d^{2}+R^{2}}}-\frac{R^{2}}{d\sqrt{d^{2}+R^{2}}}-\frac{2R^{3}}{d^{3}}\left(\frac{R}{\sqrt{d^{2}+R^{2}}}-1\right)\right]^2\\\label{longtimes_hemisphere2}
 &C^{(2)}(t\gg \tau_V)/J^2\approx\frac{\pi^2}{4}  \left[\frac{R }{d}\left(\frac{7 R}{\sqrt{d^2+R^2}}-6\right)+\frac{8 d}{\sqrt{d^2+R^2}}+\frac{2 R^3 }{d^3}\left(\frac{R}{\sqrt{d^2+R^2}}-1\right)+6 \log \left(\frac{d+R}{d}\right)-8\right]^2.
\end{align} 
We where not able to calculate the integral for $m=1$ analytically.
Finally, we provide numerical results for $\tilde{C}^{(m)}$.
Substituting the diffusion propagator \eqref{Hmisphere_prop} into \eqref{corr_general} we arrive at
\begin{align}\label{corr_hemisphere_temporal_decay}
\tilde{C}^{(m)}=\frac{1}{V}\sum_{l=0}^{\infty}\sum_{k=1}^{\infty}\frac{4\pi\left[1+\left(-1\right)^{l+m}\right]}{3j_{l}^{2}\left(\tilde{\nu}_{k,l}\right)\left(1-\left(\frac{l}{\tilde{\nu}_{k,l}}\right)^{2}\right)}\left[\int \frac{d^3r}{r^3}j_{l}\left(\frac{\tilde{\nu}_{k,l}}{R}r\right)Y_2^{(m)*}\left(\theta,\phi\right)Y_{l}^{(m)}\left(\theta,\phi\right)\right]^2\exp\left(-\frac{\tilde{\nu}_{k,l}^{2}}{\tau_{V}}t\right)
\end{align}

\begin{figure}
		\subfloat[]{\includegraphics[width=0.32\textwidth]{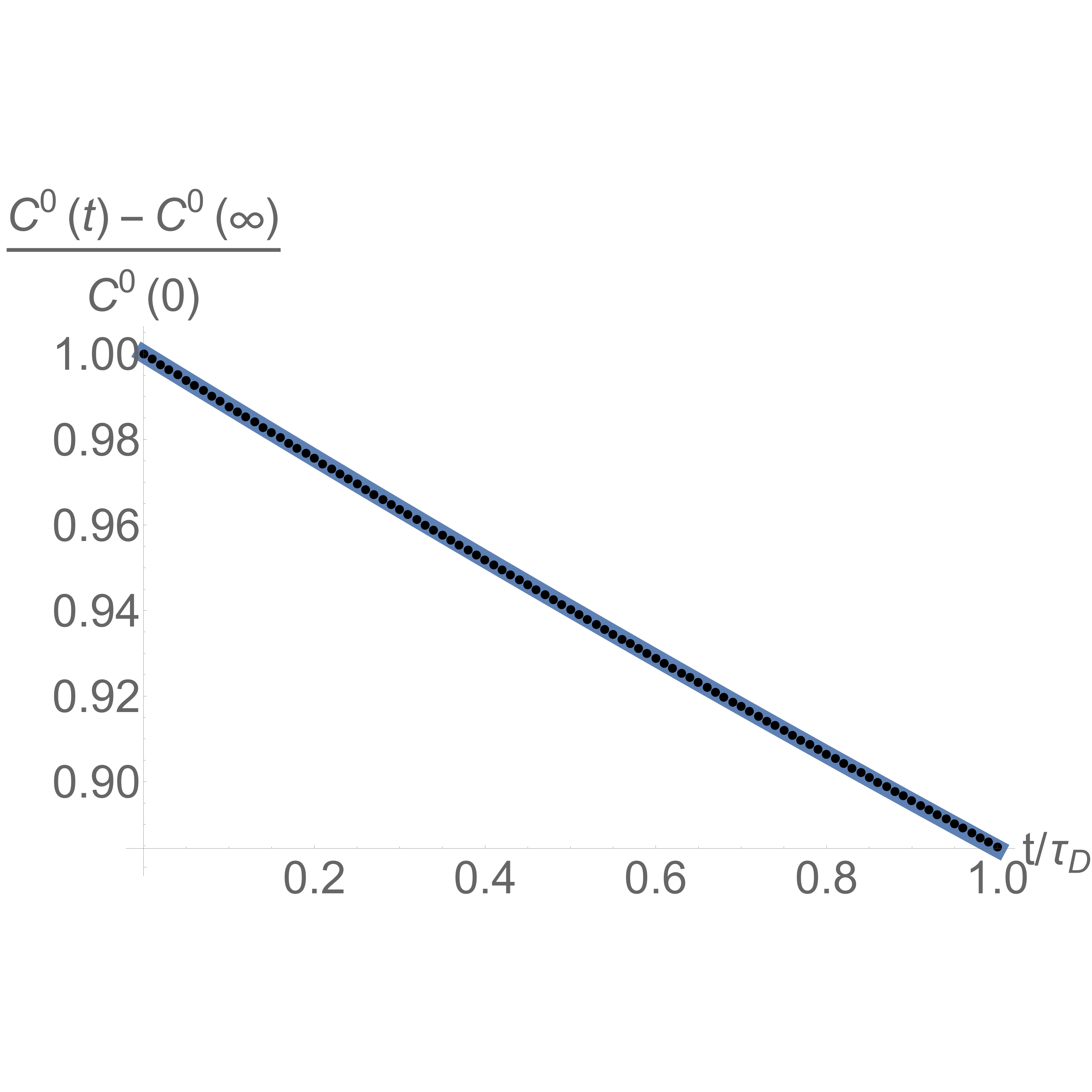} \label{hemisphere200short}}
	\subfloat[]{\includegraphics[width=0.32\textwidth]{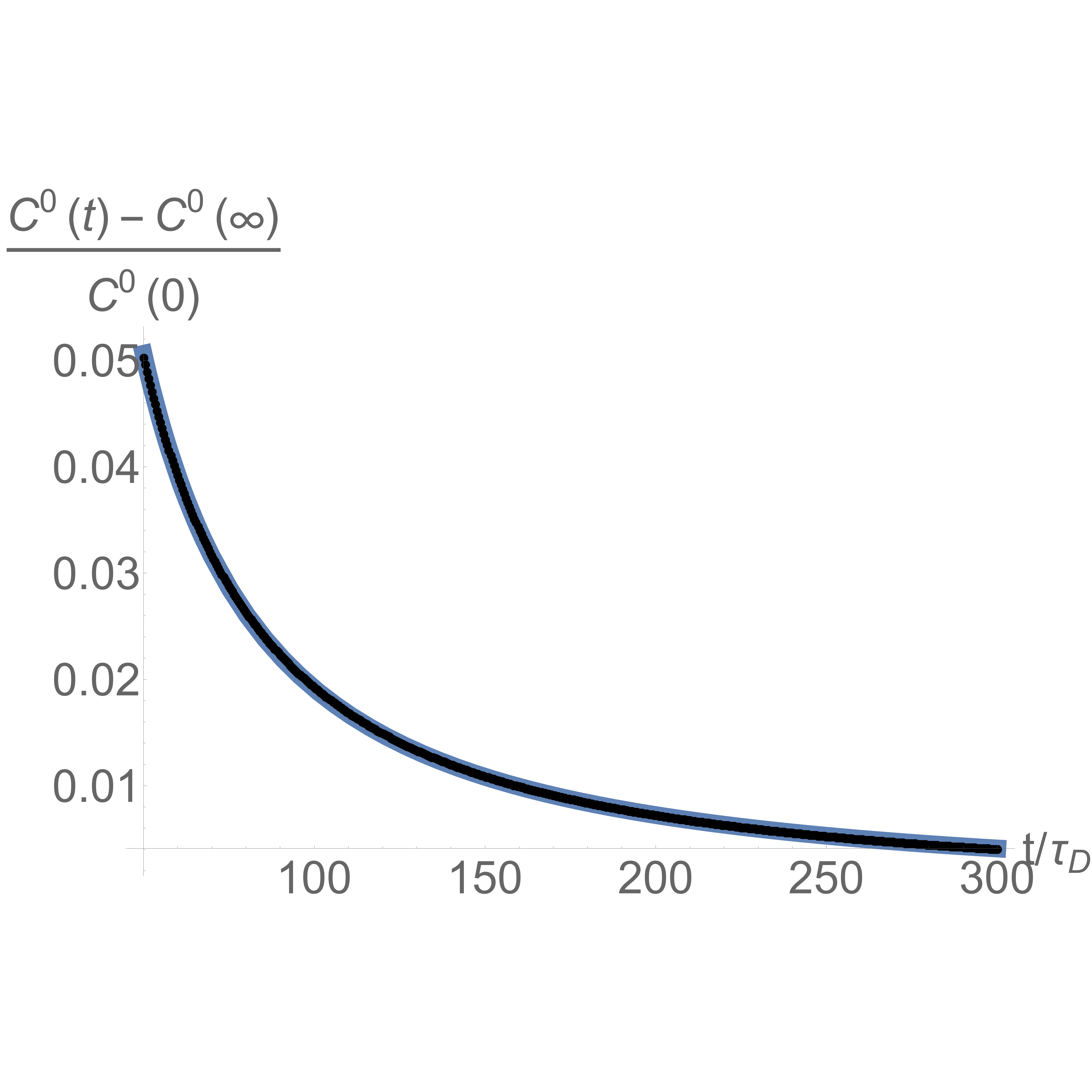} \label{hemisphere200inter}}
	\subfloat[]{\includegraphics[width=0.32\textwidth]{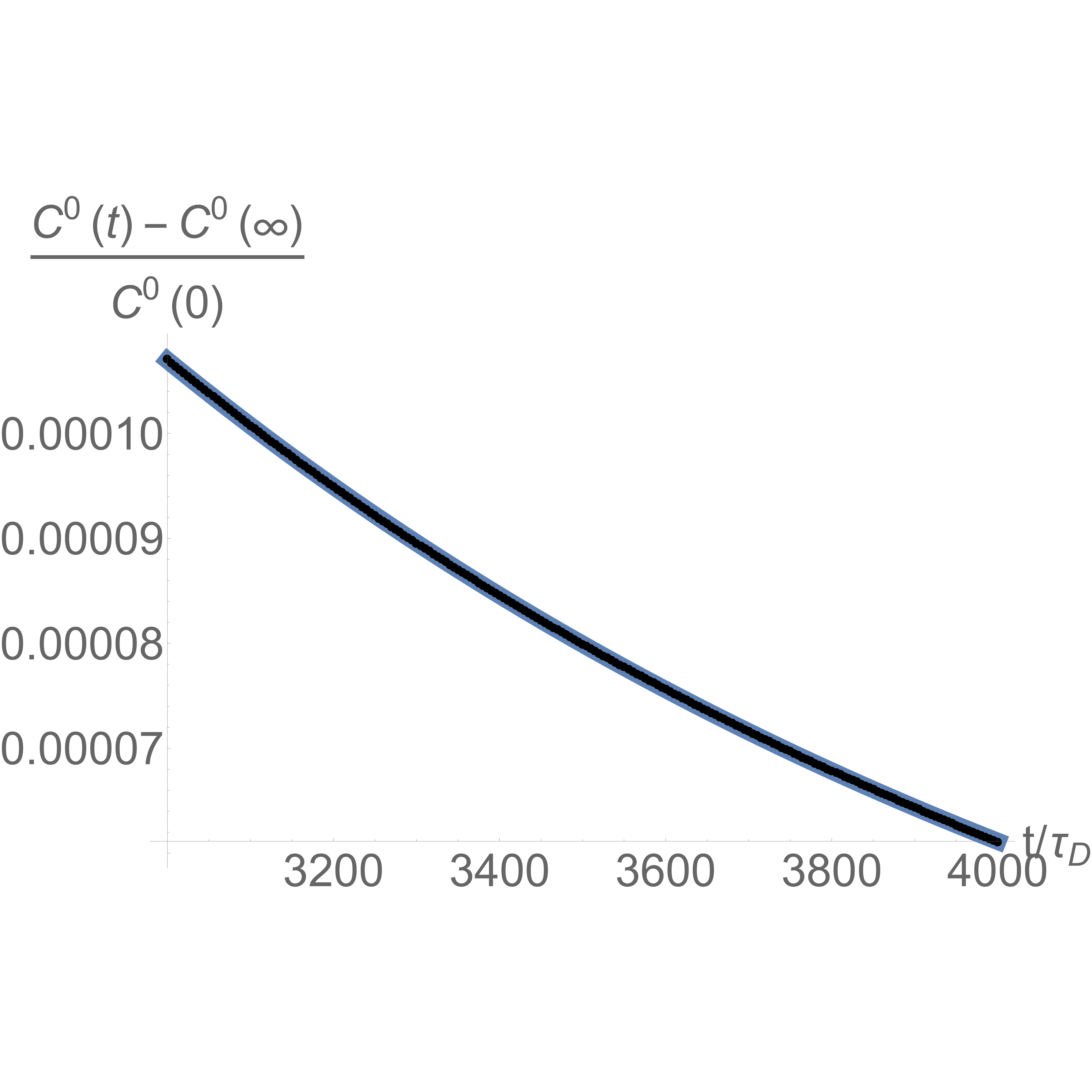} \label{hemisphere200long}}
	\hfill
	\subfloat[]{\includegraphics[width=0.32\textwidth]{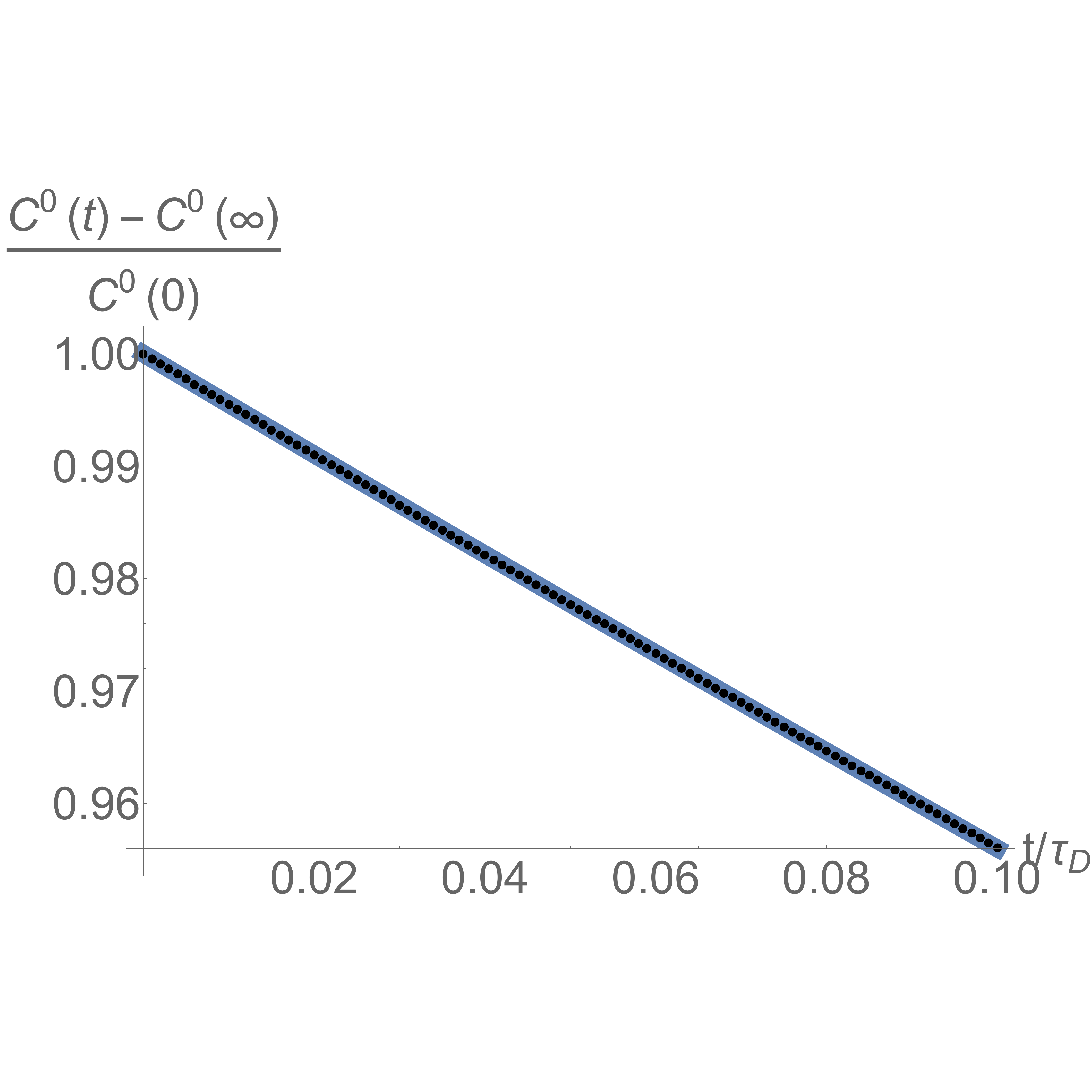} \label{hemisphere100short}}
	\subfloat[]{\includegraphics[width=0.32\textwidth]{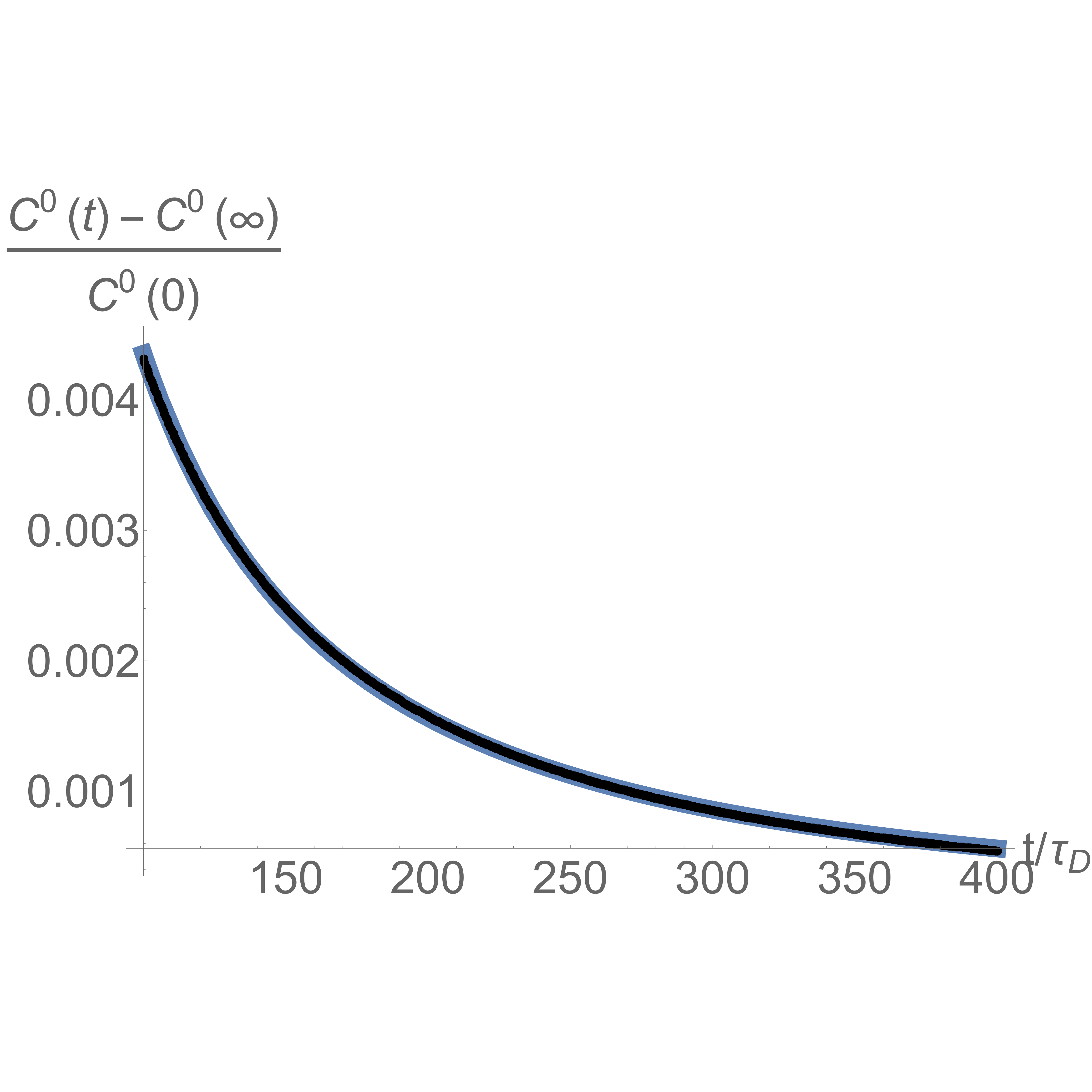} \label{hemisphere100inter}}
	\subfloat[]{\includegraphics[width=0.32\textwidth]{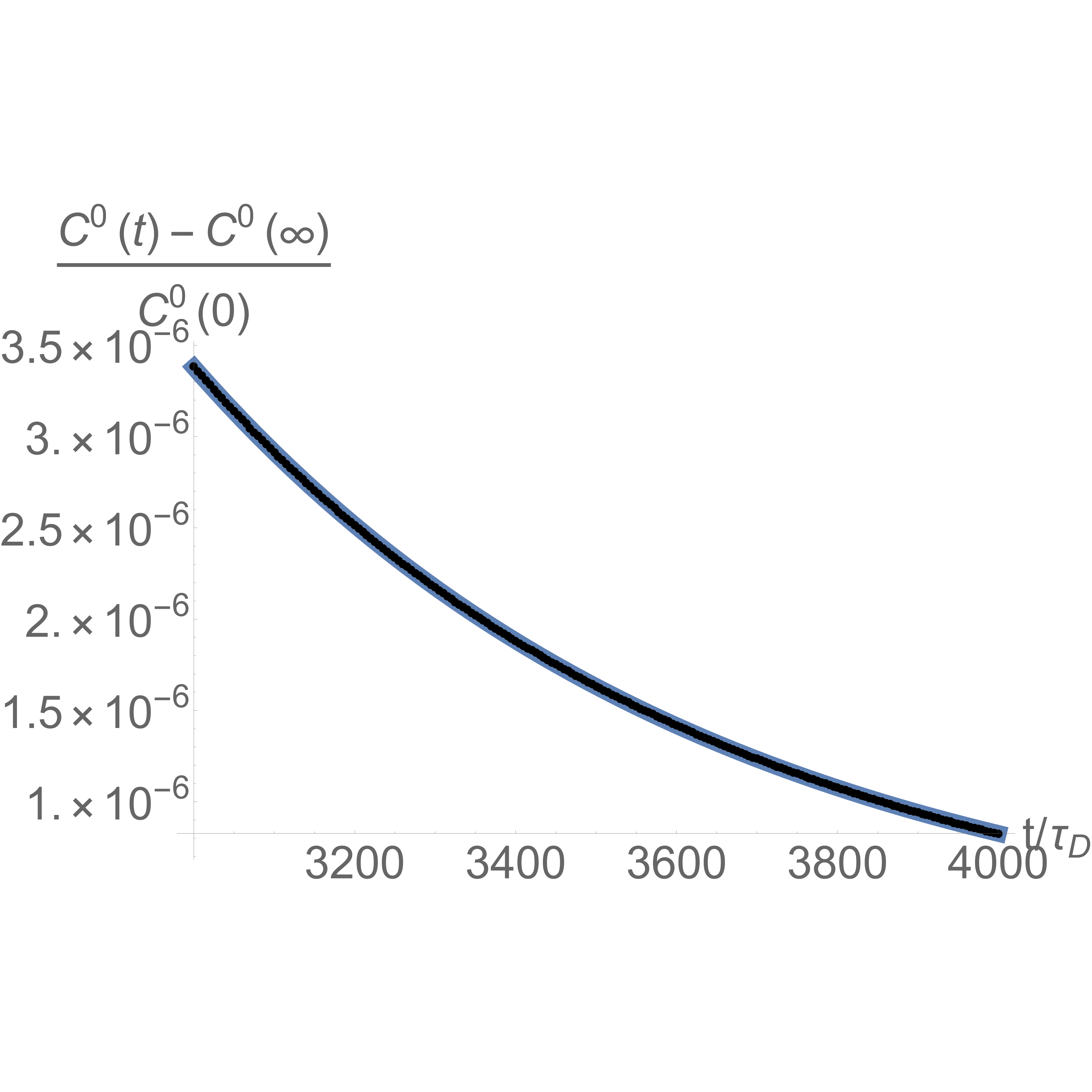} \label{hemispherer100long}}
	\hfill
	\subfloat[]{\includegraphics[width=0.32\textwidth]{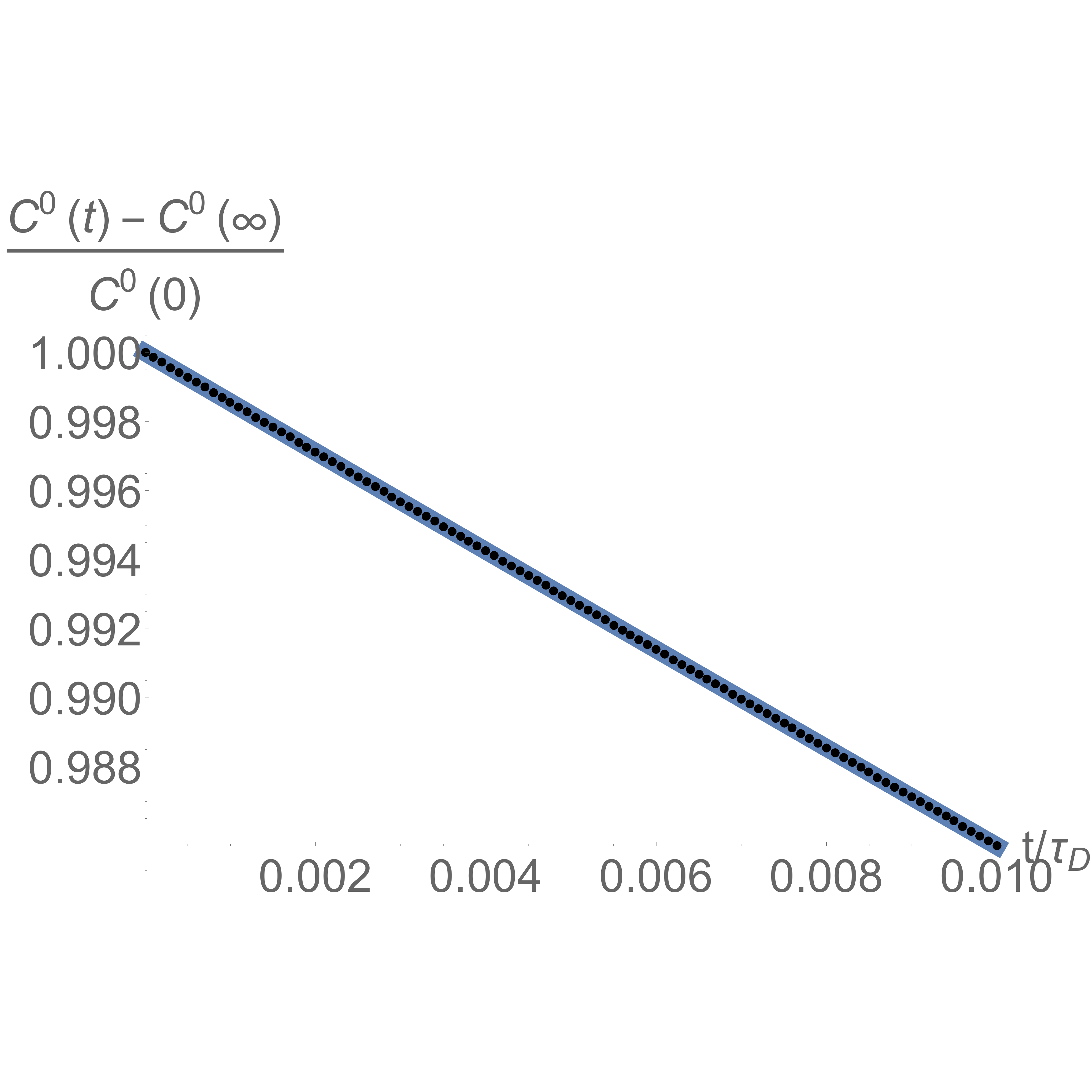} \label{hemisphere50short}}
	\subfloat[]{\includegraphics[width=0.32\textwidth]{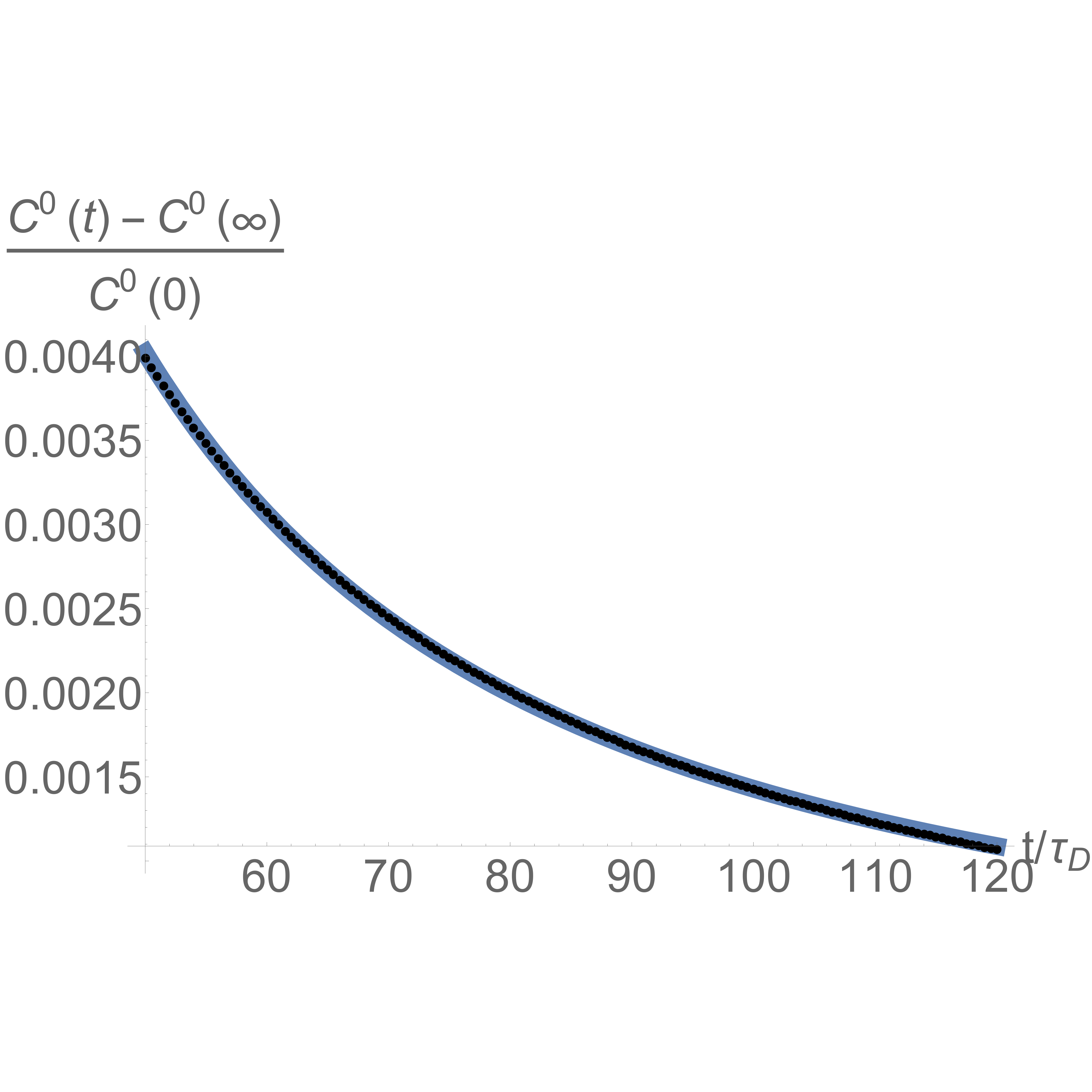} \label{hemisphere50inter}}
	\subfloat[]{\includegraphics[width=0.32\textwidth]{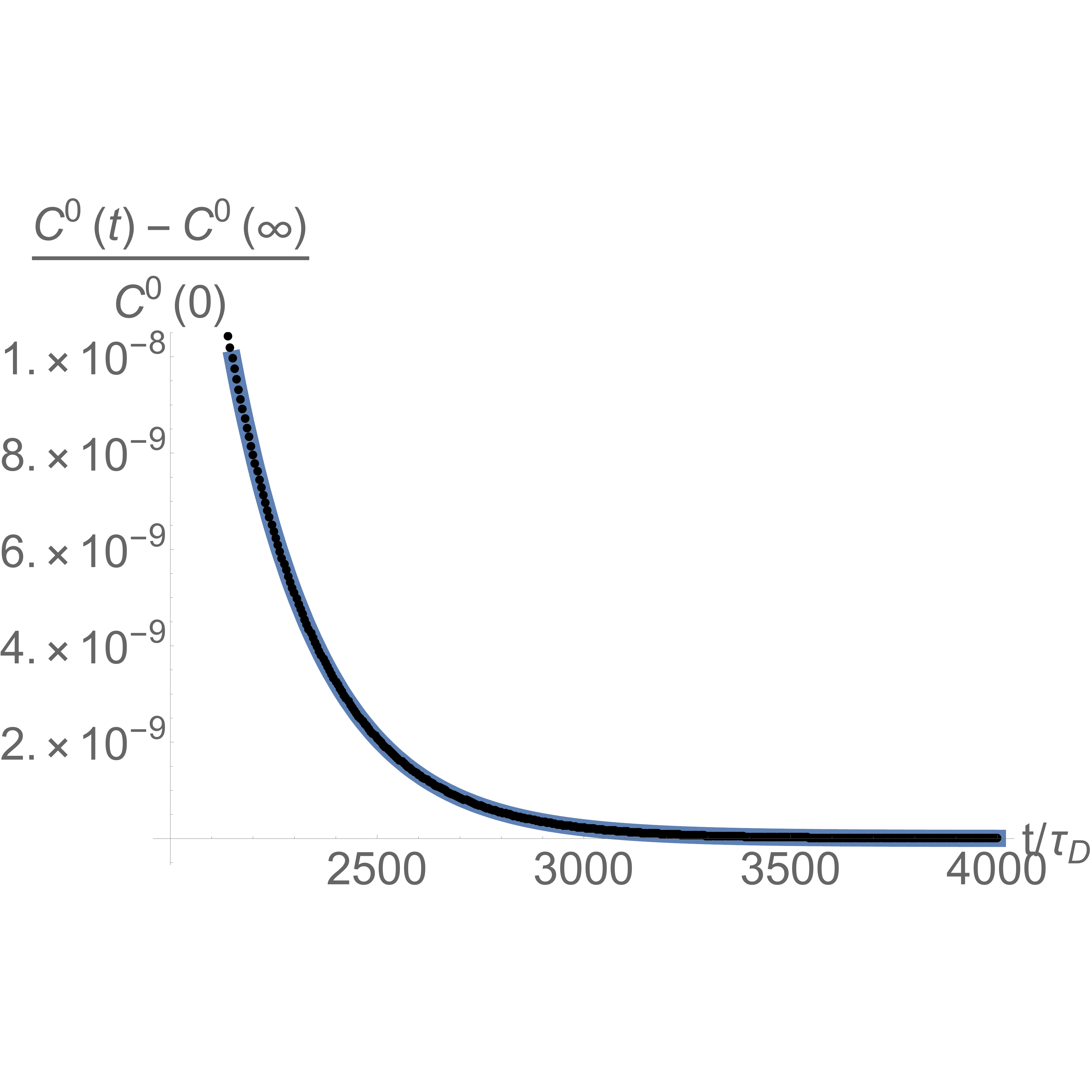} \label{hemisphere50long}}
	\caption{The correlation function in a \textit{hemispherical} geometry \eqref{corr_hemisphere_temporal_decay}  (black points) for $m=0$ in the three time regimes - $t\ll\tau_D$ (left column), $\tau_D\ll t\ll \tau_V$ (central column) and $\tau\gg\tau_V$ (right column) . The correlation was calculate for $d=1\ \textrm{nm}, \ D=0.5\ \frac{\textrm{nm}^2}{\mu\textrm{s}}$ and three different cylinder sizes - $R=200 \ \textrm{nm}$ (top row), $R=100 \ \textrm{nm}$ (central row) and $R=50 \ \textrm{nm}$ (bottom row). The short and long times limit of the correlation were fitted to exponential functions and the intermediate times to a power law (blue lines). 
	The decay rate in short times is inversely proportional to the cylinder's volume.
	The power in the intermediate regime fits the expected $-1.5$ scaling predicted in previous works, 
	and the decay rate in long times fits the slowest decaying mode of \eqref{corr_hemisphere_temporal_decay}. \label{corr_hemisphere_m0}}
\end{figure}

We evaluated the series \eqref{corr_hemisphere_temporal_decay} for $m=0$ by calculating the first terms ($l\in[0,30],k\in[1,30]$). The integral was calculated numerically for $d=\textrm{nm}$ and $R=200\ \textrm{nm},\ 100\ \textrm{nm},\ \& \ 50\ \textrm{nm}$. The
The results where fitted in the regimes $t\ll\tau_D$ and $t\gg\tau_V$ to exponential functions and for  $\tau_D \ll t\ll \tau_V$ to a power law. The results for $m=0$ are presented in  Fig. \ref{corr_hemisphere_m0}. Note, that the series converges slowly for large volumes and in practice an increasing number of terms might be needed to get a reasonable estimation at times approaching $t=0$. 
The zeroes of the derivative of the spherical Bessel function where calculated numerically using standard root finding methods with the analytical estimate \cite{Bessel_Zeros} as the initial guess.  
 \subsubsection{Full sphere}
\begin{figure}
		\subfloat[]{\includegraphics[width=0.32\textwidth]{SR200nm_exponential_short_q0p004.pdf} \label{sphere200short}}
	\subfloat[]{\includegraphics[width=0.32\textwidth]{SR200nm_power_inter_q0p43.pdf} \label{sphere200inter}}
	\subfloat[]{\includegraphics[width=0.32\textwidth]{SR200nm_exponential_long_q0p0001.pdf} \label{sphere200long}}
	\hfill
	\subfloat[]{\includegraphics[width=0.32\textwidth]{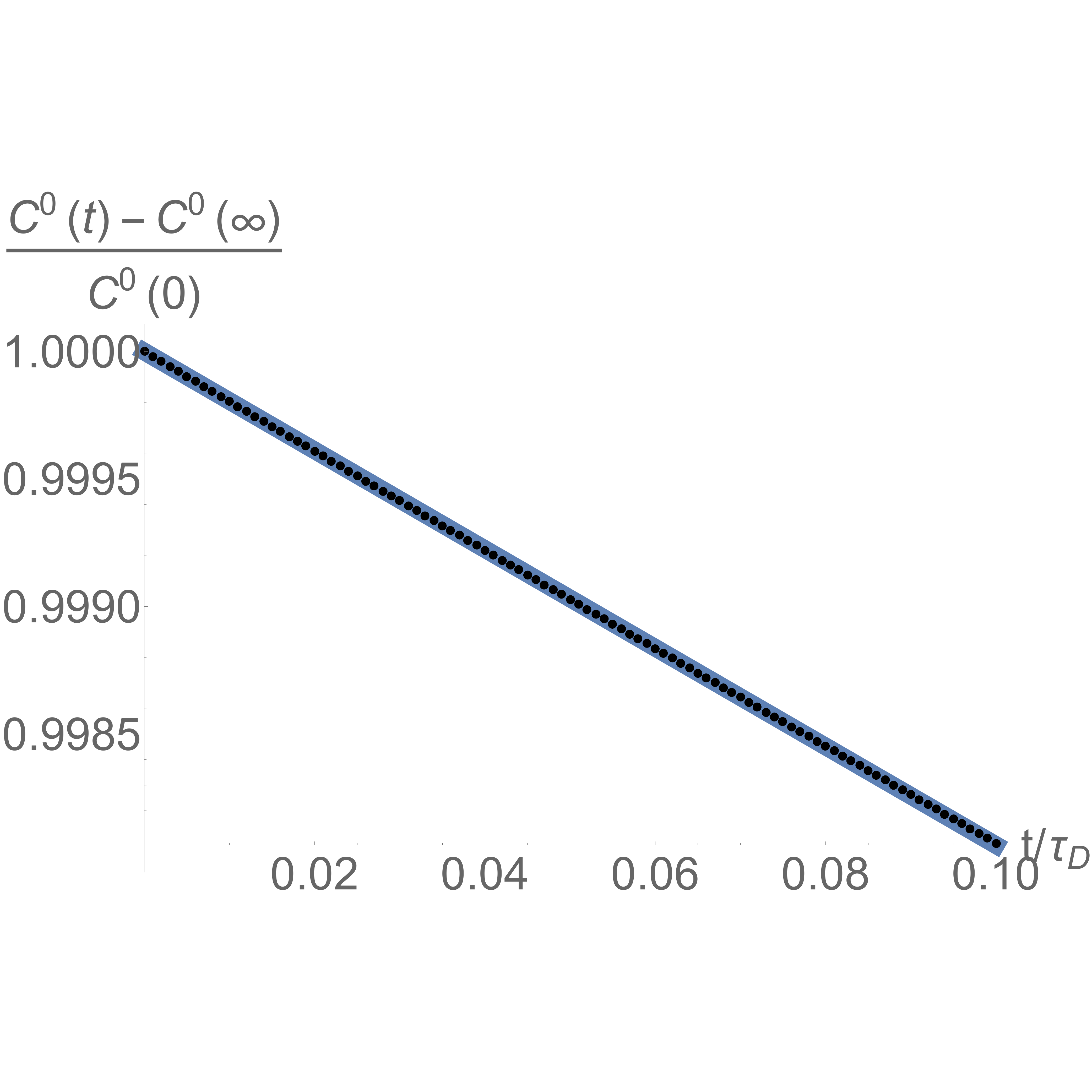} \label{sphere100short}}
	\subfloat[]{\includegraphics[width=0.32\textwidth]{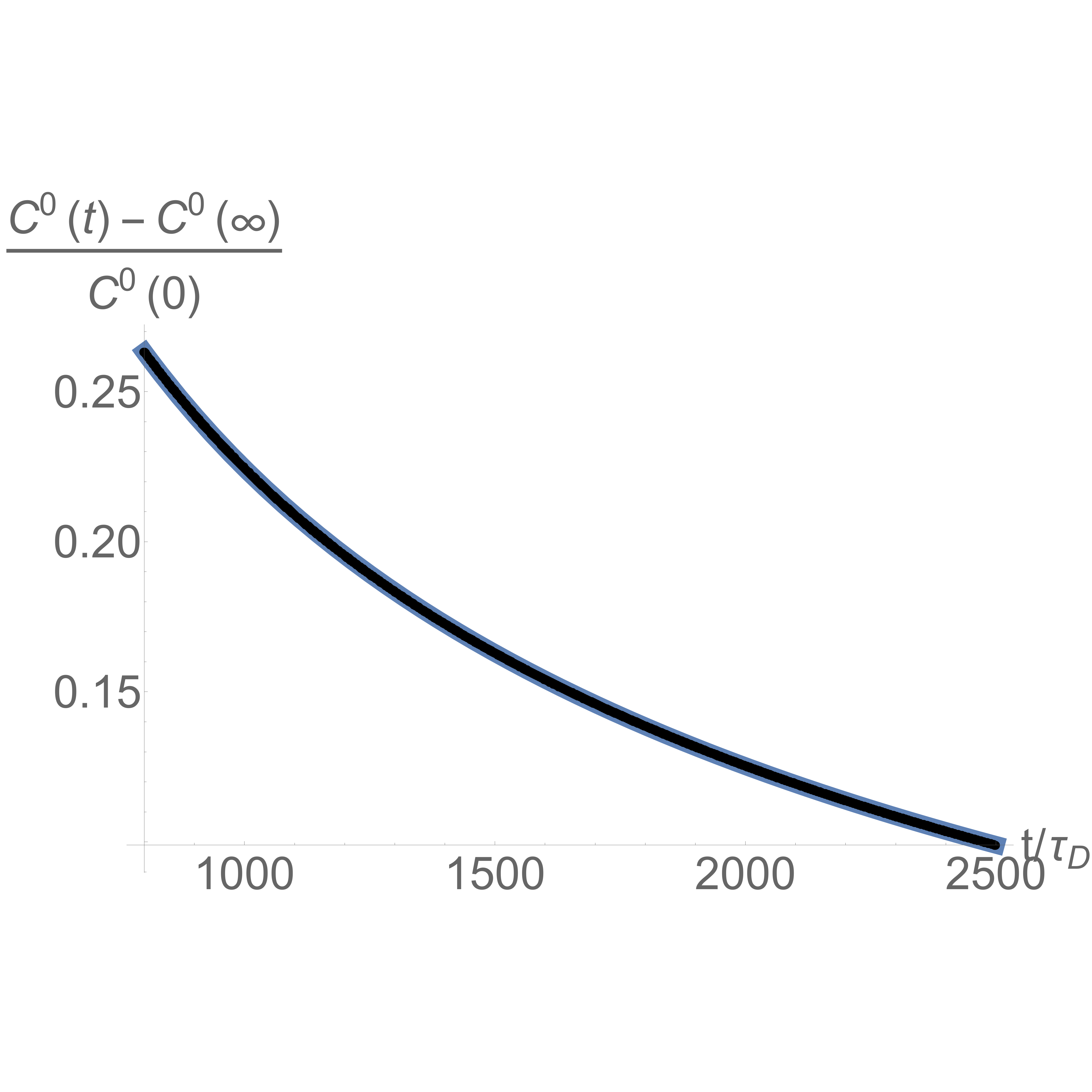} \label{sphere100inter}}
	\subfloat[]{\includegraphics[width=0.32\textwidth]{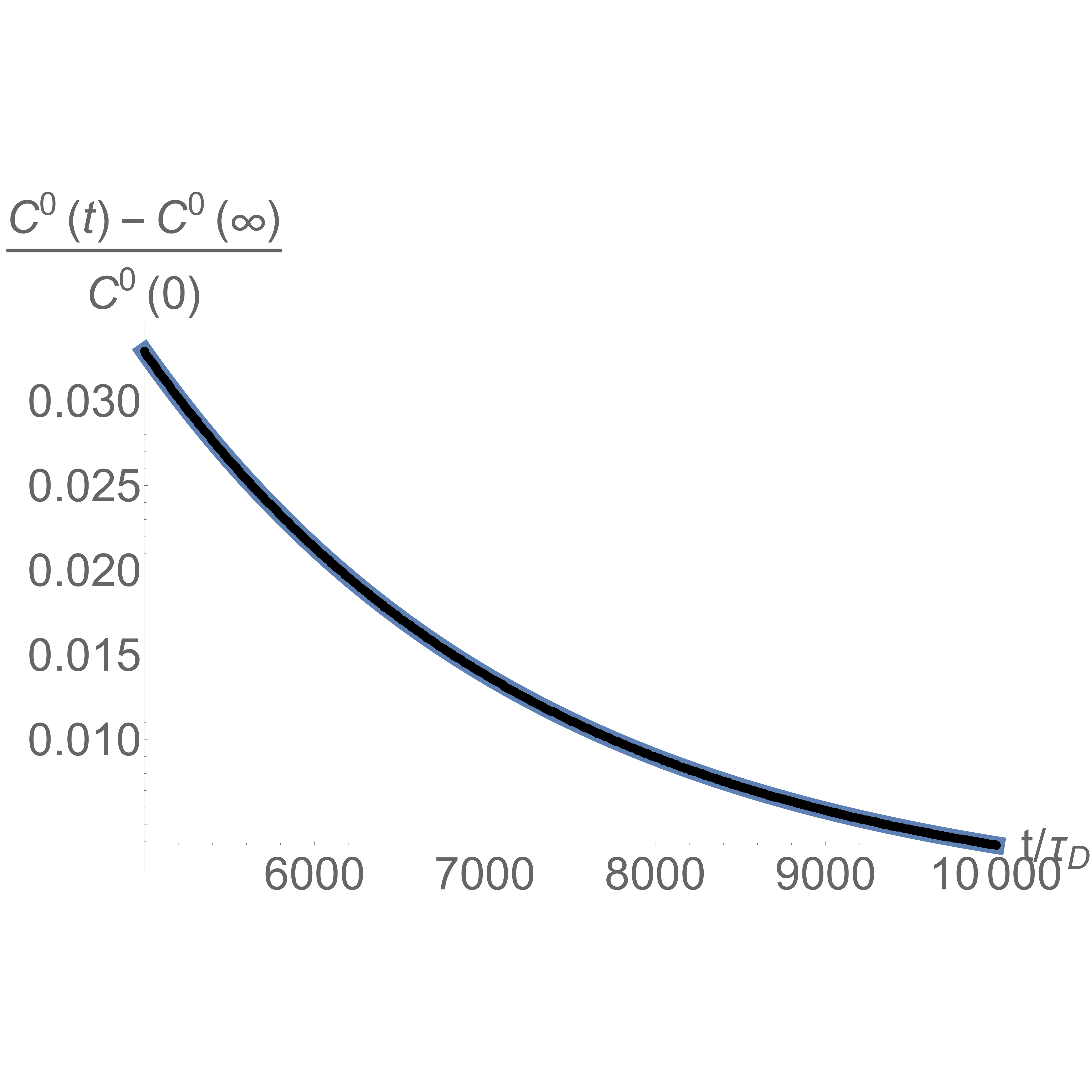} \label{spherer100long}}
	\hfill
	\subfloat[]{\includegraphics[width=0.32\textwidth]{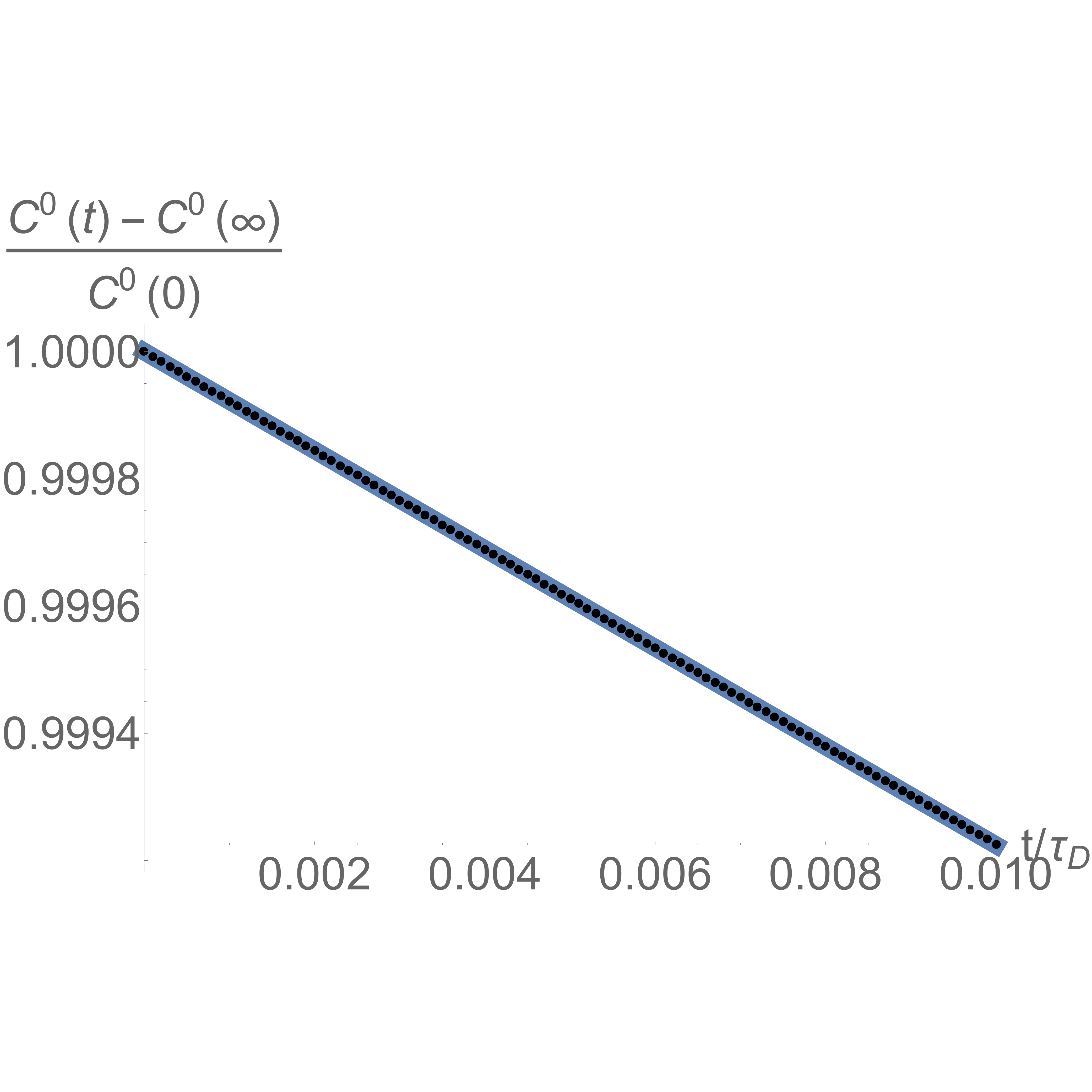} \label{sphere50short}}
	\subfloat[]{\includegraphics[width=0.32\textwidth]{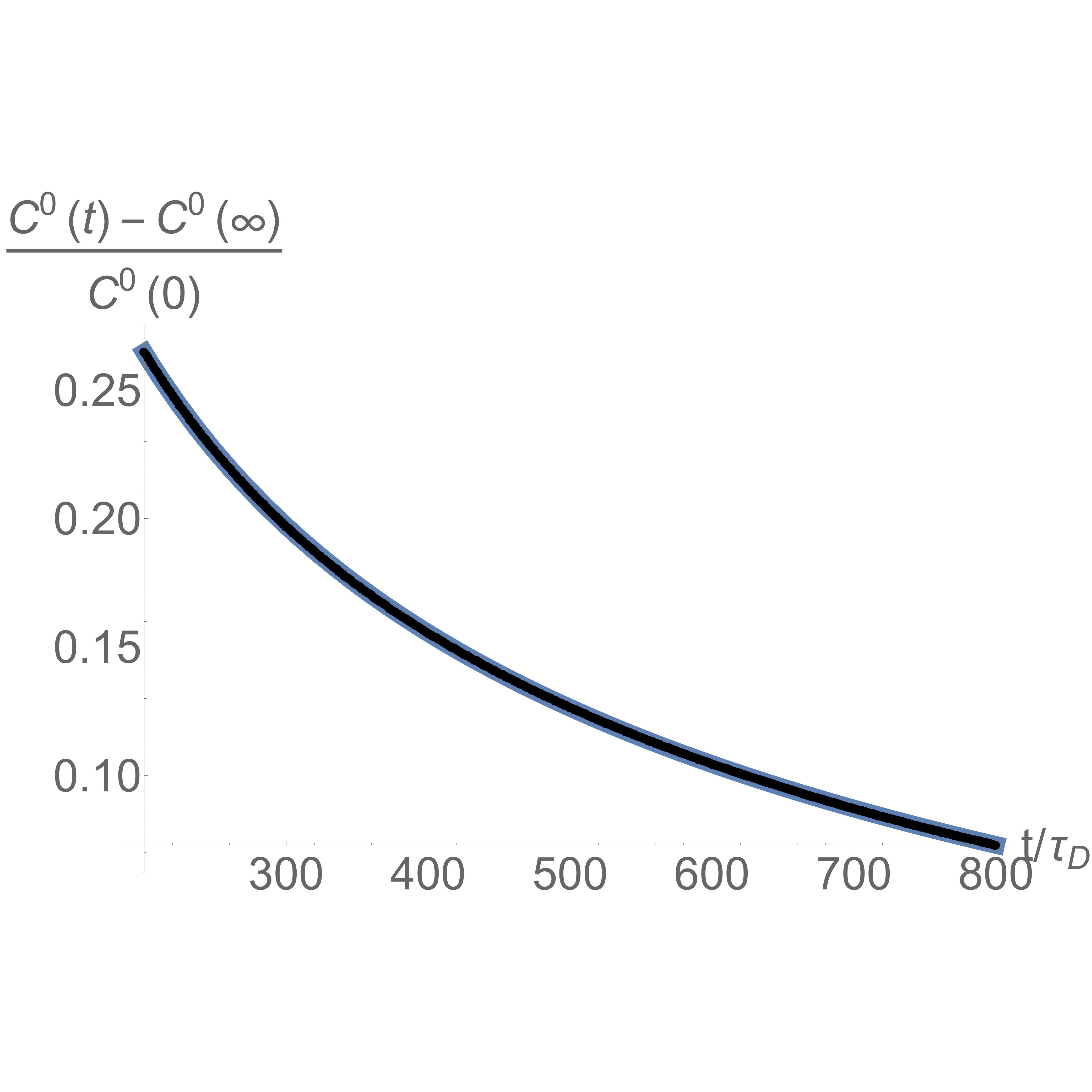} \label{sphere50inter}}
	\subfloat[]{\includegraphics[width=0.32\textwidth]{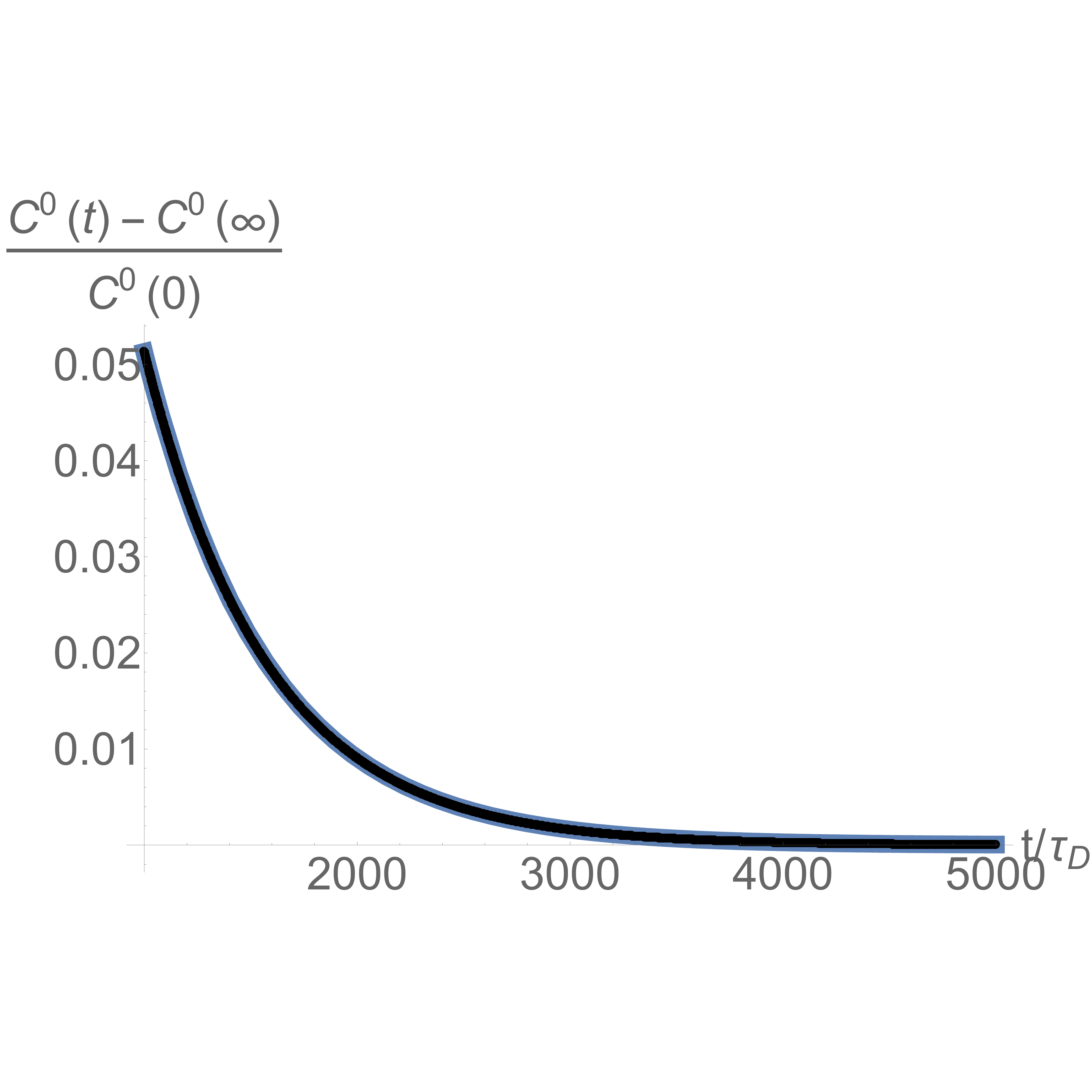} \label{sphere50long}}
	\caption{The correlation function in a \textit{spherical} geometry \eqref{corr_sphere_temporal_decay}  (black points) for $m=0$ in the three time regimes - $t\ll\tau_D$ (left column), $\tau_D\ll t\ll \tau_V$ (central column) and $\tau\gg\tau_V$ (right column) . The correlation was calculate for $d=1\ \textrm{nm}, \ D=0.5\ \frac{\textrm{nm}^2}{\mu\textrm{s}}$ and three different cylinder sizes - $R=200 \ \textrm{nm}$ (top row), $R=100 \ \textrm{nm}$ (central row) and $R=50 \ \textrm{nm}$ (bottom row). The short and long times limit of the correlation were fitted to exponential functions and the intermediate times to a power law (blue lines). 
	The decay rate in short times is inversely proportional to the cylinder's volume and the decay rate in long times fits the slowest decaying mode of \eqref{corr_sphere_temporal_decay}. \label{corr_sphere_m0}}
	The power in the intermediate regime fits the expected $-0.5$ scaling unlike the one predicted in previous works. This change might be related to the curvature of the plane near the NV, which truncates the effective iteration volume.
\end{figure}

  We first solve \eqref{Brms_general} for each value of $m$ individually:
 \begin{align}\label{Brms_sphere0}
 C^{(0)}(0)/J^2=&\frac{\pi}{8d^{3}(d+R)^{5}(d+2R)^{3}}\left[d^{3}\left(-d^{5}-8d^{4}R-16d^{3}R^{2}+16d^{2}R^{3}+80dR^{4}+64R^{5}\right)\tanh^{-1}\left(\frac{R}{d+R}\right)\right.\\\nonumber
 &\left.+d^{7}R+7d^{6}R^{2}+52d^{5}R^{3}+190d^{4}R^{4}+320d^{3}R^{5}+256d^{2}R^{6}+96dR^{7}+16R^{8}\right],
 \end{align}
 \begin{align}\label{Brms_sphere1}
 C^{(1)}(0)/J^2=&\frac{\pi}{32d^{3}(d+R)^{5}(d+2R)^{3}}\left[-d^{3}\left(3d^{5}+24d^{4}R+168d^{2}R^{3}+84d^{3}R^{2}+192dR^{4}+96R^{5}\right)\tanh^{-1}\left(\frac{R}{d+R}\right)\right.\\\nonumber
 &\left.+3d^{7}R+21d^{6}R^{2}+64d^{5}R^{3}+110d^{4}R^{4}+136d^{3}R^{5}+136d^{2}R^{6}+80dR^{7}+16R^{8}\right],
 \end{align}
 \begin{align}
 \label{Brms_sphere2}
 C^{(2)}(0)/J^2=&\frac{\pi}{128d^{3}(d+R)^{5}(d+2R)^{3}}\left[d^{3}\left(15d^{5}+120d^{4}R+384d^{3}R^{2}+624d^{2}R^{3}+528dR^{4}+192R^{5}\right)\tanh^{-1}\left(\frac{R}{d+R}\right)\right.\\\nonumber
 &\left.-15d^{7}R-105d^{6}R^{2}-284d^{5}R^{3}-370d^{4}R^{4}-224d^{3}R^{5}-32d^{2}R^{6}+32dR^{7}+16R^{8}\right]
 \end{align}
  The limit $d\ll R,L$ recovers the half-plane $B_{rms}^2$  of \cite{microfludics}.
  We continue by calculating the long times limit \eqref{corr_longtimes_general}.
  \begin{align}\label{longtimes_sphere0}
  &C^{(0)}(t\gg \tau_V)/J^2\approx\frac{64 \pi ^2 R^6}{9 V (d+R)^6}\\\label{longtimes_sphere2}
  &C^{(2)}(t\gg \tau_V)/J^2\approx\frac{4\pi ^2}{V} \left[ \tanh ^{-1}\left(\frac{R}{d+R}\right)-\frac{ R \left(3 d^2+6 d R+4 R^2\right)}{3 (d+R)^3}\right]^2.
  \end{align} 
  We where not able to calculate the integral for $m=1$ analytically.
  Finally, we provide numerical results for $\tilde{C}^{(m)}$.
  Substituting the diffusion propagator \eqref{Sphere_prop} into \eqref{corr_general} we arrive at
  \begin{align}\label{corr_sphere_temporal_decay}
  \tilde{C}^{(m)}=\frac{1}{V}\sum_{l=0}^{\infty}\sum_{k=1}^{\infty}\frac{8\pi}{3j_{l}^{2}\left(\tilde{\nu}_{k,l}\right)\left(1-\left(\frac{l}{\tilde{\nu}_{k,l}}\right)^{2}\right)}\left[\int \frac{d^3r}{r^3}j_{l}\left(\frac{\tilde{\nu}_{k,l}}{R}r\right)Y_2^{(m)*}\left(\theta,\phi\right)Y_{l}^{(m)}\left(\theta,\phi\right)\right]^2\exp\left(-\frac{\tilde{\nu}_{k,l}^{2}}{\tau_{V}}t\right)
  \end{align}
We evaluated the series \eqref{corr_sphere_temporal_decay} for $m=0$ by calculating the first terms ($l\in[0,30],k\in[1,30]$). The integral was calculated numerically for $d=\textrm{nm}$ and $R=200\ \textrm{nm},\ 100\ \textrm{nm},\ \& \ 50\ \textrm{nm}$. 
The results where fitted in the regimes $t\ll\tau_D$ and $t\gg\tau_V$ to exponential functions and for  $\tau_D \ll t\ll \tau_V$ to a power law. The results for $m=0$ are presented in  Fig. \ref{corr_sphere_m0}. Note, that the series converges slowly for large volumes and in practice an increasing number of terms might be needed to get a reasonable estimation at times approaching $t=0$.

\section{MD Simulations}
We corroborated our analytical calculations by performing molecular dynamics (MD) simulations for two of the three geometries: cylindrical and spherical. 

\subsection{Cylinder}
\begin{figure}
		\centering
		\includegraphics[width=0.9\textwidth]{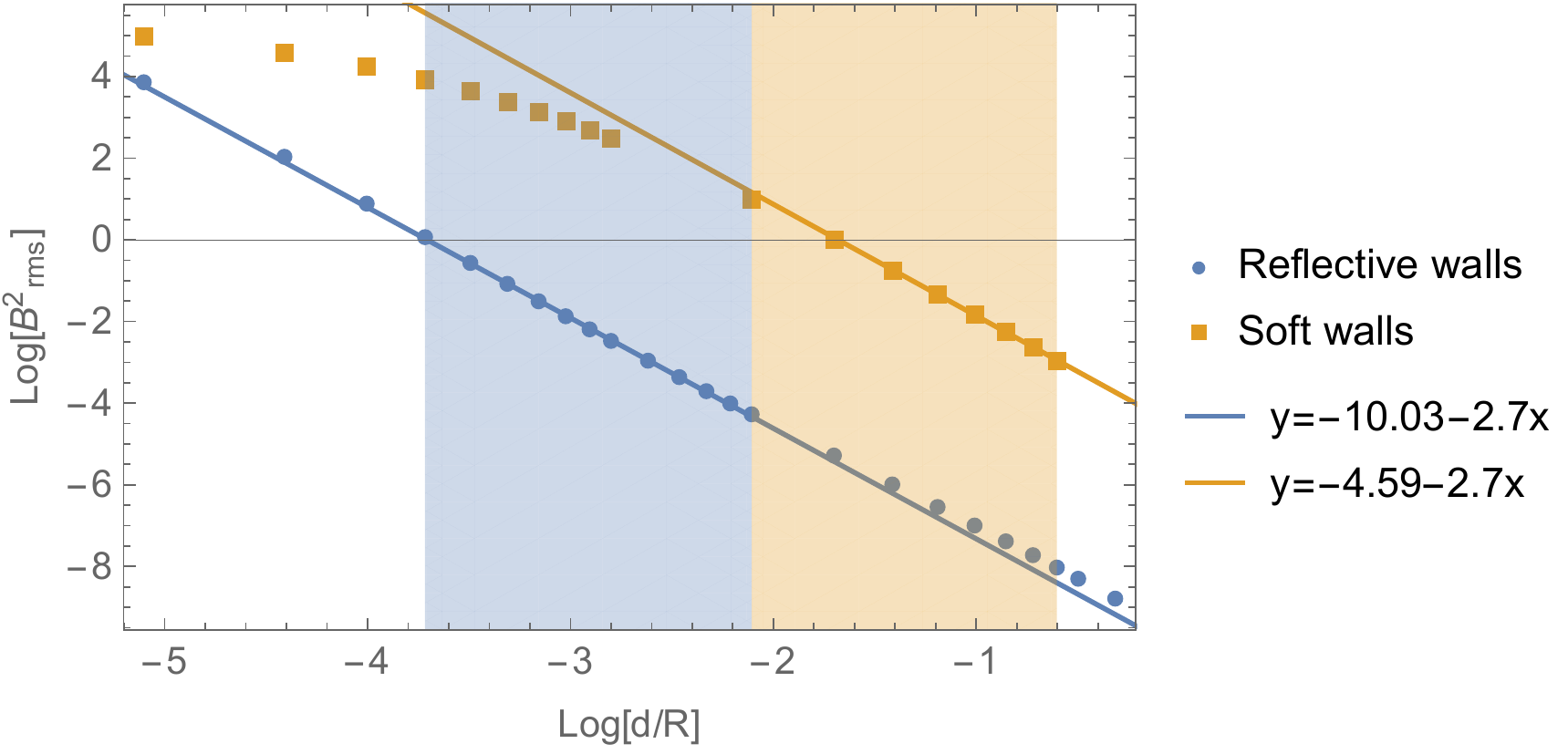} 
	\caption{The correlation of the magnetic field as a function of the NV's depth fitted to a linear model on a logarithmic scale. The fit of each set-up is based on the points within the colored region. Both simulations display a decay of $d^{-2.7}$, which is similar to the $d^{-3}$ scaling expected at small values of $d/R$. Interestingly, the reflective walls follow this scaling for almost the entire range of examined NV depths, whereas the soft walls display this scaling only at $d/R\geq0.12$. Shallower NV's don't display the expected increase in $B_{rms}$ due to the interaction with the walls.\label{fig:Brms_vs_d_simulations}}
\end{figure}

\begin{figure}
		\centering
		\includegraphics[width=0.9\textwidth]{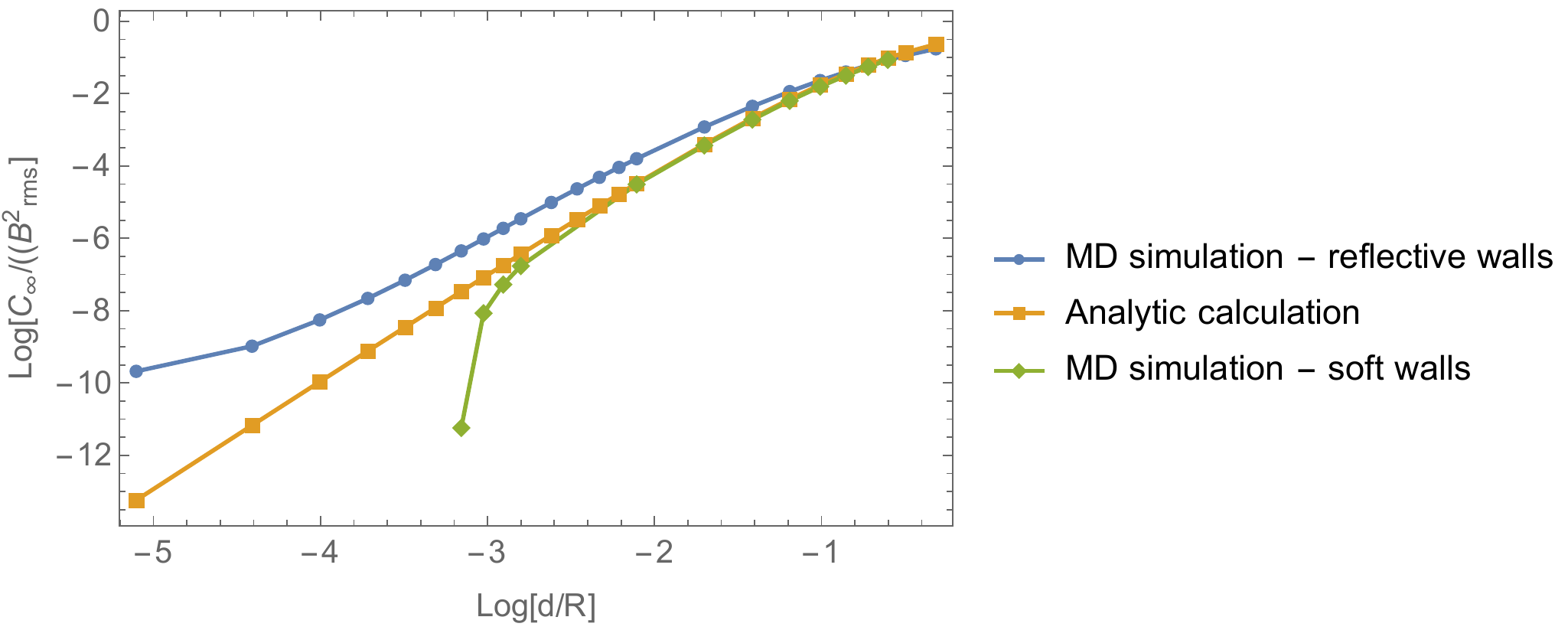} 
	\caption{The asymptotic value of the correlation function at $t\rightarrow\infty$ normalized by the correlation at $t=0$. All the results converge for large values of $d/R$, while they exhibit a large disagreement at small values. The agreement at large values is attributed to a better sampling of the parameter space, while the deviation at small values is due to the interactions between the nuclei, the nuclei and the surface (in the case of soft walls) and limited integration time. \label{fig:Cinf_vs_d}}
\end{figure}

The simulation set-up contains $N$ particles, which are confined to a cylinder
of radius $R$ and height $L$. The particles interact via
the Lennard-Jones (LJ) potential $4\epsilon[(\sigma/r)^{12}-(\sigma/r)^{6}]$
with an interaction cutoff distance of $r_{c}=2.5\sigma$. We performed two distinct simulation. In the first,
specular reflections are applied on the top and bottom walls of the cylinder and the Lennard-Jones 9/3 potential is applied at the curved walls to confine the particles to the interior of the cylinder, while in the other the Lennard-Jones 9/3 potential is applied over the entire cylindrical boundary.
The system 
is initialized into a thermal state at temperature $T$ by running
a Langevin dynamics simulation until the system reaches thermal equilibrium.
Each particle is assigned a random value of spin $I_{z}\in\{-1,1\}$.
After initializing the system, we ran a deterministic molecular dynamics
simulation, integrating Newton's laws using the Velocity-Verlet
method with step size $\Delta t=0.005\sqrt{\epsilon/(m\sigma^{2})}$.
During the simulation we computed and stored the $z$ component of the
magnetic field induced by the particles at the position of the NV
\beq
B(t)=\sum_{j=1}^{N}\frac{1}{r_{i}^{3}(t)}\left[3\cos^{2}(\theta_{i}(t))-1\right]I_{z}^{i}.
\eeq
where $r_{i}$ is the distance between particle $i$ and the NV center
and $\theta_{i}$ is the angle between $r_{i}$ and the NV quantization
axis. The NV is placed along the axis of rotational symmetry of the
cylinder, at a distance $d$ below the bottom of the cylinder. 
In both set-ups we took $R=L=16.44$ in LJ units. Since our analytic model does not include interaction we first checked the dependency of $B_{rms}$ and $C^0(\infty)/C^0(0)$ with respect to $d$. The results are presented in Figs. \ref{fig:Brms_vs_d_simulations} and \ref{fig:Cinf_vs_d} respectively. 

\begin{figure}
	\subfloat[]{\includegraphics[width=0.32\textwidth]{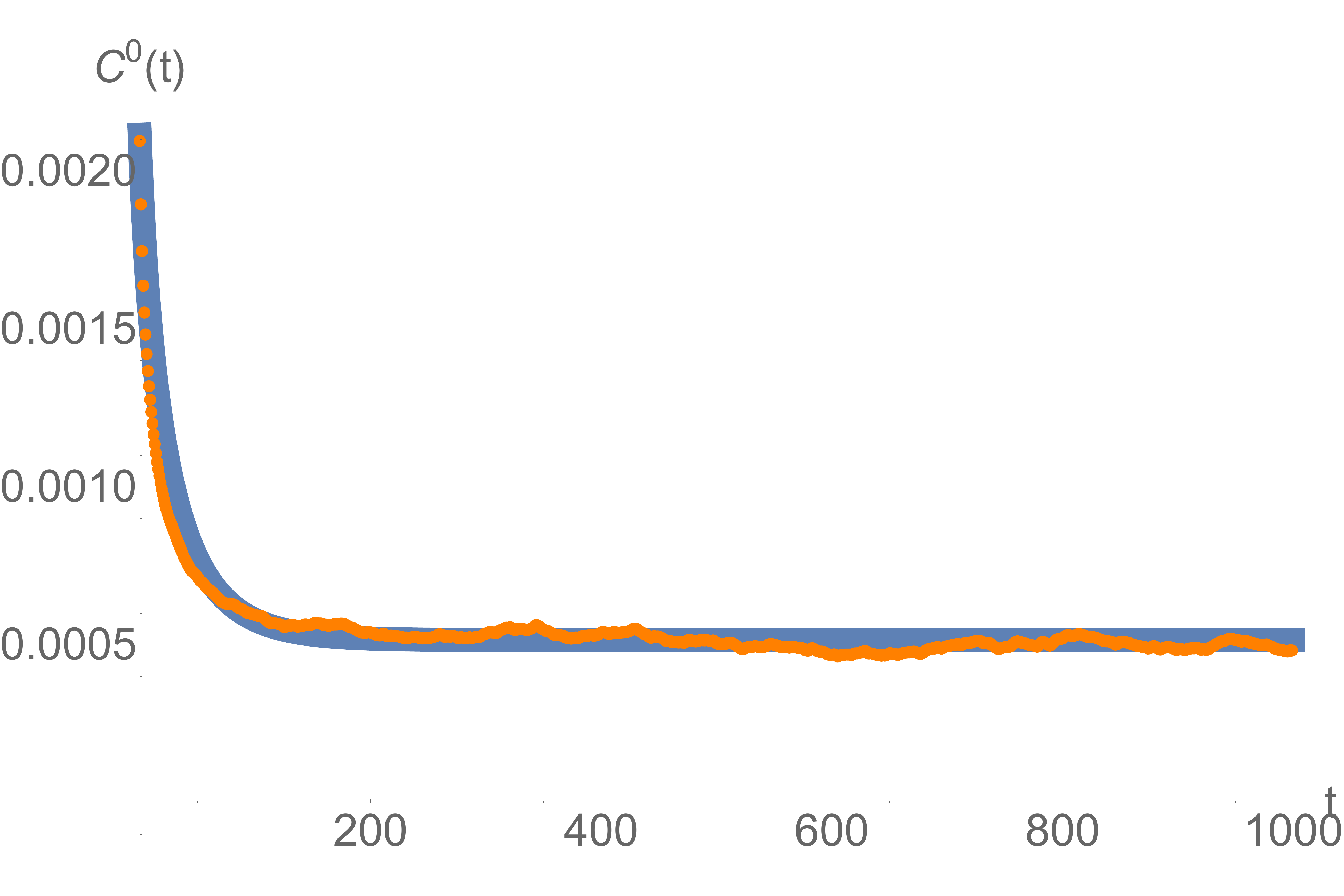} \label{Cylinder_sim_d7}}
	\subfloat[]{\includegraphics[width=0.32\textwidth]{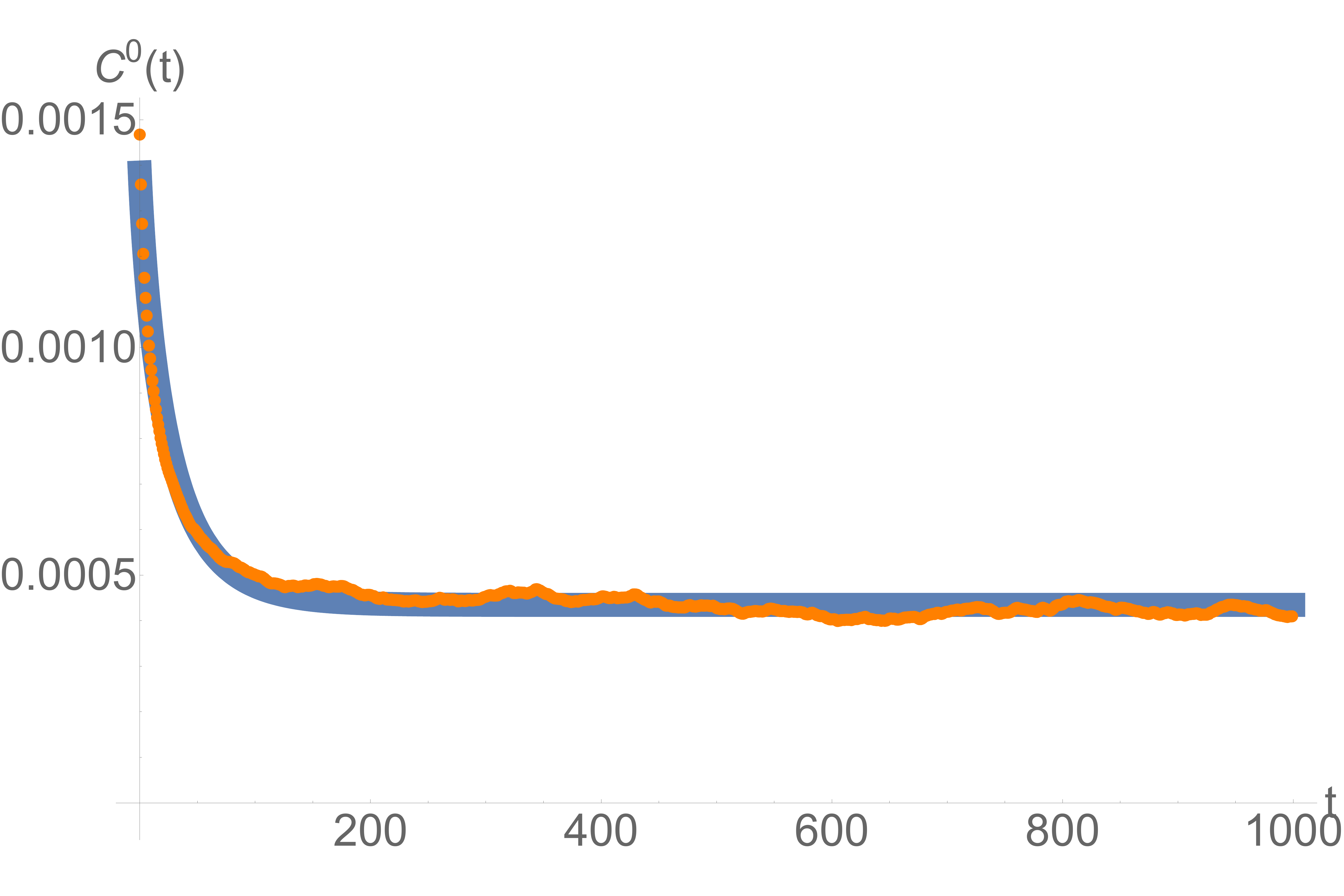} \label{Cylinder_sim_d8}}
	\subfloat[]{\includegraphics[width=0.32\textwidth]{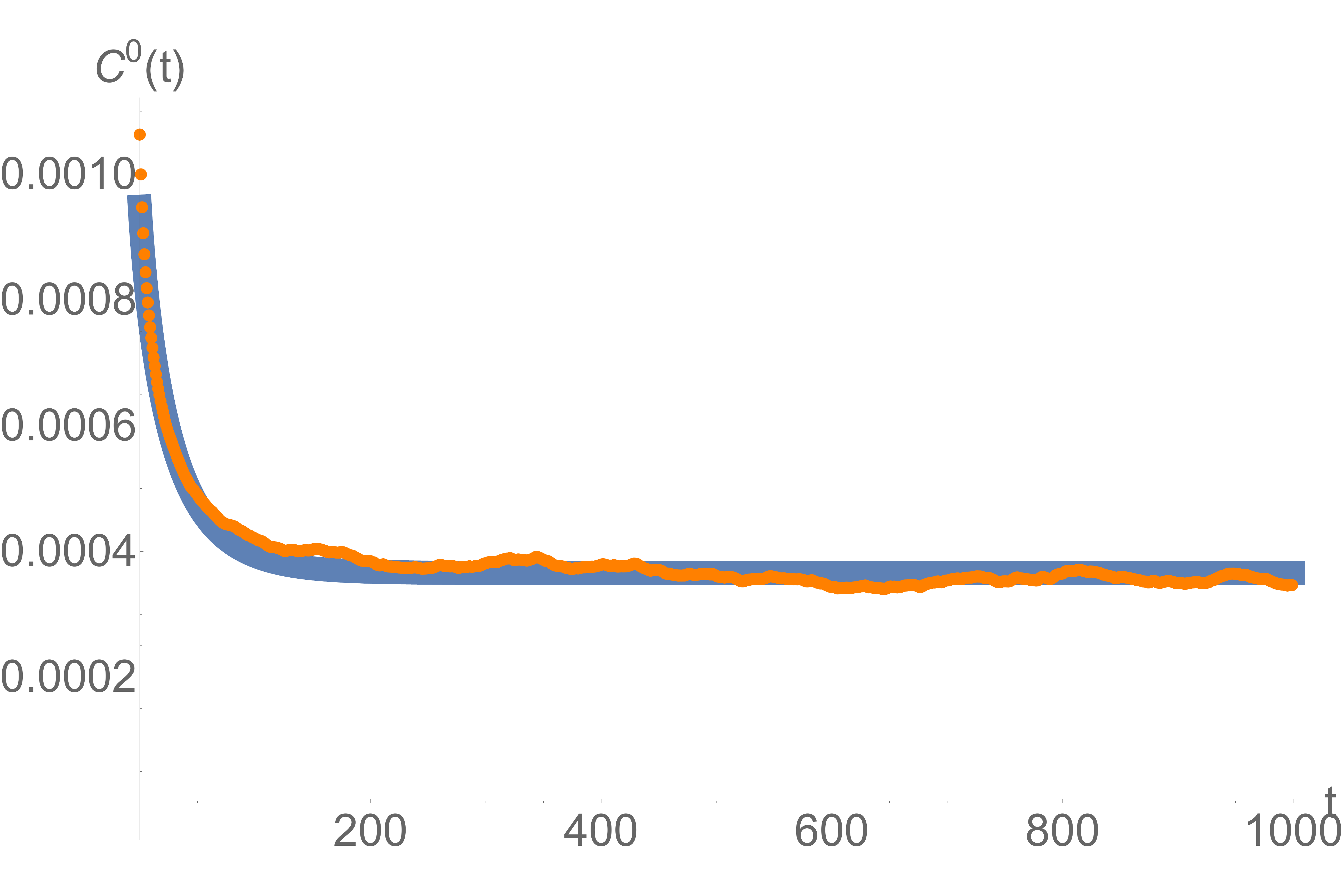} \label{Cylinder_sim_d9}}
	\caption{The magnetic field correlation function in a \textit{cylindrical} geometry with $R=L=16.44\sigma$ and (a) $d=7\sigma$ (b) $d=8\sigma$ (c) $d=9\sigma$. The correlation is given in arbitrary units, whereas the time is given in LJ units. The analytic curve was estimated by numerically calculating the first terms of Eq. \eqref{correlation_cylinder_temporal_decay} ($k,s\in[1,25]$) and adding the asymptotic constant given by Eq. \eqref{longtimes_cylinder0}. The simulation agrees with the analytic results apart from the value at $t=0$. This is because the limited integration time of the simulation is insufficient for estimating the power spectrum at $\omega=0$, which translates to an error in the estimation of $B_{rms}$. } \label{Cylinder_sim_fit}
\end{figure}

Both simulations present a scaling of $d^{-2.7}$ at certain regions of $d/R$. Since the expected scaling is $d^{-3}$ we expect our analytic results to to fit these regions.
An interesting feature of the soft walls simulation is that the $B_{rms}$ does not increase as expected for shallow NVs ($d/R\ll1$). This is due to the interaction with the walls, which repels the nuclei outside of the effective interaction region causing the $B_{rms}$ to decrease. The normalized asymptotic constants of both simulations agree with the analytic results at large values of $d/R$. This is expected because in this regime the correlation decays rapidly to the constant value, allowing for a better sampling within the limited integration time.
The disagreement for shallower NV's in the reflective walls simulation can be attributed to the underestimate of $B_{rms}$ discussed above. In the soft walls simulation it is more probably a result of the interaction with the walls that prevents nuclei from re-entering the effective interaction region, since the difference occurs at $d<\sigma$.
Since the results of the reflective boundary better match the expected behavior of the correlation at $t=0$ and $t\rightarrow\infty$ we henceforth continue only with this simulation.
We fitted the analytic results to the simulation by rescaling the simulation's results to match the analytic asymptotic constant and by choosing an appropriate diffusion coefficient. The result are presented in Fig. \ref{Cylinder_sim_fit}.

\subsection{Sphere}

\begin{figure}
	\subfloat[]{\includegraphics[width=0.32\textwidth]{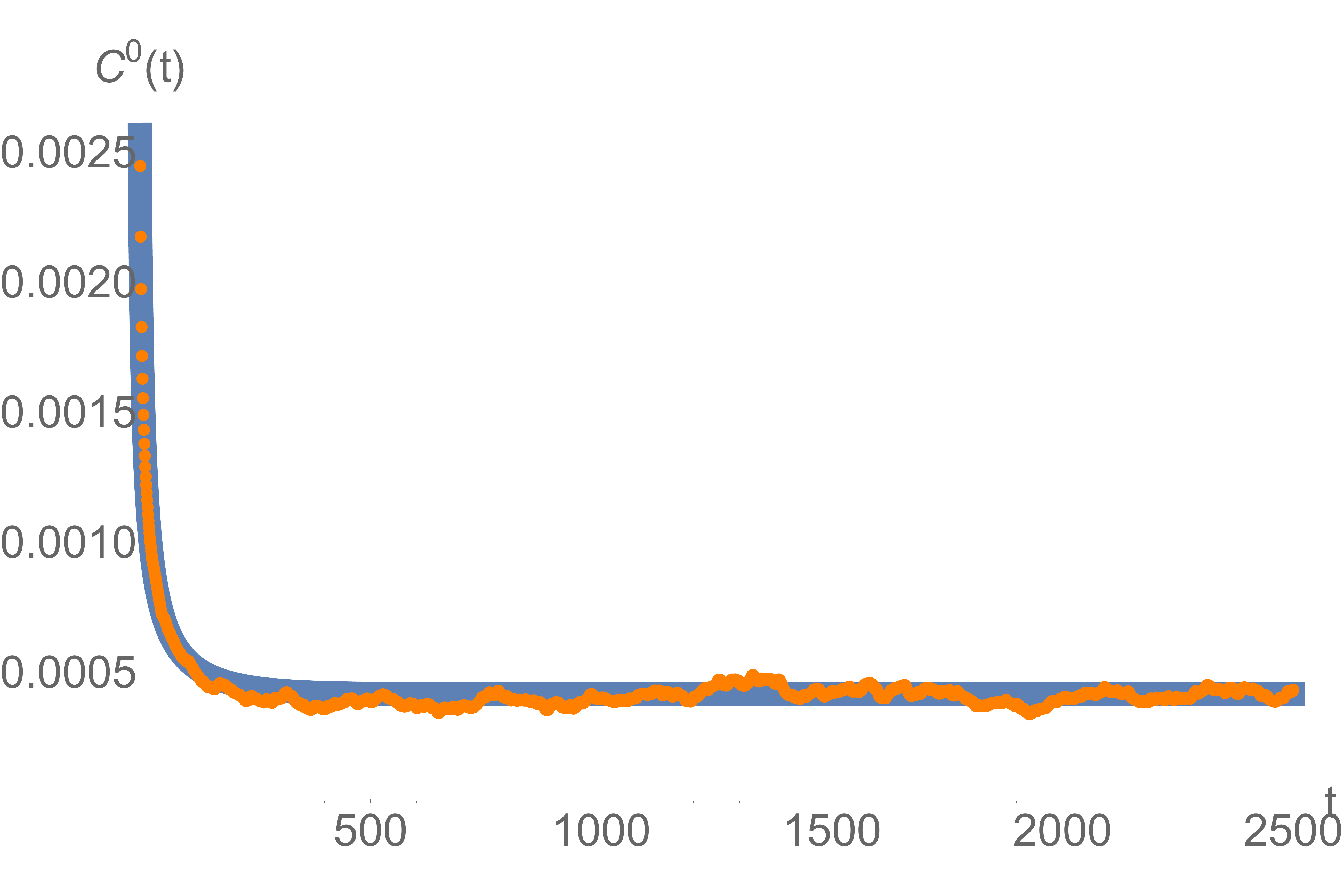} \label{Sphere_sim_d7}}
	\subfloat[]{\includegraphics[width=0.32\textwidth]{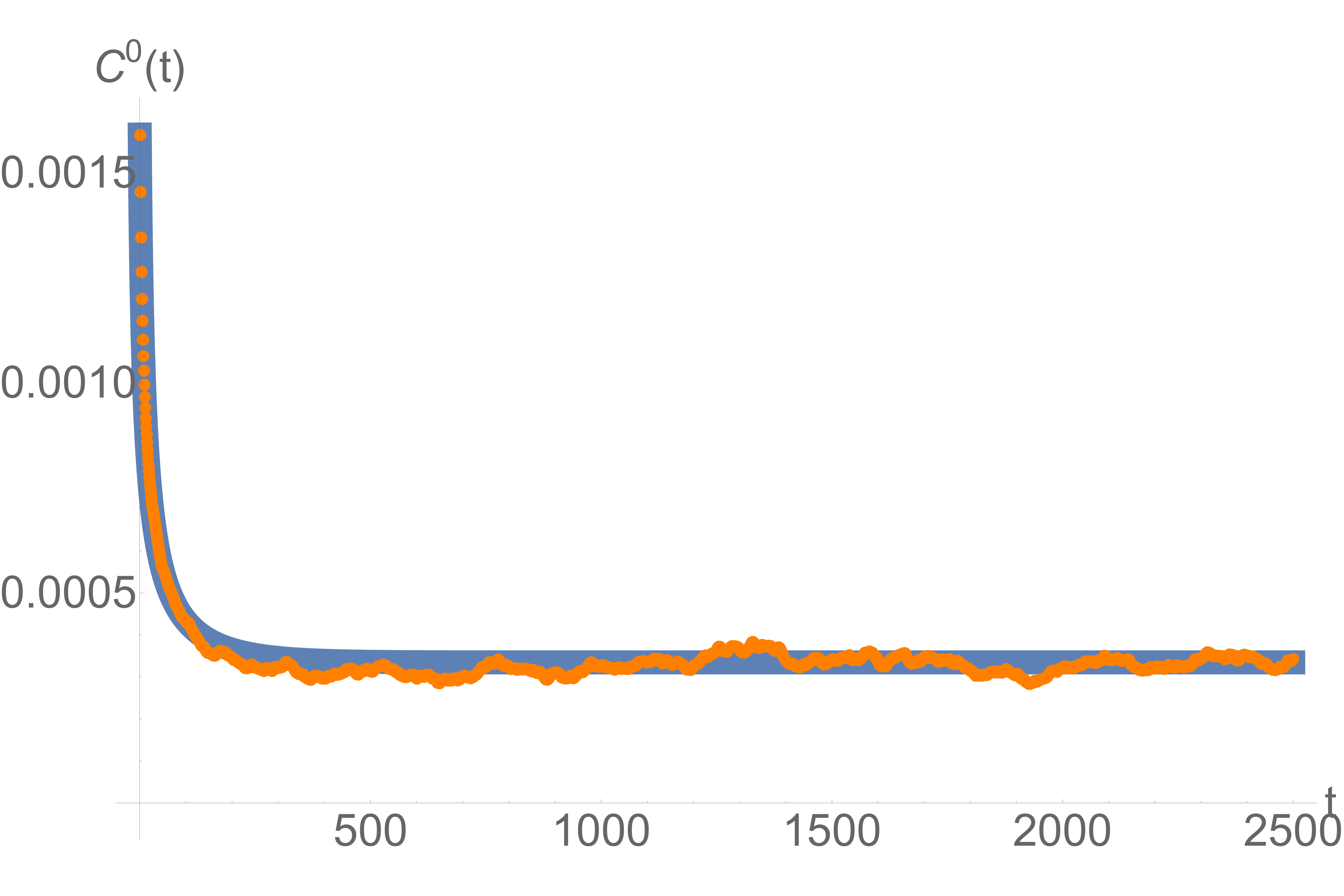} \label{Sphere_sim_d8}}
	\subfloat[]{\includegraphics[width=0.32\textwidth]{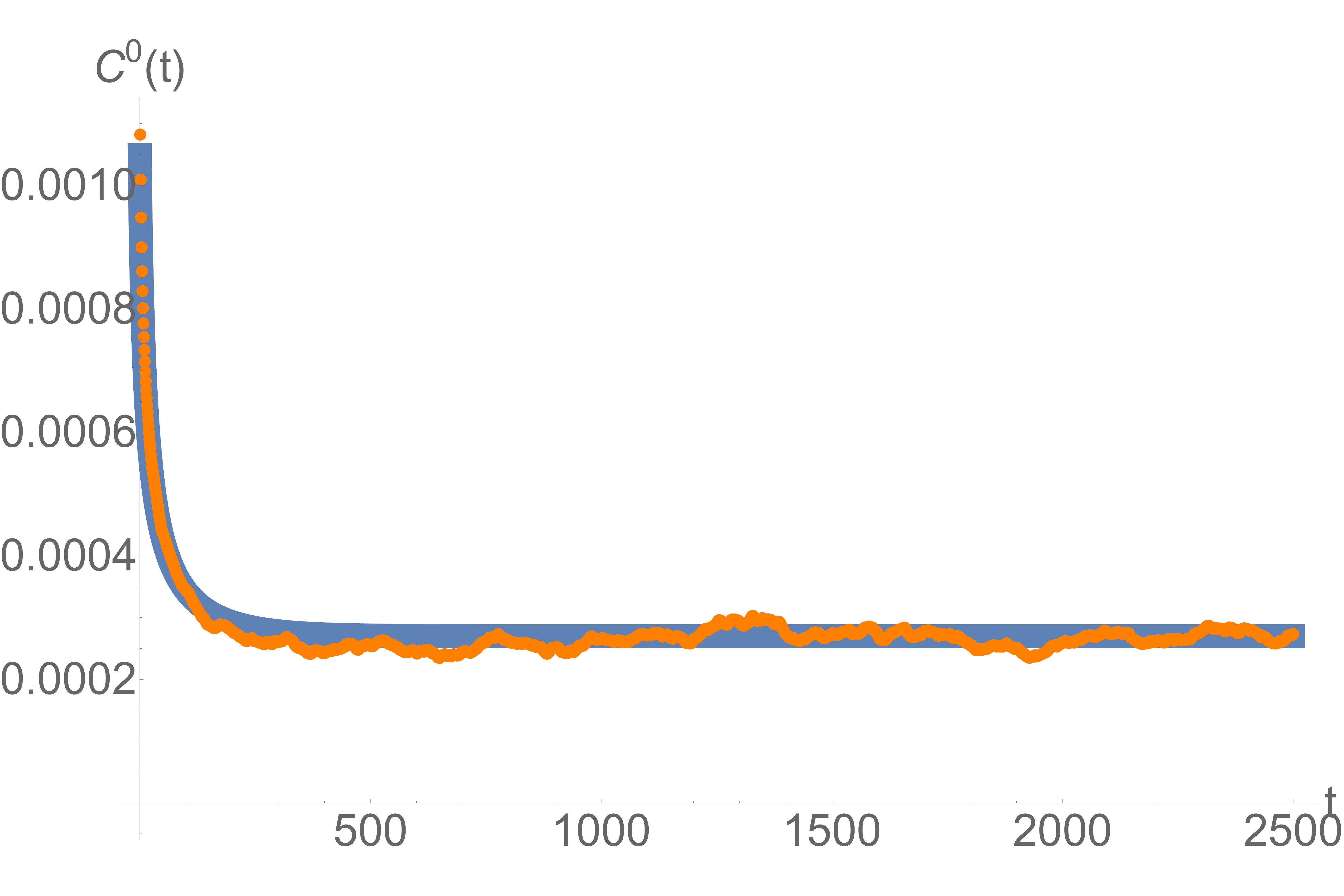} \label{Sphere_sim_d9}}
	\caption{The magnetic field correlation function in a \textit{spherical} geometry with $R=L=20.47\sigma$ and (a) $d=7\sigma$ (b) $d=8\sigma$ (c) $d=9\sigma$.  
The correlation is given in arbitrary units, whereas the time is given in LJ units. The analytic curve was estimated by numerically calculating the first terms of Eq. \eqref{corr_sphere_temporal_decay} ($l\in[0,30],k\in[1,30]$) and adding the asymptotic constant given by Eq. \eqref{longtimes_sphere0}. } \label{fig:Sphere_sim_fit}
\end{figure}

The spherical geometry simulations follow a very similar protocol to the cylindrical geometry simulations. 
The number of particles was set to $N=28371$ and the radius of the sphere was set to $R=20.47$  resulting in particle density of $\rho=0.79\sigma^{-3}$. The temperature, integration timestep and the total integration time was identical to the cylindrical geometry simulation. The LJ 9/3 potential is applied across the whole surface of the sphere to confine the particles to its interior. A comparison between the simulation results and the analytic prediction is presented at Fig. \ref{fig:Sphere_sim_fit}.
The analytic curves were estimated by numerically calculating the first terms of Eq. \eqref{corr_sphere_temporal_decay} ($l\in[0,30],k\in[1,30]$) and adding the asymptotic constant given by Eq. \eqref{longtimes_sphere0}. The MD simulation results agree well with the analytic predictions.

